\documentclass[11pt]{article}

\usepackage{authblk}

\usepackage[left=3.5cm, right=3.5cm, top=3cm, bottom=3.5cm]{geometry}
\usepackage{leftidx}
\usepackage{graphicx}
\usepackage[round]{natbib}
\usepackage{parskip}

\usepackage{algorithm}
\usepackage{algorithmic}

\usepackage{amsfonts}
\usepackage{amsmath}

\usepackage{latexsym}

\usepackage{subfigure}

\begin{document}

\newtheorem{lemma}{Lemma}[section]
\newtheorem{prop}{Proposition}[section]
\newtheorem{theo}{Theorem}[section]
\newtheorem{corr}{Corollary}[section]
\newtheorem{defi}{Definition}[section]

\newcommand{\Perp}{\perp \! \! \! \perp}

\newcommand{\bx}{{\bf X}}
\newcommand{\by}{{\bf Y}}
\newcommand{\bz}{{\bf Z}}
\newcommand{\PP}{\mathbb{P}}
\newcommand{\EE}{\mathbb{E}}
\newcommand{\eps}{\varepsilon}
\newcommand{\Var}{\mathrm{Var}}
\newcommand{\Cov}{\mathrm{Cov}}
\newcommand{\argmin}{\mathop{\arg \min}\limits}
\newcommand{\argmax}{\mathop{\arg \max}\limits}
\newcommand{\R}{\mathbb{R}}
\newcommand{\rank}{\mathrm{rank}}
\newcommand{\hrho}{\hat{\rho}}
\newcommand{\hSigma}{\hat{\Sigma}}
\newcommand{\bep}{{\boldsymbol \eps}}

\title{Correlated variables in regression:\\
clustering and sparse estimation} 

\author[1]{Peter B\"uhlmann}
\author[1]{Philipp R\"utimann}
\author[1]{Sara van de Geer}
\author[2]{Cun-Hui Zhang}
\affil[1]{Seminar for Statistics, ETH Zurich}
\affil[2]{Department of Statistics, Rutgers University}

\maketitle 

\begin{abstract}
We consider estimation in a high-dimensional linear model with strongly
correlated variables. We propose to cluster the
variables first and do subsequent sparse estimation such as
the Lasso for
cluster-representatives or the
group Lasso based on the structure from the clusters. Regarding the first
step, we present a novel and bottom-up agglomerative clustering algorithm based
on canonical correlations, and we show that it finds an optimal
solution and is statistically consistent. We also present some theoretical
arguments that canonical 
correlation based clustering leads to a better-posed compatibility constant
for the design matrix which ensures identifiability and an oracle
inequality for the group Lasso. Furthermore, we discuss circumstances where
cluster-representatives and using the Lasso as subsequent estimator leads to
improved results for prediction and  
detection of variables. We complement the theoretical analysis with various 
empirical results. 
\end{abstract}


\section{Introduction}

High-dimensional regression is used in many fields of
applications nowadays where the number of covariables $p$ greatly exceeds
sample size $n$, i.e., $p \gg n$. We focus here on the simple yet useful
high-dimensional linear model 
\begin{eqnarray}\label{mod1}
\by = \bx \beta^0 + {\boldsymbol \eps},
\end{eqnarray}
with univariate $n \times 1$ response vector $\by$, $n \times p$ design
matrix $\bx$, $p \times 1$ true underlying coefficient vector $\beta^0$ and
$n \times 1$ error vector ${\boldsymbol \eps}$. Our primary goal is to do variable
screening for the active
set, i.e., the support of  $\beta^0$, denoted by $S_0 = \{j;\ \beta^0_j \neq 0,\
j=1,\ldots p\}$: we want to have a statistical procedure
$\hat{S}$ such that with high probability, $\hat{S} \supseteq S_0$ (and
$|\hat{S}|$ not too large). In the 
case where $p \gg n$, the obvious difficulties are due to
(near) non-identifiability. While some positive
results have been shown under some assumptions on the design $\bx$, see
the paragraph below, high 
empirical correlation between variables or near linear dependence among a
few variables remain as notorious problems which are 
often encountered in many applications. Examples 
include genomics where correlation and the degree of linear dependence is
high within a group of genes sharing the same biological pathway
\citep{SC03}, or genome-wide association studies where SNPs are highly
correlated or linearly dependent within segments of the DNA sequence
\citep{balding06}.  

An important line of research to infer the active set $S_0$ or for variable
screening has been
developed in the past using the Lasso \citep{T96} or versions thereof
\citep{zou06,Meinshausen:05,zouli08}.  
Lasso-type methods have proven to be successful in a range of
practical problems. From a theoretical perspective, their properties for
variable selection and screening have been established assuming various
conditions on the design matrix $\bx$, such as the neighborhood stability
or irrepresentable condition \citep{mebu06,zhaoyu06}, and various forms of
``restricted'' eigenvalue conditions, see
\cite{vandeGeer:07a, zhang2008sparsity, MY08, brt09, van2009conditions,
  sunzhang11}. Despite of these positive findings, situations where high
empirical correlations between covariates or near linear dependence
among a few covariables occur cannot be
handled well with the Lasso: the Lasso tends to select only one variable
from the group of correlated or nearly linearly dependent variables, even
if many or all of these variables belong to the active set $S_0$. The
elastic net \citep{zouhastie05}, OSCAR \citep{bonre08} and ``clustered
Lasso'' \citep{she10} have been proposed to 
address this problem but they do not explicitly take correlation-structure
among the variables into account and still exhibit difficulties
when groups of variables are nearly linearly dependent. A sparse Laplacian
shrinkage estimator has  
been proposed \citep{huangetal11} and proven to select a correct set of
variables  under certain regularity conditions. 
However, the sparse Laplacian shrinkage estimator is geared toward the case
where highly correlated variables have similar predictive effects (which we
do not require here) and its selection consistency theorem  necessarily
requires a uniform lower bound for the nonzero signals above an inflated
noise level due to model uncertainty.   

We take here the point of view that we want to avoid false negatives, i.e.,
to avoid not selecting an active variable from $S_0$: the price to pay for
this is an increase in false positive selections. From a practical point of
view, it 
can be very useful to have a selection method $\hat{S}$ which includes
\emph{all} variables from a group of nearly linearly independent variables
where at least one of them is active. Such a procedure is often a good
screening method, when measured by $|\hat{S} \cap S_0|/|S_0|$ as a
function of $|\hat{S}|$. The desired
performance can 
be achieved by clustering or grouping the variables first and then selecting
whole clusters instead of single variables. 

\subsection{Relation to other work and new contribution}
The idea of clustering or grouping variables and then pursuing model
fitting is not new, of course. Principal component regression
\citep[cf.]{kendall57} is among the earliest proposals, and \cite{H00} have
used principal component analysis in order to find a set of highly
correlated variables where the clustering can be supervised by a response
variable. Tree-harvesting 
\citep{HTBB01} is another proposal which uses supervised learning methods 
to select groups of predictive variables formed by hierarchical
clustering. An algorithmic approach, simultaneously performing the
clustering and supervised model fitting, was proposed by \cite{DB04}, and also
the OSCAR method \citep{bonre08} does such simultaneous grouping
and supervised model fitting. 

Our proposal differs from previous work in various aspects. 
We primarily propose to use canonical correlation for
clustering the variables 
as this reflects the notion of linear dependence among variables: and it is
exactly this notion of linear dependence which causes the identifiability
problems in the linear model in (\ref{mod1}). Hence, this is conceptually a
natural strategy for clustering variables when having the aim to address
identifiability problems with variable selection in the linear model
(\ref{mod1}). We present in Section \ref{subsec.cancluster} an agglomerative
hierarchical clustering method using canonical correlation, and we prove
that it finds the finest clustering which satisfies the criterion function
that between group canonical correlations are smaller than a threshold and
that it is statistically consistent. Furthermore, we
prove in Section \ref{subsec.theoryCGL} that the construction of groups
based on canonical correlations leads to well-posed behavior of the group
compatibility constant of the design matrix $\bx$ which ensures
identifiability and an oracle inequality for the group Lasso
\citep{YI06}. The latter is a natural choice for estimation and cluster
selection; another possibility is to use the Lasso for cluster
representatives. We analyze both of these methods: this represents a
difference to earlier work where at the time, such high-dimensional
estimation techniques have been less or not established at all. 

We present some supporting theory in Section \ref{sec.theory},
describing circumstances where clustering and subsequent estimation
improves over standard Lasso without clustering, and the 
derivations also show the limitations of such an approach. This sheds light
for what kind of models and scenarios the commonly used two-stage approach in
practice, consisting of clustering variables first and subsequent estimation, is
beneficial. Among the favorable scenarios which we will examine for the
latter approach are: (i) high within cluster correlation and weak 
between cluster correlation with potentially many active variables per
cluster; and (ii) at most one active variable per 
cluster where the clusters are tight (high within correlation) but not
necessarily assuming low between cluster correlation. 
Numerical results which complement the theoretical results are presented in
Section \ref{sec.empres}.   

\section{Clustering covariables}\label{sec.cluster}

Consider the index set $\{1,\ldots ,p\}$ for the covariables in
(\ref{mod1}). In the sequel, we denote by $x^{(j)}$ the $j$th component of
a vector $x$ and by $X^{(G)} = \{X^{(j)};\ j \in G\}$ the group of
variables from a cluster $G \subseteq \{1,\ldots ,p\}$. The goal is to
find a partition ${\cal G}$ into disjoint clusters $G_1,\ldots ,G_q$:
${\cal G} = \{G_1,\ldots ,G_q\}$ with $\cup_{r=1}^q G_r = 
\{1,\ldots ,p\}$ and $G_r \cap G_{\ell} = \emptyset\ (r \neq \ell)$. The
partition 
${\cal G}$ should then satisfy certain criteria. 

We propose two methods for clustering the variables, i.e., finding a
suitable partition: one is a novel approach  
based on canonical correlations while the other uses standard correlation
based hierarchical clustering.  

\subsection{Clustering using canonical
  correlations}\label{subsec.cancluster}  

For a partition ${\cal G} = \{G_1,\ldots ,G_q\}$ as above, we consider:
\begin{eqnarray*}
\hat{\rho}_{\mathrm{max}}({\cal G}) =
\max\{\hat{\rho}_{\mathrm{can}}(G_r,G_{\ell});\ r, \ell \in \{1,\ldots
,q\},\ r \neq \ell \}.
\end{eqnarray*}
Here, $\hat{\rho}_{\mathrm{can}}(G_r,G_{\ell})$
denotes the empirical canonical correlation \citep[cf.]{anderson84} between the
variables from $X^{(G_r)}$ and 
$X^{(G_{\ell})}$. (The empirical canonical correlation is always non-negative).  
A clustering with $\tau$-separation between clusters is
defined as:  
\begin{eqnarray}\label{crit.cc}
\hat{{\cal G}}(\tau) = \mbox{a partition $\hat{{\cal G}}$ of
  $\{1,\ldots ,p\}$ such that}\ \hat{\rho}_{\mathrm{max}}(\hat{{\cal G}}) \le
\tau\ (0 < \tau < 1). 
\end{eqnarray}
Not all values of $\tau$ are feasible: if $\tau$ is
too small, then there is no partition which would satisfy
(\ref{crit.cc}). For this reason, we define the canonical correlation of
all the variables with the empty set of variables as zero: hence, the
trivial partition ${\cal G}_{\mathrm{single}}$ consisting of the single
cluster $\{1,\ldots ,p\}$ has $\hat{\rho}_{\mathrm{max}}({\cal 
  G}_{\mathrm{single}}) = 0$ which satisfies (\ref{crit.cc}). The fact that
$\tau$ may not be feasible (except with ${\cal G}_{\mathrm{single}}$) can be
understood from the view point that coarse partitions do not necessarily
lead to smaller values of $\hat{\rho}_{\mathrm{max}}$: for example, when $p
\gg n$ and if
$\mathrm{rank}(X^{(G_r \cup G_{\ell})}) = n$, which would 
typically happen if $|G_r \cup G_{\ell}| > n$, then
$\hat{\rho}_{\mathrm{can}}(G_r,G_{\ell}) = 1$. In general, clustering with
$\tau$-separation does not have a unique solution.  
For example, if $\hat{{\cal G}}(\tau)=\{G_1,\ldots,G_q\}$ is a clustering with
$\tau$-separation  
and $\{G_{r;k}, k=1,\ldots,q_r\}$ is a nontrivial partition of $G_r$ with 
$\max_{1\le k_1<k_2\le q_r} \hat{\rho}_{\mathrm{can}}(G_{r;k_1},G_{r;k_2})\le\tau$, then 
$\{G_1,\ldots,G_{r-1},G_{r;k}, k=1,\ldots,q_r,G_{r+1},\ldots,G_q\}$ is a
strictly finer clustering with $\tau$-separation, see also Lemma \ref{lemm1}
below. 
The non-uniqueness of clustering with $\tau$-separation 
motivates the following definition of the finest clustering with
$\tau$-separation. 
A clustering with $\tau$-separation between clusters, say $\hat{{\cal G}}(\tau)$,
is finest if every other clustering with $\tau$-separation is strictly
coarser than $\hat{{\cal 
  G}}(\tau)$, and we denote such a finest clustering with $\tau$-separation
by
\begin{eqnarray*}
\hat{{\cal G}}_{\mathrm{finest}}(\tau).
\end{eqnarray*}
The existence and uniqueness of the finest clustering with $\tau$-separation 
are provided in Theorem \ref{th1} below. 

\begin{algorithm}[b]
\begin{algorithmic}[1]
\STATE Start with the single variables as $p$ clusters (nodes at the bottom of a
tree). Set $b=0$
\REPEAT
\STATE Increase $b$ by one. Merge the two clusters having highest canonical
correlation. 
\UNTIL Criterion (\ref{crit.cc}) is satisfied. 
\end{algorithmic}
\caption{Bottom-up, agglomerative hierarchical clustering using canonical correlations.}\label{alg1}
\end{algorithm}

A simple hierarchical bottom-up agglomerative clustering (without the
need to 
define linkage between clusters) can be used as an estimator $\hat{\cal
  G}(\tau)$ which satisfies (\ref{crit.cc}) and which is finest: the
procedure is described in Algorithm \ref{alg1}.  
\begin{theo}\label{th1}
The hierarchical bottom-up agglomerative clustering Algorithm \ref{alg1} leads
to a partition $\hat{\cal
  G}(\tau)$ which satisfies (\ref{crit.cc}). If $\tau$ is not feasible with
a nontrivial partition, the solution is the   
coarsest partition $\hat{\cal
  G}(\tau) = {\cal G}_{\mathrm{single}} = \{G\}$ with $G = \{1,\ldots
,p\}$. 

Furthermore, if $\tau$ is feasible with a nontrivial partition, 
the solution ${\hat {\cal G}}(\tau) = \hat{\cal
  G}_{\mathrm{finest}}(\tau)$ is the finest clustering with
$\tau$-separation.  
\end{theo}
A proof is given in Section \ref{sec.proofs}. Theorem \ref{th1} describes
that a bottom-up greedy strategy leads to an optimal solution. 

We now present consistency of the clustering Algorithm \ref{alg1}. Denote
the population canonical correlation between 
$X^{(G_r)}$ and $X^{(G_{\ell})}$ by $\rho_{\mathrm{can}}(G_r,G_{\ell})$,
and the maximum population canonical correlation by $\max_{r 
  \neq \ell} \rho_{\mathrm{can}}(G_r,G_{\ell})$, respectively, for
a partition ${\cal G}= \{G_1,\ldots,G_q\}$ of $\{1,\ldots,p\}$. 
As in (\ref{crit.cc}), a partition ${\cal G}(\tau)$ is a population clustering
with $\tau$-separation  
if $\rho_{\max}(\cal G) \le \tau$. The finest population clustering with
$\tau$-separation,  denoted by ${\cal G}_{\mathrm{finest}}(\tau)$, is the
one which is finer 
than any other population clustering with $\tau$-separation.  
With the convention $\rho_{\max}({\cal G}_{\mathrm{single}})=0$, the
existence and uniqueness of the finest population clustering with
$\tau$-separation follows from Theorem \ref{th1}. In fact, the hierarchical
bottom-up  
agglomerative clustering Algorithm \ref{alg1} yields the finest population
clustering ${\cal G}_{\mathrm{finest}}(\tau)$ with $\tau$-separation   
if the population canonical correlation is available and used in the algorithm. 

Let ${\cal G}_{\mathrm{finest}}(\tau)$ be the finest population clustering
with $\tau$-separation 
and ${\hat {\cal G}}(\tau)$ be the sample clustering with $\tau$-separation 
generated by the hierarchical bottom-up agglomerative clustering Algorithm
\ref{alg1} based on the design matrix $\bx$. 
The following theorem provides a sufficient condition for the consistency
of ${\hat {\cal G}}(\tau)$ in the Gaussian model 
\begin{eqnarray}\label{Gaussian}
X_1,\ldots ,X_n\ \mbox{i.i.d.}\ \sim {\cal N}_p(0,\Sigma). 
\end{eqnarray}
For any given $t>0$ and positive integers $q$ and $d_1,\ldots,d_q$, define 
\begin{eqnarray}\label{eq:a1}
t_r = \sqrt{d_r/n}+\sqrt{(2/n)(t+\log(q(q+1))},\ 
\Delta_{r,\ell}^* = \frac{3(t_r\wedge t_{\ell})+(t_r\vee
  t_{\ell})}{(1-t_r)(1-t_{\ell})}.  
\end{eqnarray}
\begin{theo}\label{th-a} Consider $\bx$ from (\ref{Gaussian}) 
and ${\cal G}^0 = \{G_1,\ldots, G_q\}$ a partition of $\{1,\ldots,p\}$. 
Let $t>0$ and $d_r=\rank(\Sigma_{G_r,G_r})$. Define  
$\Delta^*_{r,\ell}$ by (\ref{eq:a1}). 
Suppose 
\begin{eqnarray}\label{th-a-1}
& &\max_{1\le r<\ell\le
  q}\big\{\rho_{\mathrm{can}}(G_r,G_{\ell})+\Delta^*_{r,\ell}\big\}\le\tau_-  
\le \tau_+ \nonumber\\
&<& \min_{1\le r\le q}\min_{\{G_{r;k}\}}\big\{\max_{k_1<k_2}\rho_{\mathrm{can}}(G_{r;k_1},G_{r;k_2}) - \Delta^*_{r,r}\big\}, 
\end{eqnarray}
where $\min_{\{G_{r;k}\}}$ is taken over all nontrivial partitions $\{G_{r;k}, k\le q_r\}$ of $G_r$. Then, 
${\cal G}^0 = {\cal G}_{\mathrm{finest}}(\tau)$ is the finest population
clustering with $\tau$-separation for all $\tau_-\le \tau \le\tau_+$, and 
\begin{eqnarray*}
\PP[{\hat{\cal G}}(\tau) = {\cal G}_{\mathrm{finest}}(\tau),\ \forall
\tau_- \le \tau \le \tau_+\Big] \ge 1 - \exp(-t).  
\end{eqnarray*}
\end{theo}

A proof is given in Section \ref{sec.proofs}. We note that $t =
\sqrt{\log(p)}$ leads to $t_r 
  \asymp \sqrt{\rank (\Sigma_{G_r,G_r})/n}+\sqrt{\log(p)/n}$: this is small
  if $\rank(\Sigma_{G_r,G_r}) = o(n)$ and $\log(p) = o(n)$,  and then,
  $\Delta^*_{r,\ell}$ is small as well (which means that the probability
  bound becomes $1 - p^{-1} \to 1\ (p \to \infty)$ (or $p \ge n \to \infty$).

\begin{figure}[!htb]
\begin{center}
\includegraphics[scale=0.35]{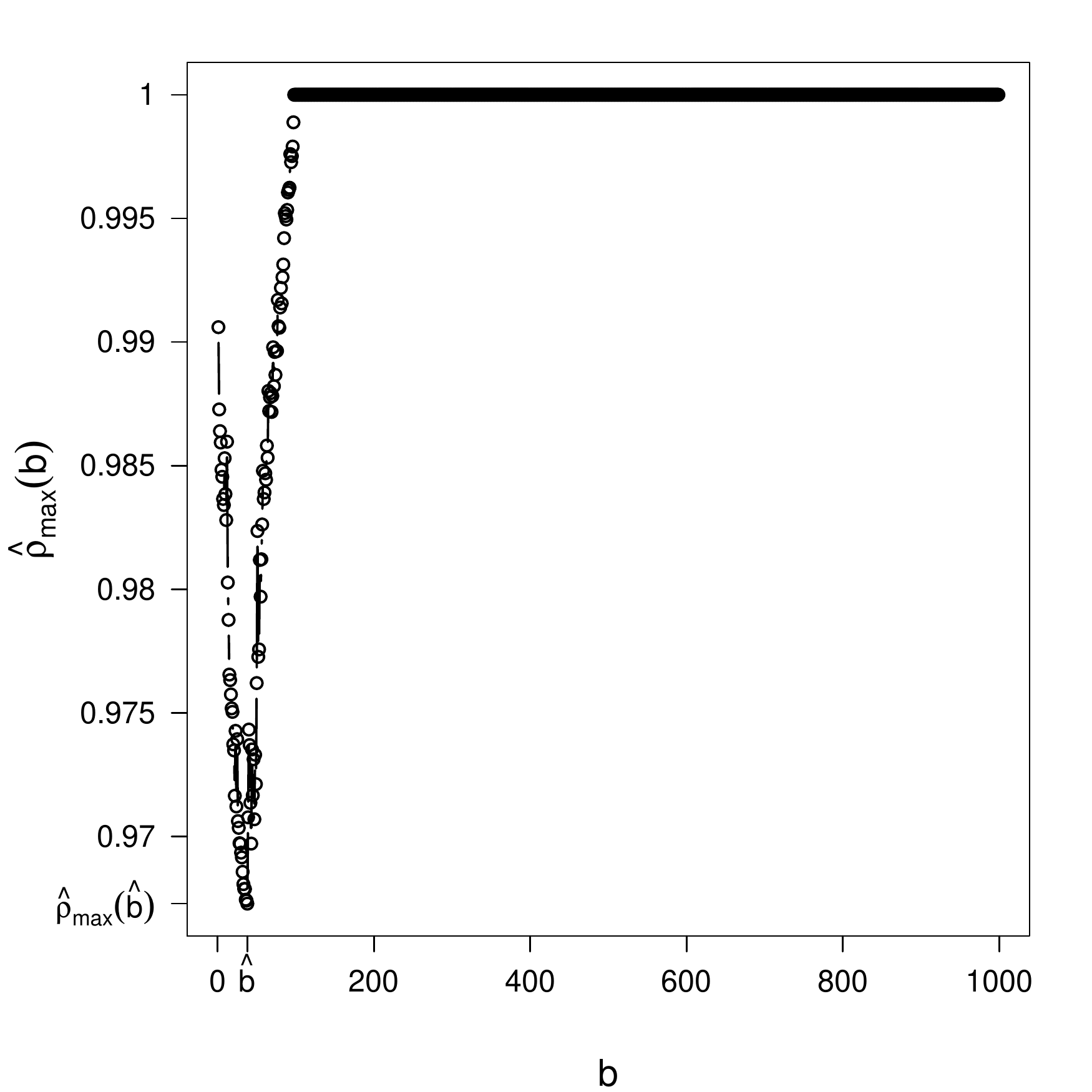}
\caption{Path of $\hat{\rho}_{\mathrm{max}}(b)$ as a function of the
iterations $b$: from real data described in Section \ref{realdata} with $p
= 1000$ and $n=71$.} 
\label{resol}
\end{center}
\end{figure}

The parameter $\tau$ in (\ref{crit.cc}) needs to be chosen. We advocate the
use of the minimal resulting $\tau$. This can be easily implemented: we run
the bottom-up agglomerative clustering Algorithm \ref{alg1} and record in every
iteration the maximal canonical correlation between clusters, denoted by
$\hat{\rho}_{\max}(b)$ where $b$ is the iteration number. We then use the
partition corresponding to the iteration
\begin{eqnarray}\label{cutoffb}
\hat{b} = \argmin_b \hat{\rho}_{\mathrm{max}}(b).
\end{eqnarray}
A typical path of $\hat{\rho}_{\mathrm{max}}(b)$ as a function of the
iterations $b$, the choice $\hat{b}$ and the corresponding minimal
$\hat{\rho}_{\max}(\hat{b})$ are shown in Figure \ref{resol}. 

We conclude that the hierarchical bottom-up agglomerative clustering
Algorithm \ref{alg1} with the rule in (\ref{cutoffb}) is fully data-driven,
and there is no need to define linkage between clusters. 

\subsection{Ordinary hierarchical clustering}\label{sec.ohclust}

As an alternative to the clustering method in Section \ref{subsec.cancluster}, we
consider in Section \ref{realdata} ordinary hierarchical agglomerative clustering
based on the dissimilarity matrix $D$ with entries $D_{r,\ell} = 1 -
|\hat{\rho}(X^{(r)},X^{(\ell)})|$, where $\hat{\rho}(X^{(r)},X^{(\ell)})$ denotes
the sample correlation between $X^{(r)}$ and $X^{(\ell)}$. We choose
average-linkage as dissimilarity between clusters. 

As a cutoff for determining the number of clusters, we proceed according to
an established principle. In every clustering iteration $b$, proceeding in an
agglomerative way, we record the new value $h_b$ of the corresponding linkage
function from the merged clusters (in iteration $b$): we then use the
partition corresponding to the iteration
\begin{eqnarray*}
\hat{b} = \argmax_b (h_{b+1} - h_b).
\end{eqnarray*}  

\section{Supervised selection of clusters}\label{sec.selection}

From the design matrix $\bx$, we infer the clusters $G_1,\ldots ,G_q$ as
described in Section \ref{sec.cluster}. We select the variables in the
linear model (\ref{mod1}) in a group-wise fashion where all variables from a
cluster $G_r$ ($r=1,\ldots ,q$) are selected or not: this is denoted by
\begin{eqnarray*}
\hat{S}_{\mathrm{cluster}} = \{r;\ \mbox{cluster $G_r$ is selected},\
r=1,\ldots ,q\}.
\end{eqnarray*}
The selected set of
variables is then the union of the selected clusters:
\begin{eqnarray}\label{select-var}
\hat{S} = \cup_{r \in \hat{S}_{\mathrm{cluster}}} G_r.
\end{eqnarray}
We propose two methods for selecting the clusters, i.e., two different
estimators $\hat{S}_{\mathrm{cluster}}$.

\subsection{Cluster representative Lasso (CRL)}

For each cluster we consider the representative
\begin{eqnarray*}
\bar{\bx}^{(r)} = \frac{1}{|G_r|} \sum_{j \in G_r} \bx^{(j)},\ r = 1,\ldots
,q,
\end{eqnarray*}
where $\bx^{(j)}$ denotes the $j$th $n \times 1$ column-vector of $\bx$.  
Denote by $\bar{\bx}$ the $n \times q$ design matrix whose $r$th column is
given by $\bar{\bx}^{(r)}$. We use the
Lasso based on the response $\by$ and the design matrix $\bar{\bx}$:
\begin{eqnarray*}
\hat{\beta}_{\mathrm{CRL}} = \argmin_{\beta} (\|\by - \bar{\bx}
\beta\|_2^2/n + \lambda_{\mathrm{CRL}} \|\beta\|_1).
\end{eqnarray*}
The selected clusters are then given by 
\begin{eqnarray*}
\hat{S}_{\mathrm{cluster,CRL}} =
\hat{S}_{\mathrm{cluster,CRL}}(\lambda_{\mathrm{CRL}}) = \{r;\ \hat{\beta}_{\mathrm{CRL},r}(\lambda_{\mathrm{CRL}}) \neq 0,\ r=1,\ldots ,q\},
\end{eqnarray*}
and the selected variables are obtained as in (\ref{select-var}), denoted as
\begin{eqnarray*}
\hat{S}_{\mathrm{CRL}} = \hat{S}_{\mathrm{CRL}}(\lambda_{\mathrm{CRL}}) = \cup_{r
  \in \hat{S}_{\mathrm{cluster,CRL}}} G_r.
\end{eqnarray*}

\subsection{Cluster group Lasso (CGL)} 
 
Another obvious way to select clusters is given by the group Lasso. We
partition the vector of 
regression coefficients according to the clusters: $\beta = (\beta_{G_1},\ldots
,\beta_{G_q})^T$, where $\beta_{G_r} = (\{\beta_j;\ j \in G_r\})^T$. The
cluster group Lasso is defined as
\begin{eqnarray}\label{grouplasso}
\hat{\beta}_{\mathrm{CGL}} = \argmin_{\beta} \|\by
- \bx \beta\|_2^2/n + \lambda_{\mathrm{CGL}} \sum_{r=1}^q w_r
\|\bx^{(G_r)}\beta_{G_r}\|_2 n^{-1/2},
\end{eqnarray}
where $w_r$ is a multiplier, typical pre-specified as $w_r =
\sqrt{|G_r|}$. It is well known that the group Lasso enjoys a group
selection property where either $\hat{\beta}_{\mathrm{CGL},G_r} \neq 0$
(all components 
are different from zero) or $\hat{\beta}_{\mathrm{CGL},G_r} \equiv 0$ (the
zero-vector). We note that the estimator in (\ref{grouplasso}) is different
from using the usual penalty $\lambda \sum_{r=1}^q w_r \|\beta_{G_r}\|_2$:
the penalty in (\ref{grouplasso}) is termed as ``groupwise prediction
penalty'' in \citet[Sec.4.5.1]{pbvdg11}: it has nice parameterization invariance
properties, and it is a much more appropriate penalty when $\bx^{(G_r)}$
exhibits strongly correlated columns. 

The selected clusters are then given by 
\begin{eqnarray*}
\hat{S}_{\mathrm{cluster,CGL}} =
\hat{S}_{\mathrm{cluster,CGL}}(\lambda_{\mathrm{CGL}}) =\{r;\ 
\hat{\beta}_{\mathrm{CGL},G_r}(\lambda_{\mathrm{CGL}}) \neq 0,\
r=1,\ldots ,q\},
\end{eqnarray*}
and the selected variables are as in (\ref{select-var}):
\begin{eqnarray*}
\hat{S}_{\mathrm{CGL}} = \hat{S}_{\mathrm{CGL}}(\lambda_{\mathrm{CGL}}) = \cup_{r
  \in \hat{S}_{\mathrm{cluster,CGL}}} G_r = \{j;\
\hat{\beta}_{\mathrm{CGL},j} \neq 0,\ j=1,\ldots ,p\},
\end{eqnarray*}
where the latter equality follows from the group selection property of the
group Lasso. 

\section{Theoretical results for cluster Lasso methods}\label{sec.theory}  

We provide here some supporting theory, first for the cluster group Lasso (CGL)
and then for the cluster representative Lasso (CRL).

\subsection{Cluster group Lasso (CGL)}\label{subsec.theoryCGL}

We will show first that the compatibility constant of the design matrix $\bx$ is
well-behaved if the canonical correlation between groups is small, i.e., a
situation which the clustering Algorithm \ref{alg1} is exactly aiming for. 

The CGL method is based on the whole design matrix $\bx$, and we can write the 
model in group structure form where we denote by $\bx^{(G_r)}$ the $n
\times |G_r|$ design matrix with variables $\{\bx^{(j)};\ j \in G_r\}$:
\begin{eqnarray}\label{repr.grouplasso}
\by = \bx \beta^0 + {\boldsymbol \eps} = \sum_{r=1}^q
\bx^{(G_r)} \beta^0_{G_r} + {\boldsymbol \eps},
\end{eqnarray}
where $\beta^0_{G_r} = (\{\beta_j^0;\ j \in G_r\})^T$. We denote by
$S_{0,\mathrm{Group}} = \{r;\ \beta_{G_r}^0 \neq 0,\ r=1,\ldots ,q\}$. 

We introduce now a few other notations and definitions. 
We let $m_r := | G_r |$, $r=1 , \ldots , q$, denote the group sizes, and
define the average group size 
$$\bar m := {1 \over q} \sum_{r=1}^q m_r  , $$ 
and the average size of active groups
$$\bar m_{S_{0, {\rm Group}}} := {1 \over | S_{0, {\rm Group}}| } \sum_{r
  \in S_{0, {\rm Group} }} m_r . $$
Furthermore, for any $S \subseteq \{ 1 , \ldots , q\}$,
we let $\bx^{(S)}$ be the design matrix containing the variables in $\cup_{r \in
  S} G_r$. Moreover, define
$$\| \beta_S \|_{2,1} := \sum_{r \in S} \| \bx^{(G_r)}  \beta_{G_r} \|_2
\sqrt {m_r/ \bar m} . $$ 
We denote in this section by 
$$ \hat \Sigma_{r,\ell} := (\bx^{(G_r)})^T \bx^{(G_{\ell})} / n , \ r, \ell
\in \{ 1 , \ldots , q \} . $$ 
We assume that each $\hat \Sigma_{r,r}$ is non-singular (otherwise one may
use generalized inverses) 
and we write, assuming here for notational clarity that $S_{0, {\rm Group}
} = \{ 1 , \ldots , s_0 \}$ ($s_0 = | S_{0, {\rm Group}} |$) is the set of
indices of the first $s_0$ groups: 
\begin{eqnarray*}
\hat R_{S_{0, {\rm Group}} }  := \left(
\begin{array}{cccc}
I & \hat{\Sigma}_{1,1}^{-1/2} \hat \Sigma_{1,2} \hat \Sigma_{2,2}^{-1/2} &
    \cdots  
    & \hat \Sigma_{1,1}^{-1/2} \hat \Sigma_{1,s_0} \hat
    \Sigma_{s_0,s_0}^{-1/2} \\
\hat \Sigma_{2,2}^{-1/2} \hat \Sigma_{2,1} \hat \Sigma_{1,1}^{-1/2} & I & \cdots & \hat \Sigma_{2,2}^{-1/2} \hat \Sigma_{2,s_0} \hat \Sigma_{s_0,s_0}^{-1/2} \\
\vdots & \vdots & \ddots & \vdots \\
\hat \Sigma_{s_0,s_0}^{-1/2} \hat \Sigma_{s_0,1} \hat \Sigma_{1,1}^{-1/2} &
\hat \Sigma_{s_0,s_0}^{-1/2} \hat \Sigma_{s_0,2} \hat \Sigma_{2,2}^{-1/2} &
\cdots & I    
\end{array}  \right).
\end{eqnarray*}
The group compatibility constant given in \cite{pbvdg11} is a value
$\phi_{0, {\rm Group}}^2 ({\bf X})$ 
that satisfies
\begin{eqnarray*}
\phi_{0, {\rm Group} }^2  ({\bf X}) \le \min \biggl \{ { \bar m_{S_0} | S_{0, {\rm Group}} |   \over \bar m} { \| \bx \beta \|_2^2 \over \| 
\beta_{S_{0, {\rm Group}}}  \|_{2,1}^2 };\ \| \beta_{S_{0, {\rm Group}}^c}
\|_{2,1} \le \| \beta_{S_{0, {\rm Group}}} \|_{2,1}  \biggr \}.
\end{eqnarray*}
The constant 3 here is related to the condition $\lambda \ge 2 \lambda_0$
in Proposition \ref{prop.grLasso} below: in general it can be taken as $(
\lambda + \lambda_0 )/(\lambda - \lambda_0)$. 
\begin{theo} \label{groupcompatibility.lemma}
Suppose that $\hat R_{S_{0, {\rm Group}}} $ has smallest eigenvalue $\hat
\Lambda_{\rm min}^2 > 0 $. Assume moreover the incoherence conditions:
\begin{eqnarray*}
& &\rho:= \max_{r \in S_0 , \ell \in S_0^c } 
{  \bar m \over \sqrt {m_r m_{\ell}}} { \hat \rho_{\rm can} (G_r,
  G_{\ell})} \le  C {   \hat \Lambda_{\rm min}^2 \bar m / \bar m_{S_{0, {\rm
        Group}}}  \over 3  
| S_{0, {\rm Group}} |  }\ \ \mbox{for some}\ 0 < C < 1,\\
& &\rho_{S_{0, {\rm Group}}} := \max_{r,\ell \in S_{0, {\rm Group}},\ r
  \neq \ell} {  \bar m \over \sqrt {m_r m_{\ell}}} { \hat \rho_{\rm can} (G_r,
  G_{\ell})} < { 1 \over  
| S_{0, {\rm Group}}| }.
\end{eqnarray*}
Then, the group Lasso compatibility holds with compatibility constant
\begin{eqnarray*}
\phi_{0, {\rm Group} }^2  ({\bf X}) &\ge&  { \biggl (  \hat \Lambda_{\rm min}^2 \bar m / \bar m_{S_{0, {\rm Group}}}   - 3 | S_{0, {\rm Group}}|  \rho \biggr )^2 /
\biggl  ({ \hat \Lambda_{\rm min}^2 \bar m^2 / \bar m_{S_{0, {\rm
        Group}}}^2 }} \biggr ) \\
&\ge & (1 - C)^2 \hat \Lambda_{\rm min}^2 \ge (1 - C)^2 (1 - |S_{0, {\rm
        Group}}| \rho_{S_{0, {\rm Group}}})\\
&>& 0.
\end{eqnarray*}
\end{theo} 
A proof is given in Section \ref{sec.proofs}. We note that small enough
canonical correlations between groups ensure the incoherence assumptions
for $\rho$ and $\rho_{S_{0, {\rm Group}}}$, and in turn that the group
Lasso compatibility condition holds. The canonical
correlation based clustering Algorithm \ref{alg1} is tailored for this
situation. 

\bigskip\noindent
\textbf{Remark 1.} One may in fact prove a group version of Corollary 7.2
in \cite {pbvdg11} which says that  under the eigenvalue and
incoherence condition of Theorem \ref{groupcompatibility.lemma}, a group irrepresentable condition holds. This in turn implies that, with large probability, the
group Lasso will have no false positive selection of groups, i.e., one has a group version of the result of Problem 7.5 in \cite{pbvdg11}. 

\bigskip
Known theoretical results can then be applied for proving an oracle
inequality of the group Lasso. 
\begin{prop}\label{prop.grLasso}
Consider model (\ref{mod1}) with fixed design $\bx$ and Gaussian error
${\boldsymbol \eps} \sim {\cal N}_n(0,\sigma^2 I)$. Let 
\begin{eqnarray*}
\lambda_0   = \sigma \frac{2}{\sqrt{n}} \sqrt{1 + \sqrt{\frac{4t + 4
      \log(p)}{m_{\mathrm{min}}}} + \frac{4t + 4
    \log(p)}{m_{\mathrm{min}}}},
\end{eqnarray*}
where $m_{\mathrm{min}} = \min_{r=1,\ldots ,q} |m_r|$. Assume that
$\hat{\Sigma}_{r,r}$ is non-singular,for all $r=1,\ldots ,q$. 
Then, for the group
Lasso estimator $\hat{\beta}(\lambda)$ in (\ref{grouplasso}) with $\lambda \ge 2 
\lambda_0$, and with probability at least $1- \exp(-t)$:
\begin{eqnarray}\label{groupadapt}
\|\bx(\hat{\beta}(\lambda) - \beta^0)\|_2^2/n + \lambda \sum_{r=1}^q
\sqrt{m_r} \|\hat{\beta}_{G_r}(\lambda) - \beta_{G_r}^0\|_2 \le 24
\lambda^2 \sum_{r \in S_{0,\mathrm{{Group}}}} m_r/\phi_{0,\mathrm{Group}}^2(\bx),
\end{eqnarray}
where $\phi_{0,\mathrm{Group}}^2(\bx)$ denotes the group compatibility
constant. 
\end{prop} 

Proof: We can invoke the result in \citet[Th.8.1]{pbvdg11}: using the
groupwise prediction penalty in 
(\ref{grouplasso}) leads to an equivalent formalization where we can
normalize to $\hat{\Sigma}_{r,r} = I_{m_r \times m_r}$. The requirement
$\lambda \ge 4 
\lambda_0$ in \citet[Th.8.1]{pbvdg11} can be relaxed to $\lambda \ge 2
\lambda_0$ since the model (\ref{mod1}) is assumed to be true, see also
\citet[p.108,Sec.6.2.3]{pbvdg11}.\hfill$\Box$  

\subsection{Linear dimension reduction and subsequent Lasso estimation}\label{sec.dimred}  
For a mean zero Gaussian random variable $Y \in \R$ and a mean zero
Gaussian random vector $X \in \R^p$, we 
can always use a random design Gaussian linear model representation: 
\begin{eqnarray}\label{mod.Grand}
& &Y = \sum_{j=1}^p \beta^0_j X^{(j)} + \eps,\nonumber\\
& &X \sim {\cal N}_p(0,\Sigma),\ \eps \sim {\cal N}(0,\sigma^2),
\end{eqnarray}
where $\eps$ is independent of $X$. 

Consider a linear dimension reduction 
\begin{eqnarray*} 
Z = A_{q \times p} X
\end{eqnarray*}
using a matrix $A_{q \times p}$ with $q < p$. We denote in the sequel by 
\begin{eqnarray*}
\mu_X = \EE[Y|X] = \sum_{j=1}^p \beta^0_j X^{(j)},\ \ \mu_Z = \EE[Y|Z] =
\sum_{r=1}^q \gamma^0_r Z^{(r)}. 
\end{eqnarray*}
Of particular interest is $Z = (\bar{X}^{(1)},\ldots ,\bar{X}^{(q)})^T$,
corresponding to the cluster representatives $\bar{X}^{(r)} = |G_r|^{-1}
\sum_{j \in G_r} X^{(j)}$.  
Due to the Gaussian assumption in (\ref{mod.Grand}), we can always
represent
\begin{eqnarray}\label{modZ}
Y = \sum_{r=1}^q \gamma^0_r Z^{(r)} + \eta = \mu_Z + \eta,
\end{eqnarray}
where $\eta$ is independent of $Z$. Furthermore, since $\mu_Z$ is the
linear projection of $Y$ on the linear span of $Z$ and also the linear
projection of $\mu_X$ on the linear span of $Z$, 
\begin{eqnarray*}
\xi^2 = \Var(\eta) = \Var(\eps + \mu_X - \mu_Z) = \sigma^2 + \EE[(\mu_X -
\mu_Z)^2].
\end{eqnarray*}

For the prediction error, when using the dimension-reduced $Z$ and for any
estimator $\hat{\gamma}$, we have, using that $\mu_X - \mu_Z$ is orthogonal
to the linear span of $Z$:
\begin{eqnarray}\label{pred.err}
\EE[(X^T\beta^0 - Z^T \hat{\gamma})^2] = \EE[(Z^T \gamma^0 - Z^T
\hat{\gamma})^2] + \EE[(\mu_X - \mu_Z)^2].
\end{eqnarray}
Thus, the total prediction error consists of an error due to estimation of
$\gamma^0$ and a squared bias term $B^2 = \EE[(\mu_X - \mu_Z)^2]$. The
latter has already appeared in the variance $\xi^2 = \Var(\eta)$ above. 

Let $\by,\ \bx$ and $\bz$ be $n$ i.i.d. realizations of the variables
$Y,X$ and $Z$, respectively. Then,
\begin{eqnarray*}
\by = \bx \beta^0 + {\boldsymbol \eps} = \bz \gamma^0 + {\boldsymbol \eta}.
\end{eqnarray*}
Consider the Lasso, applied to $\by$ and $\bz$, for estimating $\gamma^0$:
\begin{eqnarray*}
\hat{\gamma} = \mbox{argmin}_{\gamma} (\|\by - \bz \gamma\|_2^2/n + \lambda
\|\gamma\|_1).
\end{eqnarray*}
The cluster representative Lasso (CRL) is a special case with $n \times q$
design matrix $\bz = \bar{\bx}$. 
\begin{prop}\label{prop0}
Consider $n$ i.i.d. realizations from the model
(\ref{mod.Grand}). Let 
\begin{eqnarray*}
\lambda_0   = 2 \|\hat{\sigma}_{\bz}\|_{\infty} \xi \sqrt{\frac{t^2 + 2
    \log(q)}{n}}, 
\end{eqnarray*}
where $\|\hat{\sigma}_{\bz}\|_{\infty} = \max_{r=1,\ldots ,q}
\sqrt{(n^{-1} \bz^T \bz)_{rr}}$, and
$\xi^2 = \sigma^2 + \EE[(\mu_X - \mu_Z)^2]$. Then, for $\lambda \ge 2 
\lambda_0$ and with probability at least $1- 2\exp(-t^2/2)$, conditional on
$\bz$:
\begin{eqnarray*}
\|\bz(\hat{\gamma} - \gamma^0)\|_2^2/n + \lambda \|\hat{\gamma} -
\gamma^0\|_1 \le 4 \lambda^2 s(\gamma^0)/\phi_0^2(\bz),
\end{eqnarray*}
where $s(\gamma^0)$ equals the number of non-zero coefficients in
$\gamma^0$ and $\phi_0^2(\bz)$ denotes the compatibility constant of the
design matrix $\bz$ \citep[(6.4)]{pbvdg11}.
\end{prop} 
A proof is given in Section \ref{sec.proofs}. 
The choice $\lambda = 2 \lambda_0$ leads to the convergence rates: 
\begin{eqnarray}\label{oracleZ}
& &\|\bz(\hat{\gamma}(\lambda) - \gamma^0)\|_2^2/n =
O_P \Big(\frac{s(\gamma^0)\log(q)}{n\phi_0^2(\bz)} \Big),\nonumber\\
& &\|\hat{\gamma}(\lambda) - \gamma^0\|_1 =
O_P\Big(\frac{s(\gamma^0)}{\phi_0^2(\bz)}\sqrt{\frac{\log(q)}{n}}\Big).
\end{eqnarray}
The second result about the $\ell_1$-norm convergence rate implies a
variable screening property as follows: assume a so-called beta-min
condition requiring that 
\begin{eqnarray}\label{betaminZ}
\min_{r \in S(\gamma^0)}|\gamma^0_r| \ge C s(\gamma^0)
\sqrt{\frac{\log(q)}{n}}/\phi_0^2(\bz) 
\end{eqnarray}
for a sufficiently large $C >0$, then, with high probability,
$\hat{S}_{\by,\bz}(\lambda) \supseteq S(\gamma^0)$, where
$\hat{S}_{\by,\bz}(\lambda) = \{r;\ \hat{\gamma}_r(\lambda) \neq 0\}$ and
$S(\gamma^0) = \{r;\ \gamma^0_r \neq 0\}$. 

The results in Proposition \ref{prop0}, (\ref{oracleZ}) and
(\ref{betaminZ}) describe inference for $\bz \gamma^0$
and for $\gamma^0$. Their meaning for inferring $\bx \beta^0$ and for
(groups) of $\beta^0$ are further discussed in Section \ref{subsec.bias}
for specific examples, representing $\gamma^0$ in terms of $\beta^0$ and the
correlation 
structure of $X$, and analyzing the squared bias $B^2 = \EE[(\mu_X -
\mu_Z)^2]$.  
The compatibility constant $\phi_0^2(\bz)$ in Proposition \ref{prop0} is
typically (much) better behaved, if $q \ll p$, than the corresponding constant
$\phi_0^2(\bx)$ of the original design $\bx$. Bounds of $\phi_0^2(\bx)$ and
$\phi_0^2(\bz)$ in terms of their population covariance $\Sigma$ and $A
\Sigma A^T$, respectively, can be derived from \citet[Cor.6.8]{pbvdg11}. Thus, 
loosely speaking, we have to deal with a trade-off: the term
$\phi_0^2(\bz)$, coupled with a $\log(q)$- instead of a $\log(p)$-factor
(and also to a certain extent the sparsity factor $s(\gamma^0)$) are favorable for
the dimensionality reduced $\bz$. The price to pay for this is the bias
term $B^2 = \EE[(\mu_X - \mu_Z)^2]$, discussed further in Section
\ref{subsec.bias}, which appears in the variance $\xi^2$ entering the
definition of $\lambda_0$ in 
Proposition \ref{prop0} as well as in the prediction error
(\ref{pred.err}); furthermore, the detection of $\gamma^0$ instead of
$\beta^0$ can be favorable for some cases and not favorable for others,
as discussed in Section \ref{subsec.bias}. 

Finally, note that Proposition \ref{prop0} makes a statement conditional on
$\bz$: with high probability $1 - \alpha$,
\begin{eqnarray*}
\PP[\|\bz \hat{\gamma} - \bz \gamma^0\|_2^2/n \le 4 \lambda^2
s(\gamma^0)/\phi_0^2(\bz) |\bz] \ge 1 - \alpha.
\end{eqnarray*}
Assuming that $\phi_0^2(\bz) \ge \phi_0^2(A,\Sigma)$ is bounded with high
probability \citep[Lem.6.17]{pbvdg11}, we obtain (for a small but different 
$\alpha$ than above): 
\begin{eqnarray*}
\PP[\|\bz \hat{\gamma} - \bz \gamma^0\|_2^2/n \le 4 \lambda^2
s(\gamma^0)/\phi_0^2(A,\Sigma)] \ge 1 - \alpha.
\end{eqnarray*}
In view of (\ref{pred.err}), we then have for the prediction error:
\begin{eqnarray}\label{orac-bias}
\EE[\|\bx \beta^0 - \bz \hat{\gamma}\|_2^2/n] = \EE[\|\bz \hat{\gamma} - \bz
\gamma^0\|_2^2/n + \EE[(\mu_X - \mu_Z)^2],
\end{eqnarray}
where $\EE[\|\bz \hat{\gamma} - \bz \gamma^0\|_2^2/n] \approx O(\xi s(\gamma^0)
\sqrt{\log(p)/n})$ when choosing $\lambda = 2 \lambda_0$.   

\subsection{The parameter $\gamma^0$ for cluster representatives}\label{subsec.bias}

In the sequel, we consider the case where $Z = \bar{X} = (\bar{X}^{(1)},\ldots
,\bar{X}^{(q)})^T$ encodes the cluster representatives $\bar{X}^{(r)} =
|G_r|^{-1} \sum_{j \in G_r} X^{(j)}$. We analyze the coefficient vector
$\gamma^0$ and discuss it from the view-point of detection. For specific
examples, we also
quantify the squared bias term $B^2 = \EE[(\mu_X - \mu_{\bar{X}})^2]$ which
plays mainly a role for prediction. 

The coefficient $\gamma^0$ can be exactly described if the cluster
representatives are independent.
\begin{prop}\label{prop2}
Consider a random design Gaussian linear model as in (\ref{mod.Grand})
where $\Cov(\bar{X}^{(r)},\bar{X}^{(\ell)}) = 0$ for all $r \neq \ell$. Then,
\begin{eqnarray*} 
& &\gamma_r^0 = |G_r| \sum_{j \in G_r} w_j \beta^0_j,\ r=1,\ldots ,q,\\
& &w_j = \frac{\sum_{k =1}^p \Sigma_{j,k}}{\sum_{\ell \in G_r} \sum_{k=1}^p
  \Sigma_{\ell,k}},\ \ \sum_{j \in G_r} w_j =1.
\end{eqnarray*}
Moreover:
\begin{enumerate}
\item
If, for $r \in \{1,\ldots ,q\}$, $\Sigma_{j,k} \ge 0$ for all $j,k \in
G_r$, then $w_j \ge 0\ (j \in G_r)$, and $\gamma^0_r/|G_r|$ is a convex
combination of $\beta_j^0\ (j \in G_r)$. In particular, if $\beta^0_j \ge
0$ for all $j \in G_r$, or $\beta^0_j \le 0$ for all $j \in G_r$, then
\begin{eqnarray*}
|\gamma^0_r| \ge |G_r| \min_{j \in G_r}|\beta^0_j|.
\end{eqnarray*}
\item If, for $r \in \{1,\ldots ,q\}$, $\Sigma_{j,j}\equiv 1$ for all $j
  \in G_r$ and $\sum_{k \neq j} \Sigma_{j,k} \equiv \zeta$
  for all $j \in G_r$, then $w_j \equiv |G_r|^{-1}$ and 
\begin{eqnarray*}
\gamma^0_r = \sum_{j \in G_r} \beta^0_j.
\end{eqnarray*}
A concrete example is where $\Sigma$ has a block-diagonal structure with
equi-correlation within blocks: $\Sigma_{j,k} \equiv \rho_r\ (j,k \in G_r,\
j \neq k)$ with $-1/(|G_r| -1) < \rho_r < 1$ 
(where the lower bound for $\rho_r$ ensures positive definiteness of the
block-matrix $\Sigma_{G_r,G_r}$).
\end{enumerate}
\end{prop}

A proof is given in Section \ref{sec.proofs}. The assumption of
uncorrelatedness across $\{\bar{X}^{(r)};\ r=1,\ldots ,q\}$ is reasonable
if we have tight clusters 
corresponding to blocks of a block-diagonal structure of $\Sigma$. 

We can immediately see that there are benefits or
disadvantages of using the group representatives in terms of the size of
the absolute value $|\gamma_r^0|$: obviously a large value would make it
easier to detect the group $G_r$. Taking a group representative is
advantageous if all the coefficients within a group have the same
sign. However, we should be a bit careful since the size of a regression
coefficient should be placed in context to the standard deviation of the
regressor: here, the standardized coefficients are
\begin{eqnarray}\label{stand-gamma}
\gamma^0_r \sqrt{\Var(\bar{X}^{(r)})}.
\end{eqnarray}
For e.g. high positive correlation among variables within a group,
$\Var(\bar{X}^{(r)})$ is much larger than for independent variables: for
the equi-correlation scenario in statement 2. of Proposition \ref{prop2} we
obtain for the standardized coefficient
\begin{eqnarray}\label{stand-gamma2}
\gamma_r^0 \sqrt{\Var(\bar{X}^{(r)})} = \sum_{j \in G_r} \beta^0_j \sqrt{\rho
  + |G_r|^{-1} (1 - \rho)} \approx \sum_{j \in G_r} \beta^0_j,
\end{eqnarray}
where the latter approximation holds if $\rho \approx 1$. 

The disadvantages occur if rough or near cancellation among $\beta_j^0\ (j
\in G_r)$ takes place. This can cause a reduction of the
absolute value of $|\gamma^0_r|$ in comparison to $\max_{j \in
  G_r}|\beta_j^0|$: again, the scenario in statement 2. of Proposition
\ref{prop2} is most clear in the sense that the sum of $\beta^0_j\ (j \in
G_r)$ is equal to $\gamma^0_r$, and near cancellation would mean that
$\sum_{j \in G_r} \beta_j^0 \approx 0$. 

An extension of Proposition \ref{prop2} can be derived for covering the case
where the regressors $\{\bar{X}^{(r)};\ r =1,\ldots ,q\}$ are only
approximately uncorrelated.
\begin{prop}\label{prop3}
Assume the conditions of Proposition \ref{prop2} but instead of
uncorrelatedness of $\{\bar{X}^{(r)};\ r=1,\ldots ,q\}$ across $r$, we require:
for $r \in \{1,\ldots ,q\}$, 
\begin{eqnarray*}
|\mathrm{Cov}(X^{(i)},X^{(j)}|\{\bar{X}^{(\ell)};\ \ell \neq r\})| \le \nu\ \mbox{for
  all}\ j \in G_r,\ i \notin G_r.
\end{eqnarray*}
Moreover, assume that $\Var(\bar{X}^{(r)}|\{\bar{X}^{(\ell)};\ \ell \neq r\})
\ge C > 0$. Then,
\begin{eqnarray*}
\gamma_r^0 = |G_r| \sum_{j \in G_r} w_j \beta^0_j + \Delta_r,\ |\Delta_r| \le
\nu \|\beta^0\|_1/C.
\end{eqnarray*}
Furthermore, if $\mathrm{Cov}(X^{(i)},X^{(j)}|\{\bar{X}^{(\ell)};\ \ell
\neq r\}) \ge 0$ for all $j \in G_r,\ i \notin G_r$, and
$\Var(\bar{X}^{(r)}|\{\bar{X}^{(k)};\ k \neq r\}) \ge C > 0$, then:
\begin{eqnarray*}
|\gamma_r^0| \ge |G_r| \min_{j \in G_r}|\beta^0_j| + \Delta_r,\ |\Delta_r| \le
\nu \|\beta^0\|_1/ C.
\end{eqnarray*}
\end{prop}

A proof is given in Section \ref{sec.proofs}. The assumption that
$|\mathrm{Cov}(X^{(i)},X^{(j)}|\{\bar{X}^{(k)};\ k \neq r\})| \le \nu$ for
all $j \in G_r,\ i \notin G_r$ is
implied if the variables in $G_r$ and $G_{\ell}\ (r \neq \ell)$ are rather
uncorrelated. Furthermore, if we require that $\nu \|\beta^0\|_1 \ll |G_r|
\min_{j \in G_r}|\beta^0_j|$ and $C \asymp 1$ (which holds if $\Sigma_{j,j}
\equiv 1$ for all $j \in G_r$ and if the
variables within $G_r$ have high conditional correlations given
$\{\bar{X}^{(\ell)};\ \ell \neq r\}$), then: 
\begin{eqnarray*}
|\gamma^0_r| \ge |G_r| \min_{j \in G_r}|\beta^0_j| (1 + o(1)).
\end{eqnarray*}
Thus, also under clustering with only moderate independence between the
clusters, we can have beneficial behavior for the representative cluster
method. This also implies that the representative cluster method works if
the clustering is only approximately correct, as shown in Section
\ref{subsec.illustr}. 

We discuss in the next subsections two examples in more detail. 

\subsubsection{Block structure with equi-correlation}\label{subsubsec.equic}

Consider a partition with groups $G_r\ (r=1,\ldots ,q)$. The population
covariance matrix $\Sigma$ is block-diagonal having $|G_r| \times |G_r|$
block-matrices $\Sigma_{G_r,G_r}$ with equi-correlation: 
$\Sigma_{j,j} \equiv 1$ for all $j \in G_r$, and $\Sigma_{j,k} \equiv \rho_r$
for all $j,k
\in G_r\ (j \neq k)$, where $-1/(|G_r| -1) < \rho_r < 1$ (the lower bound
for $\rho_r$ ensures positive definiteness of $\Sigma_{G_r,G_r}$). This is
exactly the setting as in statement 2. of Proposition \ref{prop2}. 

The parameter $\gamma^0$ equals, see Proposition \ref{prop2}:
\begin{eqnarray*}
\gamma_r^0 = \sum_{j \in G_r} \beta^0_j.
\end{eqnarray*}

Regarding the bias, observe that 
\begin{eqnarray*}
\mu_X - \mu_{\bar{X}} = \sum_{r=1}^q (\mu_{X;r} - \mu_{\bar{X};r}),
\end{eqnarray*}
where $\mu_{X;r} = \sum_{j \in G_r} \beta^0_j X^{(j)}$ and $\mu_{\bar{X};r}
= \bar{X}^{(r)} \gamma^0_r$, and due to block-independence of $X$, we have
\begin{eqnarray*}
\EE[(\mu_X - \mu_{\bar{X}})^2] = \sum_{r=1}^q \EE[(\mu_{X;r} - \mu_{\bar{X};r})^2].
\end{eqnarray*}
For each summand, we have $\mu_{X;r} - \mu_{\bar{X};r} = \sum_{j \in G_r}
X^{(j)}(\beta^0_j - \bar{\beta}^0_r)$, where $\bar{\beta}^0_r = |G_r|^{-1} \times$\\
$\sum_{j \in G_r} \beta^0_j$. Thus,  
\begin{eqnarray*}
\EE[(\mu_{X;r} - \mu_{\bar{X};r})^2] &=& \sum_{j \in G_r} (\beta^0_j -
\bar{\beta}^0_r)^2 + 2 \rho_r \sum_{j,k \in G_r; j \neq k} (\beta^0_j -
\bar{\beta}^0_r)(\beta^0_k - \bar{\beta}^0_r)\\
&=& (1 - \rho_r) \sum_{j \in G_r}
(\beta^0_j - \bar{\beta}^0_r)^2.
\end{eqnarray*}
Therefore, the squared bias equals
\begin{eqnarray}\label{biasadd1}
B^2= \EE[(\mu_{X;r} - \mu_{\bar{X};r})^2] = \sum_{r=1}^q (1 - \rho_r) \sum_{j \in G_r}
(\beta^0_j - \bar{\beta}^0_r)^2.
\end{eqnarray}
We see from the formula that the bias is small if there is little
variation of $\beta^0$ within the groups $G_r$ or/and the $\rho_r$'s are
close to one. The latter is what we obtain with tight clusters: there is a
large within and small between groups correlation. Somewhat
surprising is the fact that the bias is becoming large if $\rho_r$ tends to
negative values (which is related to the fact that detection becomes bad
for negative values of $\rho_r$, see (\ref{stand-gamma2})). 

Thus, in summary: in comparison to an
estimator based on $\bx$, using the cluster 
representatives $\bar{\bx}$ and subsequent estimation leads to equal (or
better, due to smaller dimension) prediction error if all $\rho_r$'s are
close to 1, regardless of $\beta^0$. When
 using the cluster representative Lasso, if $B^2 = \sum_{r=1}^q (1 - \rho_r)
 \sum_{j \in G_r} (\beta^0_j - \bar{\beta}^0_r)^2 = O(s(\gamma^0)
 \log(q)/n)$, then the 
squared bias has no disturbing effect on the prediction error as can be
seen from (\ref{oracleZ}). 

With respect to detection, there can
be a substantial gain for inferring the cluster $G_r$ if $\beta^0_j\ (j \in
G_r)$ have all the same sign, and if $\rho_r$ is close to 1. Consider the
active groups 
\begin{eqnarray*}
S_{0,\mathrm{Group}} = \{r;\ \beta^0_{G_r} \neq 0,\ r=1,\ldots ,q\}.
\end{eqnarray*}
For the current model and assuming that $\sum_{j \in G_r}\beta^0_j \neq 0\
(r=1,\ldots ,q)$ (i.e., no exact cancellation of coefficients within groups),
$S_{0,\mathrm{Group}} = S(\gamma^0) = \{r;\ \gamma^0_r \neq 0\}$. In view of
(\ref{stand-gamma2}) and (\ref{betaminZ}), the screening property for
groups holds if  
\begin{eqnarray*}
\min_{r \in S_{0,\mathrm{Group}}} |\sum_{j \in G_r} \beta^0_j \sqrt{\rho
  + |G_r|^{-1} (1 - \rho)}| \ge  C \frac{s(\gamma^0)}{\phi_0^2(\bar{\bx})}
\sqrt{\frac{\log(q)}{n}}.
\end{eqnarray*}
This condition holds even if the non-zero $\beta^0_j$'s are very small but
their sum within a group is sufficiently strong. 

\subsubsection{One active variable per cluster}\label{subsubsec.oneact}

We design here a model with at most one active variable in each
group. Consider a low-dimensional $q \times 1$ variable $U$ and perturbed
versions of $U^{(r)}\ (r=1,\ldots ,q)$ which constitute the $p \times 1$
variable $X$: 
\begin{eqnarray}\label{mod.corr2}
& &X^{(r,1)} = U^{(r)},\ r=1,\ldots ,q,\nonumber\\
& &X^{(r,j)} = U^{(r)} + \delta^{(r,j)},\ j=2,\ldots,m_r,\ r=1,\ldots ,q,\nonumber\\
& &\delta^{(r,2)},\ldots ,\delta^{(r,m_r)}\ \mbox{i.i.d.}\ {\cal
  N}(0,\tau_r^2)\ \mbox{and independent among}\ r=1,\ldots ,q.
\end{eqnarray}
The index $j=1$ has no specific meaning, and the fact that one covariate is
not perturbed is discussed in Remark 2 below. The purpose is to have at
most one active variable in every cluster by assuming a low-dimensional
underlying linear model   
\begin{eqnarray}\label{mod.corr}
Y = \sum_{r=1}^q \tilde{\beta}^0_r U^{(r)} + \eps,
\end{eqnarray}
where $\eps \sim {\cal N}(0,\sigma^2)$ is independent of $U$. Some of the
coefficients in $\tilde{\beta}^0$ might be zero, and hence, some of the
$U^{(r)}$'s might be noise covariates.
We construct a $p \times 1$ variable $X$ by stacking the variables
$X^{(r,j)}$ as follows: $X^{(\sum_{\ell=1}^{r-1}m_{\ell} + j)} = X^{(r,j)}\ (j=1,\ldots 
  ,m_r)$ for $r=1,\ldots ,q$. Furthermore, we use an augmented vector of
the true regression coefficients 
\begin{eqnarray*}
\beta^0_{j} = \left \{ \begin{array}{ll}
\tilde{\beta}^0_r & \mbox{if $j = \sum_{\ell=1}^{r-1} m_{\ell} +1,\ r=1,\ldots
  ,q$},\\
0 & \mbox{otherwise}.
\end{array}
\right.
\end{eqnarray*}
Thus, the model in (\ref{mod.corr}) can be represented as
\begin{eqnarray*}
Y = \sum_{j=1}^p \beta^0_j X^{(j)} + \eps,
\end{eqnarray*}
where $\eps$ is independent of $X$. 

\medskip
\textbf{Remark 2.} Instead of the model in
(\ref{mod.corr2})--(\ref{mod.corr}), we could consider 
\begin{eqnarray*}
& &X^{(r,j)} = U^{(r)} + \delta^{(r,j)},\ j=1,\ldots,m_r,\ r=1,\ldots ,q,\nonumber\\
& &\delta^{(r,1)},\ldots ,\delta^{(r,m_r)}\ \mbox{i.i.d.}\ {\cal
  N}(0,\tau_r^2)\ \mbox{and independent among}\ r=1,\ldots ,q,
\end{eqnarray*}
and
\begin{eqnarray*}
Y = \sum_{j=1}^p \beta^0_j X^{(j)} + \eps,
\end{eqnarray*}
where $\eps$ is independent of $X$.
Stacking the variables $X^{(r,j)}$ as before, the covariance matrix $\Cov(X)$ has
$q$ equi-correlation blocks but the blocks are generally dependent if
$U^{(1)},\ldots ,U^{(q)}$ are correlated, i.e., $\Cov(X)$ is not of
block-diagonal form. If $\Cov(U^{(r)},U^{(\ell)}) = 0\ (r \neq \ell)$, we are
back to the model in Section \ref{subsubsec.equic}. For more general
covariance structures of $U$, an analysis of the model seems rather
cumbersome (but see Proposition \ref{prop3}) while the analysis of the
``asymmetric'' model (\ref{mod.corr2})-(\ref{mod.corr}) remains rather
simple as discussed below.

\medskip
We assume that the clusters $G_r$ are corresponding to the variables
$\{X^{(r,j)};\ j=1,\ldots ,m_r\}$ and thus, $m_r = |G_r|$. In contrast to
the model in Section \ref{subsubsec.equic}, we do not assume uncorrelatedness or
quantify correlation between clusters. The
cluster representatives are 
\begin{eqnarray*}
& &\bar{X}^{(r)} = m_r^{-1} \sum_{j \in G_r} X^{(r,j)} = U^{(r)} +
W^{(r)},\\
& &W^{(r)} \sim {\cal N}(0,\tau_r^2 \frac{m_r-1}{m_r^2})\ \mbox{and
  independent among}\ r=1,\ldots ,q.
\end{eqnarray*}
As in (\ref{modZ}), the dimension-reduced model
is written as $Y = \sum_{r=1}^q \gamma_r^0 \bar{X}^{(r)} + \eta$. 

For the bias, we immediately find:
\begin{eqnarray}\label{bias2}
B^2 &=& \EE[(\mu_X - \mu_{\bar{X}})^2]\nonumber\\
&\le&\EE[(U^T \tilde{\beta}^0 - \bar{X}
\tilde{\beta}^0)^2] = \EE|W^T \tilde{\beta}^0|^2 \le s_0 \max_j |\beta^0_j|^2
\max_r \frac{m_r - 1}{m_r^2} \tau_r^2.
\end{eqnarray}
Thus, if the cluster sizes $m_r$ are large and/or the perturbation noise
$\tau_r^2$ is small, the squared bias $B^2$ is small. 

Regarding detection, we make use of the following result.
\begin{prop}\label{prop0a}
For the model in (\ref{mod.corr}) we have:
\begin{eqnarray*}
& &\|\tilde{\beta}^0 - \gamma^0\|_2^2 \le 2
B^2/\lambda_{\mathrm{min}}^2(\Cov(U))
 = 2 \EE|W^T \tilde{\beta}^0|^2/\lambda_{\mathrm{min}}^2(\Cov(U))\\
&\le& 2 s_0 \max_j |\beta^0_j|^2 \max_r \frac{m_r - 1}{m_r^2}
\tau_r^2/\lambda_{\mathrm{min}}^2(\Cov(U)), 
\end{eqnarray*}
where $\lambda_{\mathrm{min}}^2(\Cov(U))$ denotes the minimal eigenvalue of
$\Cov(U)$. 
\end{prop}

A proof is given in Section \ref{sec.proofs}. Denote by $\tilde{S}_0 =
\{r;\ \tilde{\beta}^0_r \neq 0\}$. We then have:
\begin{eqnarray}\label{detect1}
&\mbox{if}&\ \min_{r \in \tilde{S}_0} |\tilde{\beta}^0_r| > 2 \sqrt{2 s_0
\max_j |\beta^0_j| \max_r \frac{m_r - 1}{m_r^2} \tau_r^2 /\lambda_{\mathrm{min}}^2(\Cov(U))},\nonumber\\
&\mbox{then:} &\ \min_{r \in \tilde{S}_0} |\gamma^0_r| >
\sqrt{2 s_0 \max_j |\beta^0_j| \max_r
  \frac{m_r - 1}{m_r^2} \tau_r^2/\lambda_{\mathrm{min}}^2(\Cov(U))}.
\end{eqnarray}
This follows immediately: if the implication would not hold, it would
create a contradiction to Proposition \ref{prop0a}. 

We argue now that 
\begin{eqnarray}\label{taum}
\max_r \frac{\tau_r^2}{m_r} =O( \log(q)/n)
\end{eqnarray}
is a sufficient condition to achieve prediction error and detection as in
the $q$-dimensional model (\ref{mod.corr}). Thereby, we implicitly
assume that $\max_j|\beta_j^0| \le C < \infty$ and
$\lambda_{\mathrm{min}}^2(\Cov(U)) \ge L > 0$. Since we have that
$s(\gamma^0) \ge s_0$ (excluding the pathological case for a particular
combination of $\beta^0$ and $\Sigma = \Cov(X)$), using (\ref{bias2}) the
condition (\ref{taum}) implies  
\begin{eqnarray*}
B^2 = O(s_0 \log(q)/n) \le O(s(\gamma^0) \log(q)/n)
\end{eqnarray*}
which is at most of the order of the prediction error in (\ref{oracleZ}),
where $\phi_0^2(\bz) = \phi_0^2(\bar{\bx})$ can be lower-bounded by the
population version $\phi_0^2(\Cov(\bar{X})) \ge \phi_0^2(\Cov(U))$
\citep[Cor.6.8]{pbvdg11}. For detection, we note that (\ref{taum}) implies
that 
the bound in (\ref{detect1}) is at most 
\begin{eqnarray*}
\min_{r \in \tilde{S}_0} |\gamma^0_r| > \sqrt{2 s_0 \max_j |\beta^0_j| \max_r
  \frac{m_r - 1}{m_r^2} \tau_r^2/\lambda_{\mathrm{min}}^2(\Cov(U))} \le 
O(s(\gamma^0) \sqrt{\log(q)/n}).
\end{eqnarray*}
The right-hand side is what we require as beta-min condition in
(\ref{betaminZ}) for group screening such that with high probability
$\hat{S} \supseteq S(\gamma^0) \supseteq S_0$ (again excluding a particular
constellation of $\beta^0$ and $\Sigma$). 

The condition (\ref{taum}) itself is fulfilled if $m_r \asymp n/\log(q)$,
i.e., when the cluster sizes are large, or if $\tau_r^2 = O(\log(q)/n)$,
i.e., the clusters are tight. An example where the model
(\ref{mod.corr2})--(\ref{mod.corr}) with (\ref{taum}) seems reasonable is
for genome-wide association studies with SNPs where $p \approx 10^6$, $n
\approx 1000$ and $m_r$ can be in the order of $10^3$ and hence $q \approx
1000$ when e.g. using the order of magnitude of number of target SNPs
\citep{carlson04}. Note that this is a scenario where the group sizes $m_r
\gg n$ where the cluster group Lasso seems inappropriate. 

The analysis in Sections \ref{sec.dimred} and \ref{subsec.bias} about the
bias $B^2$ and the parameter $\gamma^0$ has immediate
implications for the cluster representative Lasso (CRL), as discussed in
the next section.  

\subsection{A comparison}\label{subsec.compare} 

We compare now the results of the cluster representative Lasso (CRL), the
cluster group Lasso (CGL) and the plain Lasso, at least on a ``rough scale''. 

For the plain Lasso we have: with high probability, 
\begin{eqnarray}\label{rateLasso}
& &\|\bx(\hat{\beta}_{\mathrm{Lasso}} -
\beta^0)\|_2^2/n = O \Big(\frac{\log(p)s_0}{n  
\phi_0^2(\bx)} \Big),\nonumber\\
& &\|\hat{\beta}_{\mathrm{Lasso}} - \beta^0\|_1 = O
\Big(\frac{s_0}{\phi_0^2(\bx)} \sqrt{\frac{\log(p)}{n}} \Big),
\end{eqnarray}
which both involve $\log(p)$ instead of $\log(q)$ and more importantly, the
compatibility constant $\phi_0^2(\bx)$ of the $\bx$ design matrix instead of
$\phi_0^2(\bar{\bx})$ of the matrix $\bar{\bx}$. If $p$ is large, then  
$\phi_0^2(\bx)$ might be close to zero; furthermore, it is exactly in situations
like model (\ref{mod.corr2}) and (\ref{mod.corr}), having a few ($s_0 \le
q$) active variables and noise covariates being highly correlated with the
active variables, which leads to very small values of $\phi_0^2(\bx)$, see
\citet{vdgled11}. For variable screening $\hat{S}_{\mathrm{Lasso}}(\lambda)
\supseteq S_0$ with high probability, the corresponding (sufficient)
beta-min condition is 
\begin{eqnarray}\label{betaminLasso}
\min_{j \in S_{0}}|\beta^0_{j}| \ge C s_0
  \sqrt{\frac{\log(p)}{n}}/\phi_0^2(\bx) 
\end{eqnarray}
for a sufficiently large $C>0$. 
 
For comparison with the cluster group Lasso (CGL) method, we assume for
simplicity equal
group sizes $|G_r|= m_r \equiv m$ for all $r$ and $\log(q)/m \le 1$, i.e., the
group size is sufficiently large. We then obtain: with
high probability,     
\begin{eqnarray}\label{rateGroup}
& &\|\bx(\hat{\beta}_{\mathrm{CGL}} - \beta^0)\|_2^2/n = O \Big(
\frac{|S_{0,\mathrm{Group}}| m}{n \phi_{0,\mathrm{Group}}^2(\bx)} \Big),\nonumber\\
& &\sum_{r=1}^q \|\hat{\beta}_{G_r} - \beta_{G_r}^0\|_2 =
O \Big(\frac{|S_{0,\mathrm{Group}}| \sqrt{m}}{\sqrt{n}\phi_{0,\mathrm{Group}}^2(\bx)} \Big). 
\end{eqnarray}
For variable screening $\hat{S}_{\mathrm{CGL}}(\lambda)
\supseteq S_0$ with high probability, the corresponding (sufficient)
beta-min condition is 
\begin{eqnarray*}
\min_{r \in S_{0,\mathrm{Group}}} \|\beta_{G_r}^0\|_2 \ge C
\frac{|S_{0,\mathrm{Group}}| \sqrt{m}}{\sqrt{n} \phi_{0,\mathrm{Group}}^2(\bx)}  
\end{eqnarray*}
for a sufficiently large $C>0$. 
The compatibility constants $\phi_0^2(\bx)$ in (\ref{rateLasso}) and
$\phi_{0,\mathrm{Group}}^2(\bx)$ in (\ref{rateGroup}) 
are not directly comparable, but see Theorem \ref{groupcompatibility.lemma}
which is in favor of the CGL method. ``On the rough scale'', we can
distinguish two cases: if the group-sizes are large with only
none or a few active variables per group, implying $s_0 = |S_0| \approx
|S_{0,\mathrm{Group}}|$, the Lasso is better than the CGL method because the CGL
rate involves $|S_{0,\mathrm{Group}}| m$ or $|S_{0,\mathrm{Group}}|
\sqrt{m}$, respectively, instead of the sparsity $s_0$
appearing in the rate for the standard Lasso; for the case where we have
either none or many active variables within groups, the CGL method is
beneficial, mainly for detection, since $|S_{0,\mathrm{Group}}| m \approx
|S_0| = s_0$ but $|S_{0,\mathrm{Group}}| \sqrt{m} < |S_0| = s_0$. The behavior in the
first case is to be
expected since in the group Lasso representation (\ref{repr.grouplasso}),
the parameter vectors $\beta_{G_r}^0$ are very sparse within groups, and this
sparsity is not exploited by the group Lasso. A sparse group Lasso method
\citep{friedetal10} would address this issue. On the other hand, the CGL
method has the advantage that it works without bias, in contrast to the
CRL procedure. Furthermore, the CGL can also lead to good detection if many
$\beta_j^0$'s in a group are small in absolute value. For detection of the
group $G_r$, we only need that $\|\beta_{G_r}^0\|_2$ is sufficiently
large: the signs of the coefficients of $\beta_{G_r}$ can be
different and (near) cancellation does not happen. 

For the cluster representative Lasso (CRL) method, the range of scenarios,
with good performance of CRL, is more restricted. The method works
well and is superior over the plain Lasso (and group Lasso) if the bias $B^2$
is small and the detection is well-posed in terms of the dimension-reduced
parameter $\gamma^0$. More precisely, if
\begin{eqnarray*}
B^2 = O \Big(\frac{s(\gamma^0) \log(q)}{n \phi_0^2(\bar{\bx})} \Big),
\end{eqnarray*}
the CRL is better than plain Lasso for prediction since the corresponding
oracle inequality for the CRL becomes, see (\ref{orac-bias}): with high
probability, 
\begin{eqnarray*}
\|\bar{\bx} \hat{\gamma} - \bx \beta^0\|_2^2/n + \lambda \|\hat{\gamma} -
\gamma^0\|_1 \le  O \Big(\frac{s(\gamma^0) \log(q)}{n \phi_0(\bar{\bx})} \Big).
\end{eqnarray*}
We have given two
examples and conditions ensuring that the bias is small, namely
(\ref{biasadd1}) and (\ref{taum}). The latter condition (\ref{taum}) is
also sufficient for 
better screening property in the model from Section
\ref{subsubsec.oneact}. For the equi-correlation model in Section
\ref{subsubsec.equic}, the success of
screening crucially depends on whether 
the coefficients from active variables in a group nearly cancel or add-up
(e.g. when having the same sign), see Propositions
\ref{prop2} and \ref{prop3}.   
The following Table \ref{tab1} recapitulates the findings. 
\begin{table}
\begin{center}
\begin{tabular}{l|ccc}
model & assumption & predict. & screen. \\
\hline
equi-corr. blocks (Sec. \ref{subsubsec.equic}) & small value in
(\ref{biasadd1}) & + & NA \\ 
equi-corr. blocks (Sec. \ref{subsubsec.equic}) & e.g. same sign for $\beta^0_j\ (j \in G_r)$ &
NA & + \\ 
$\le 1$ var. per group (Sec. \ref{subsubsec.oneact}) & (\ref{taum}) & + & + 
\end{tabular}
\caption{Comparison of cluster representative Lasso (CRL) with plain
  Lasso in terms of prediction and variable screening. The symbol ``+''
  encodes better theoretical results for the CRL in comparison to Lasso; an
  ``NA'' means that no comparative statement can be 
made.}\label{tab1}
\end{center}
\end{table}

Summarizing, both the CGL and CRL are useful and can be substantially
better than plain Lasso in terms of prediction and detection in presence of
highly correlated variables. If the cluster sizes are smaller than sample
size, the CGL
method is more broadly applicable, in the sense of consistency but not
necessarily efficiency, as it does
not involve the bias term 
$B^2$ and constellation of signs or of near cancellation of coefficients in
$\beta_{G_r}^0$ is not an issue. For group sizes which are larger than sample
size, the CGL is not appropriate: one would need to take a version of the
group Lasso with regularization within groups
\citep{meieretal09,friedetal10}. 
The CGL method benefits when using canonical
correlation based clustering as this improves the compatibility constant,
see Theorem \ref{groupcompatibility.lemma}.   
The CRL method is particularly suited for problems where the
variables can be grouped into tight clusters and/or the cluster sizes are
large. There is gain if there is at most one active
variable per cluster and the clusters are tight, otherwise the prediction
performance is influenced by 
the bias $B^2$ and detection is depending on whether the coefficients within
a group add-up or exhibit near cancellation.
If the variables are not very highly
correlated within large groups, the difficulty is to estimate these groups,
and in case of correct grouping, as assumed in the theoretical results
above, the CRL method may still perform (much) better than plain Lasso.

\subsection{Estimation of the clusters}

The theoretical derivations above assume that the groups $G_r\ (r=1,\ldots
,q)$ correspond to the correct clusters. For the canonical
correlation based clustering as in Algorithm \ref{alg1}, Theorem \ref{th-a}
discusses consistency in finding the true underlying population
clustering. For hierarchical clustering, the issue is much simpler. 

Consider the $n \times p$ design matrix $\bx$ as in (\ref{Gaussian}) and
assume for simplicity that $\Sigma_{j,j} = 1$ for
all $j$. It is well-known that  
\begin{eqnarray}\label{estsigma}
\max_{j,k} |\hat{\Sigma}_{j,k} - \Sigma_{j,k}| = O_P(\sqrt{\log(p)/n}),
\end{eqnarray}
where $\hat{\Sigma}$ is the empirical covariance matrix
\citep[cf. p.152]{pbvdg11}. 
Tightness and separation of the true clusters is ensured by:
\begin{eqnarray}\label{clustass}
& &\min\{|\Sigma_{j,k}|;\ j,k\ \in G_r\ (j \neq k),\ r=1,\ldots q\}\nonumber\\
&>&
\max\{|\Sigma_{j,k}|:\ j \in G_r,\ k \in G_{\ell},\ r,\ell=1,\ldots ,q\ (r \neq
\ell)\}.
\end{eqnarray}
Assuming (\ref{clustass}) and using (\ref{estsigma}), a standard clustering
algorithm, using e.g. single-linkage and dissimilarity $1 -
|\hat{\Sigma}_{j,k}|$ between variables $X^{(j)}$ and $X^{(k)}$, will
consistently find the true clusters if $\log(p)/n \to 0$. 

In summary, and rather obvious: the higher the correlation within 
and uncorrelatedness between clusters, the better we can estimate the true
underlying grouping. In this sense, and following the arguments in Sections
\ref{subsec.theoryCGL}--\ref{subsec.compare}, strong correlation within clusters
``is a friend'' when using cluster Lasso methods, while it is ``an enemy''
(at least for variable screening and selection) for plain Lasso. 

\subsection{Some first illustrations}\label{subsec.illustr}

We briefly illustrate some of the points and findings mentioned above for
the CRL and the plain Lasso.
Throughout this subsection, we show the results from
a single realization of each of different models. More systematic simulations are
shown in Section \ref{sec.empres}. We analyze scenarios with $p = 1000$
and $n=100$. Thereby, the covariates are generated as in model
(\ref{mod.corr2}) where $U \sim {\cal N}_q(0,I)$ with $q=5$ and $\tau =
0.5$, and thus, $\Cov(X) = \Sigma$ is of block-diagonal structure. The
response is as in the linear model (\ref{mod1}) with ${\boldsymbol \eps} \sim {\cal
  N}_n(0,I)$. We consider the following.

\emph{Correct and incorrect clustering.} The correct clustering consists of
$q=5$ clusters each having $m_r = |G_r| \equiv 200$ variables,
corresponding to the 
structure in model (\ref{mod.corr2}). An 
incorrect clustering was constructed as 5 clusters where the first half
(100) of
the variables in each constructed cluster correspond to the first half
(100) of the variables 
in each of the true 5 clusters, and the remaining second half (100) of the
variables in the constructed clusters are chosen randomly from the total of
500 remaining variables. We note that $m_r \equiv 200 > n = 100$ and thus,
the CGL method is inappropriate (see e.g. Proposition \ref{prop.grLasso}).

\begin{figure}[!htb]
\centerline{ 
\subfigure[]{
\includegraphics[scale=0.48]{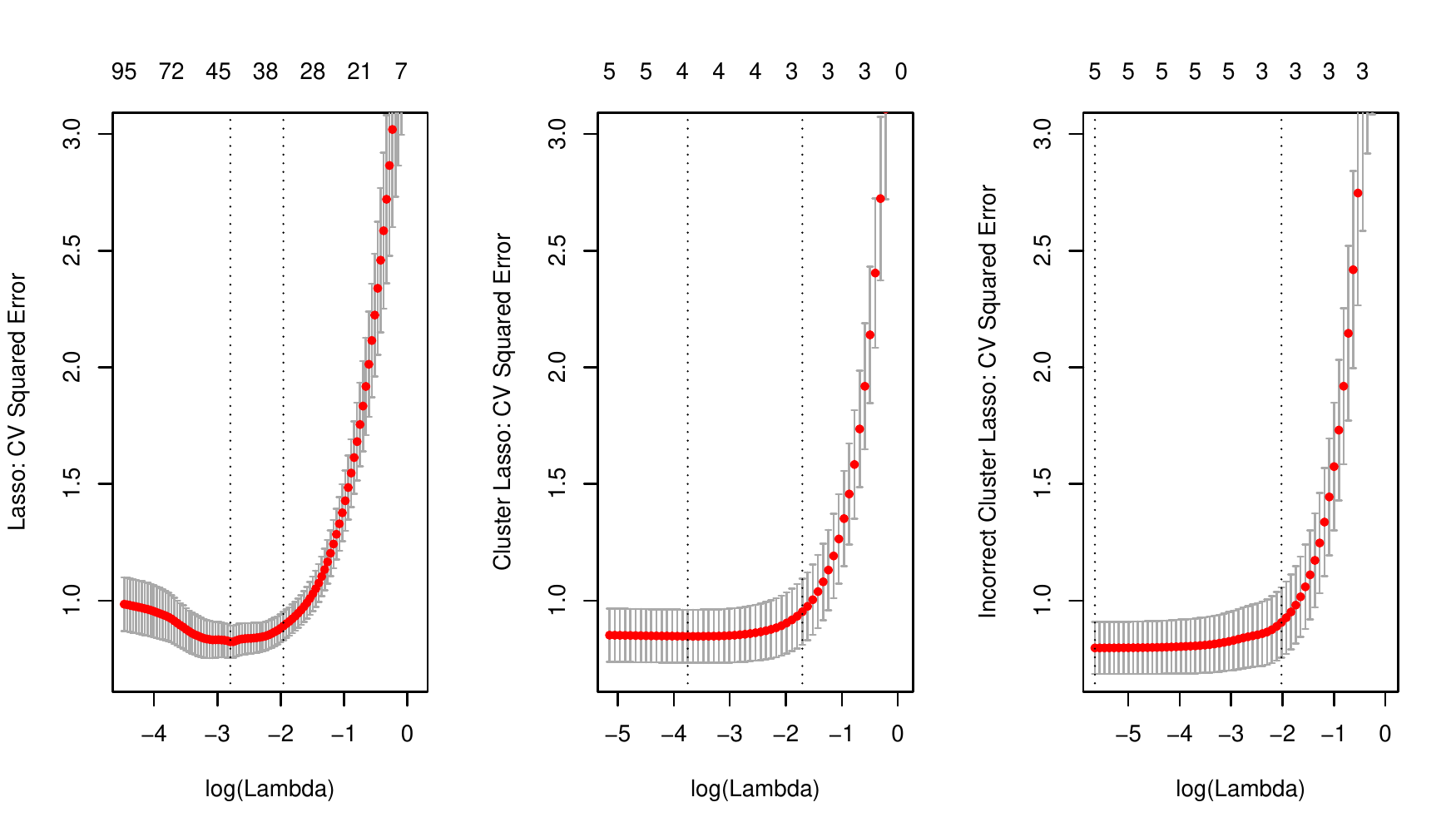}}
\subfigure[]{
\includegraphics[scale=0.48]{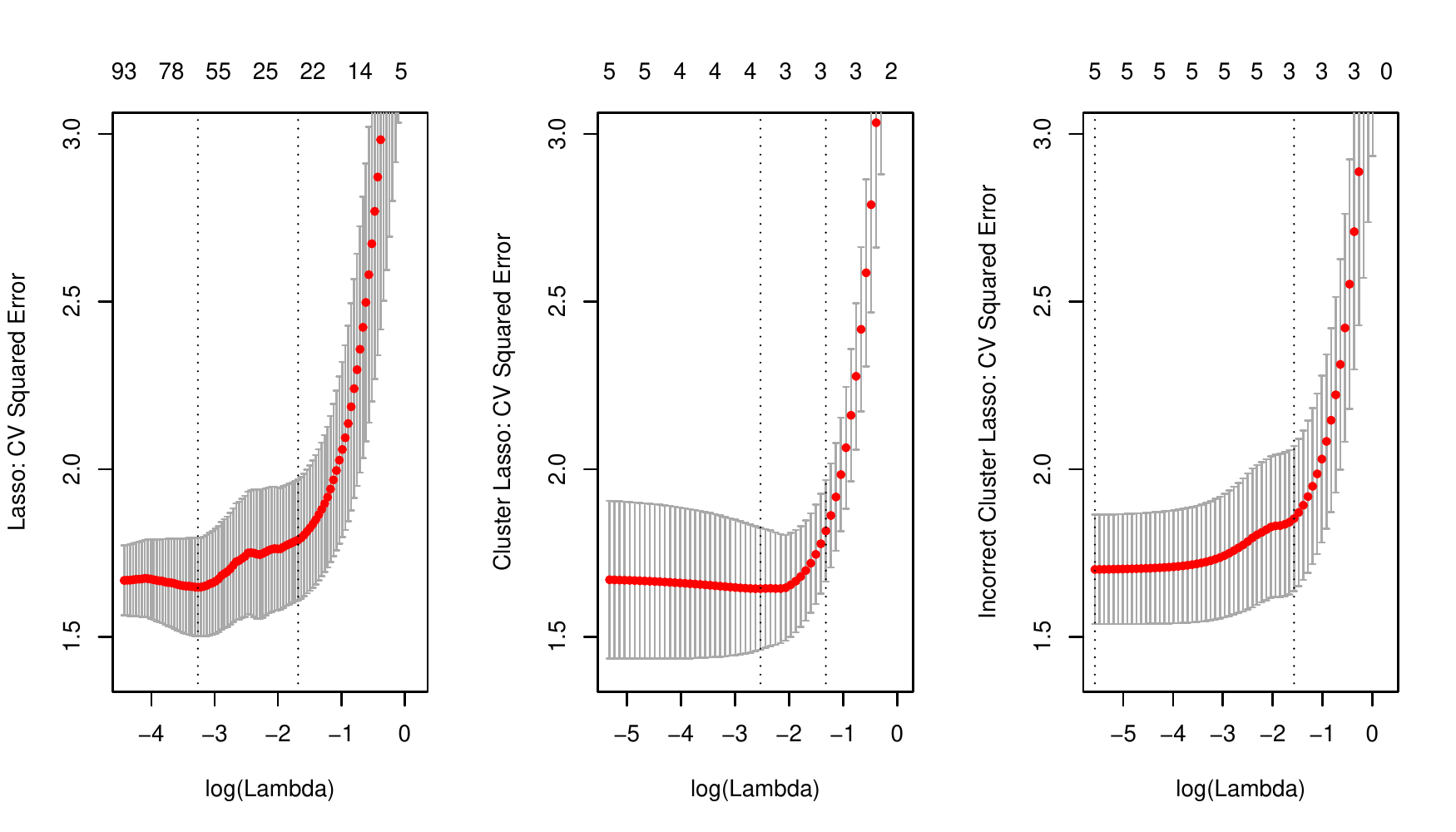}}
}
\centerline{ 
\subfigure[]{
\includegraphics[scale=0.48]{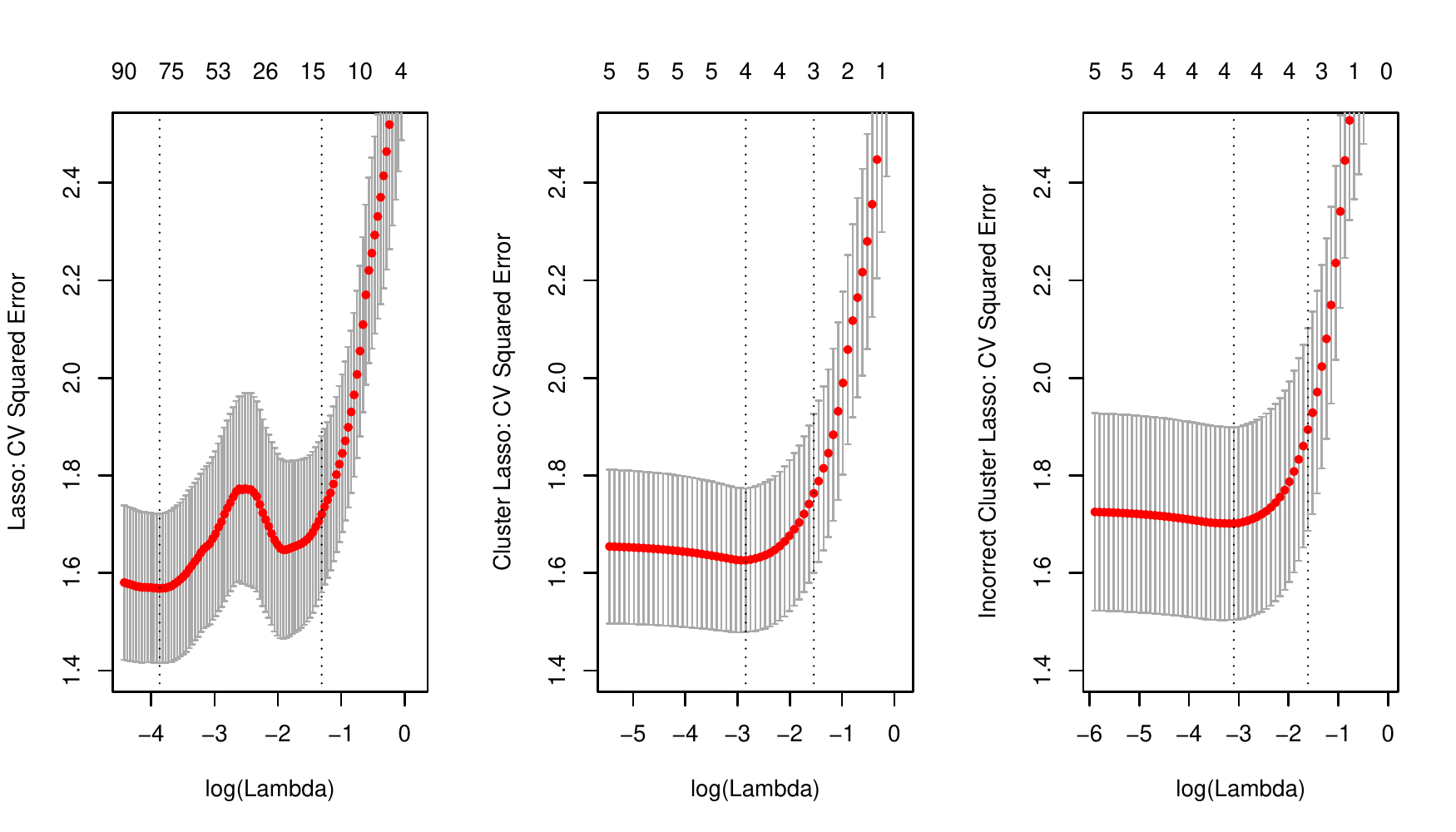}}
\subfigure[]{
\includegraphics[scale=0.48]{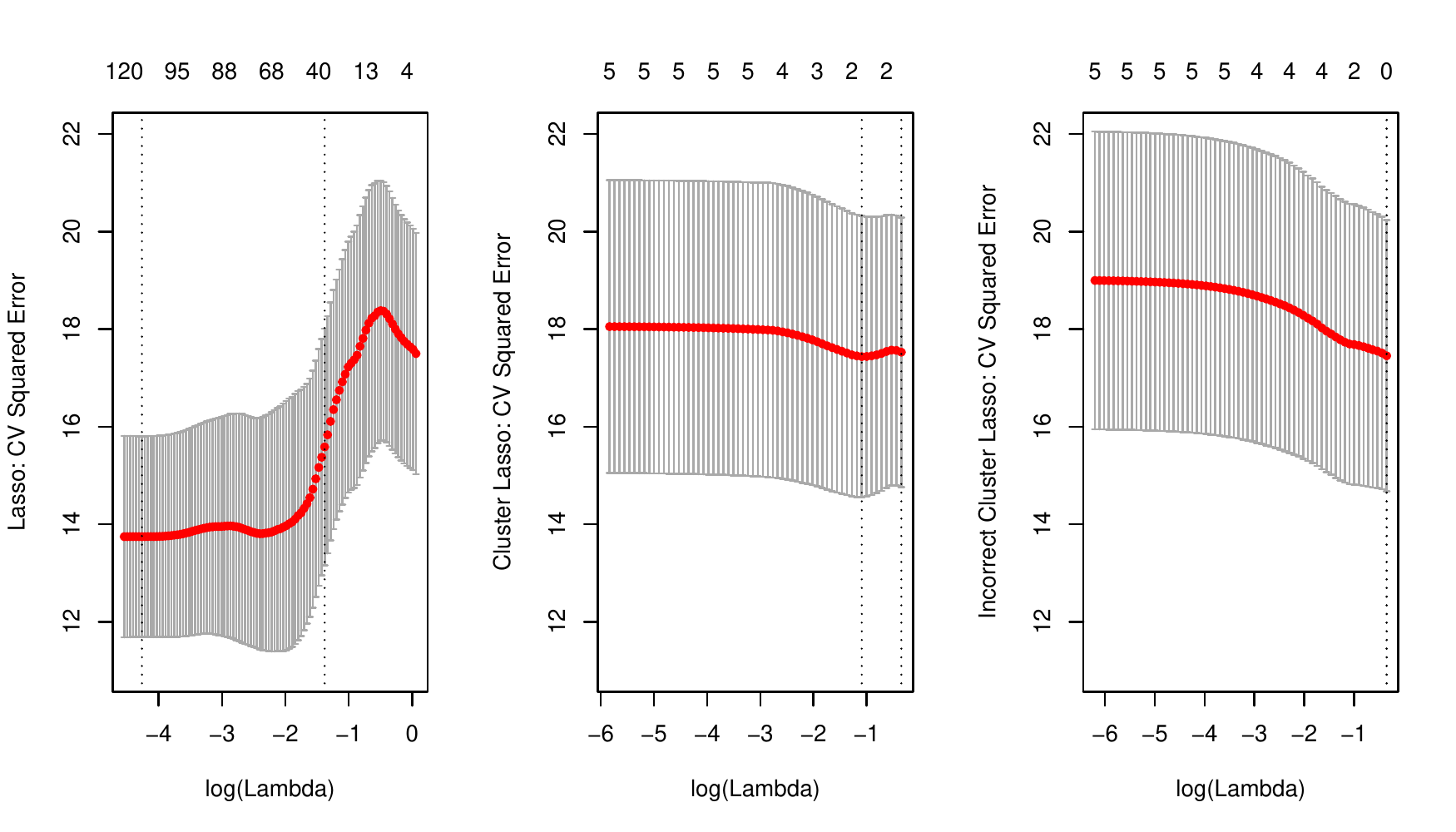}}
}
\caption{Left, middle and right panel of each subfigure: Lasso, cluster
  representative Lasso with correct clustering, and cluster representative
  Lasso with incorrect clustering as described in Section
  \ref{subsec.illustr}: 10-fold CV 
  squared error (y-axis) versus $\log(\lambda)$ (x-axis). Grey bars
  indicate the region 10-fold CV
  squared error $+/-$ estimated standard error (s.e.) of 10-fold CV squared
  error. The left vertical bar indicates the minimizer of the CV error and
  the right vertical bar corresponds to the largest value such that the CV
  error is within one standard deviation (s.d.) of the minimum. The numbers on
  top of each plot report 
  the number of selected variables (in case of the cluster representative
  Lasso, the number of selected representatives): the number of active groups is
  always equal to 3, and the number of active variables is 3 for (a) and 12
  for (b)-(d). Subfigures (a)-(d) correspond to the scenarios (a)-(d) in
  Section \ref{subsec.illustr}.}\label{fig-add}
\end{figure}

\emph{Active variables and regression coefficients.} We always consider 3
active groups (a group is called active if there is at least one active
variables in the group). The scenarios are as follows:
\begin{enumerate}
\item[(a)] One active variable within each of 3 active groups, namely $S_0
  = \{1,201,401\}$. The regression coefficients are $\beta^0_1 = -1,\
  \beta^0_{201} = -1,\ \beta^0_{401} = 1$;
\item[(b)] 4 active variables within each of 3 active groups, namely $S_0 =
  \{1,2,3,4,201,202,$\\
$203,204,401,402,403,404\}$. The regression coefficients
  are $\beta^0_j \equiv 0.25$ for $j \in S_0$;
\item[(c)] as in (b) but with regression coefficients $\{\beta^0_j;\ j \in
  S_0\}\ \mbox{i.i.d.}\ \sim \mathrm{Unif}([-0.5,0.5]$; 
\item[(d)] as in (b) but with exact cancellation of
  coefficients: $\beta^0_1 = \beta^0_3 = 2$, $\beta^0_2 = \beta^0_4 = -2$,
  $\beta^0_{201} = \beta^0_{203} = 2$, $\beta^0_{202} = \beta^0_{204} =
  -2$, $\beta^0_{401} = \beta^0_{403} = 2$, $\beta^0_{402} = \beta^0_{404}
  = -2$. 
\end{enumerate}
For the scenario in (d), we had to choose large coefficients, in absolute
value equal to 2, in order to see clear differences, in favor of the plain
Lasso. Figure \ref{fig-add}, using the \textrm{R}-package \texttt{glmnet}
\citep{friedetal09} shows the results.  

The results seem rather robust against approximate cancellation of
coefficients (Subfigure \ref{fig-add}(c)) and incorrect clustering (right
panels in Figure \ref{fig-add}). Regarding the latter, the number of
chosen clusters is worse than for correct clustering, though. A main
message of the results in Figure \ref{fig-add} is that the predictive
performance (using cross-validation) is a good indicator whether the group
representative Lasso (with correct or incorrect clustering) works. 

We can complement the rules from Section \ref{sec.cluster} for determining the
number of clusters as follows: take the representative cluster Lasso method
with the largest clusters (the least 
refined partition of $\{1,\ldots ,p\}$) such that predictive performance is
still reasonable (in comparison to the best achieved performance where we
would always consider plain Lasso among the competitors as well). In the
extreme case of Subfigure \ref{fig-add}(d), this 
rule would choose the plain Lasso (among the alternatives of correct
clustering and incorrect clustering) which is indeed the least refined
partition such that predictive performance is still reasonable. 

\section{Numerical results}\label{sec.empres}

In this section we look at three different simulation settings and a pseudo
real data example in order to empirically compare the proposed cluster
Lasso methods with plain Lasso.

\subsection{Simulated data}\label{simulationstudy}

Here, we only report results for the CRL and CGL methods where the
clustering of the variables is based on canonical
  correlations using Algorithm \ref{alg1} (see Section
  \ref{subsec.cancluster}). The corresponding results using ordinary
  hierarchical clustering, based on correlations and with average-linkage
  (see Section \ref{sec.ohclust}), are almost exactly the same because for
  the considered simulation settings, both clustering methods produce
  essentially the same partition.

We simulate data from the linear model in (\ref{mod1}) with fixed design
$\bx$, ${\boldsymbol \eps} \sim {\cal N}_n(0,\sigma^2 I)$ with $n=100$ and
$p=1000$. We 
generate the fixed design matrix $\bx$ once, from a multivariate normal
distribution $\mathcal{N}_p(0,\Sigma)$ with different structures for
$\Sigma$, and we then keep it fixed. We consider various scenarios, but the
sparsity or size of the active set is always equal to $s_0=20$. 

In order to compare the various methods, we look at two performance
measures for prediction and variable screening. For each model, our
simulated data consists of a training and an independent test set. The
models were fitted on the training data and we computed the test set mean
squared error $n^{-1} \sum_{i=1}^n \EE[(Y_{\mathrm{test}}(X_i) -
\hat{Y}(X_i))^2]$, denoted by MSE. For variable screening we consider  
the true positive rate as a measure of
performance, i.e., $|\hat{S} \cap S_0| / |S_0|$ as a function of
$|\hat{S}|$.  

For each of the methods we choose a suitable grid of values for the tuning
parameter. All reported results are based on 50 simulation runs. 

\subsubsection{Block diagonal model}\label{subsec.blockdiagonal}

Consider a block diagonal model
where we simulate the covariables $X \sim
\mathcal{N}_p(0,\Sigma_A)$ where $\Sigma_A$ is a block diagonal matrix.  We
use a $10\times 10$ matrix $\Gamma$, where 
\begin{equation*}
\Gamma_{i,j}=
\begin{cases}
  1,  & i=j,\\
  0.9, & \text{else}.
\end{cases}
\end{equation*}
The block-diagonal of $\Sigma_A$ consists of 100 such block matrices
$\Gamma$. Regarding the regression parameter $\beta^0$, we consider the
following configurations:  
\begin{itemize}

\item[(Aa)] $S_0 = \{1,2,\ldots ,20\}$ and 
for any $j \in S_0$ we sample
$\beta^0_j$ from the set $\{2/s_0, 4/s_0, \ldots ,2\}$ without
replacement (anew in each simulation run).

\item[(Ab)]  $S_0 = \{1, 2, 11, 12, 21, 22, \dots , 91, 92\}$ and 
for any $j \in S_0$ we sample
$\beta^0_j$ from the set $\{2/s_0, 4/s_0, \ldots ,2\}$ without
replacement (anew in each simulation run).

\item[(Ac)] $\beta^0$ as in (Aa) but we switch the sign of half and randomly
  chosen active parameters (anew in each simulation run). 

\item[(Ad)] $\beta^0$ as in (Ab) but we switch the sign of half and randomly
  chosen active parameters (anew in each simulation run). 

\end{itemize}

The set-up (Aa) has all the active variables in
the first two blocks of highly correlated variables. In the
second configuration (Ab), the first two variables of each of the
first ten blocks are active. Thus, in (Aa), half of the active variables appear
in the same block while in the other case (Ab), the 
active variables are distributed among ten blocks. The remaining two
configurations (Ac) and (Ad) are modifications in terms of
random sign changes. The models (Ab) and (Ad) come closest to the model
(\ref{mod.corr})-(\ref{mod.corr2}) considered for theoretical purposes: the
difference is that the former models have two active variables per active block
(or group) while the latter model has only one active variable per active
group.  

\begin{figure}[!htb]
        \centerline{ 
          \subfigure[(Aa),
          $\sigma=3$]{\includegraphics[scale=0.2]{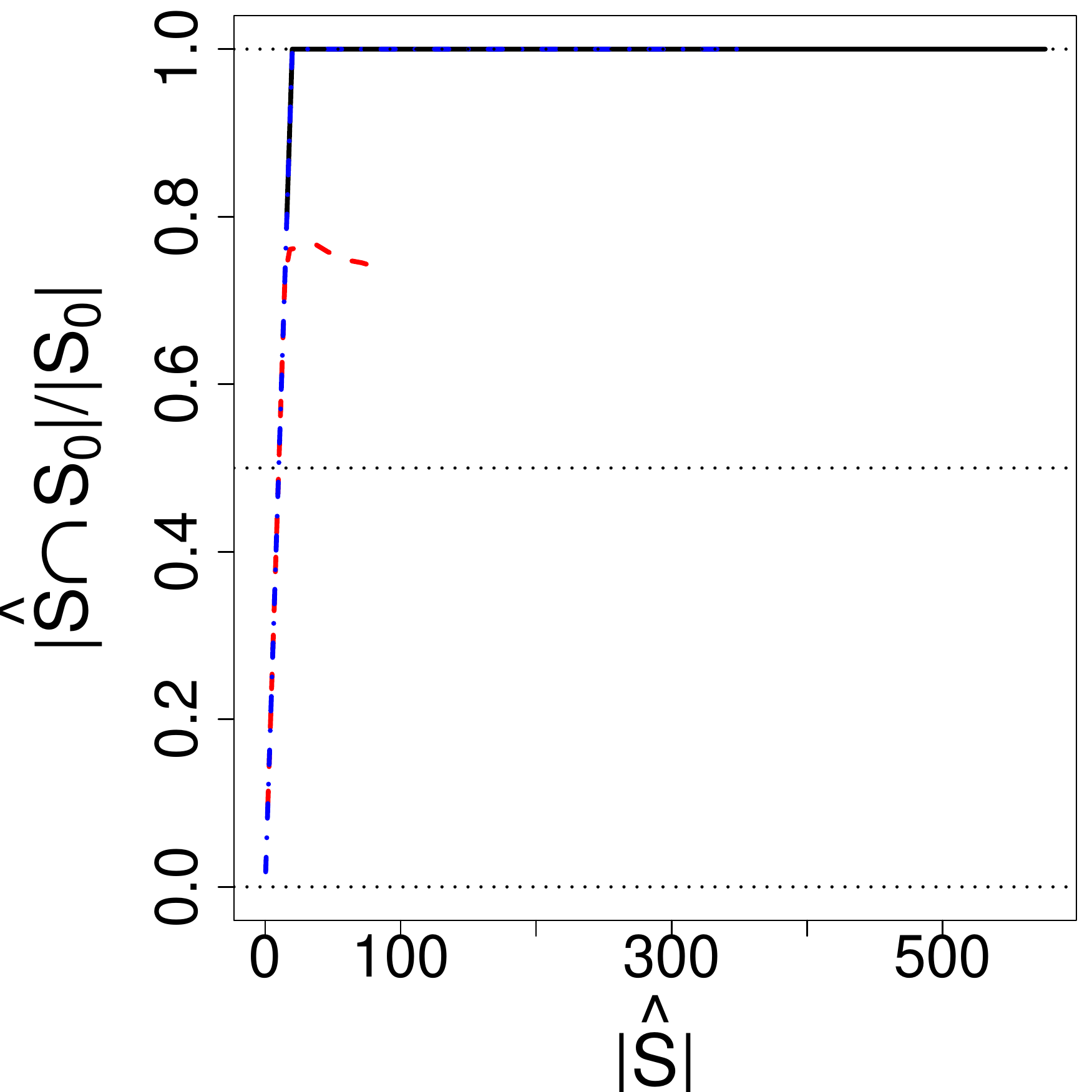}} 
          \subfigure[(Ab),
          $\sigma=3$]{\includegraphics[scale=0.2]{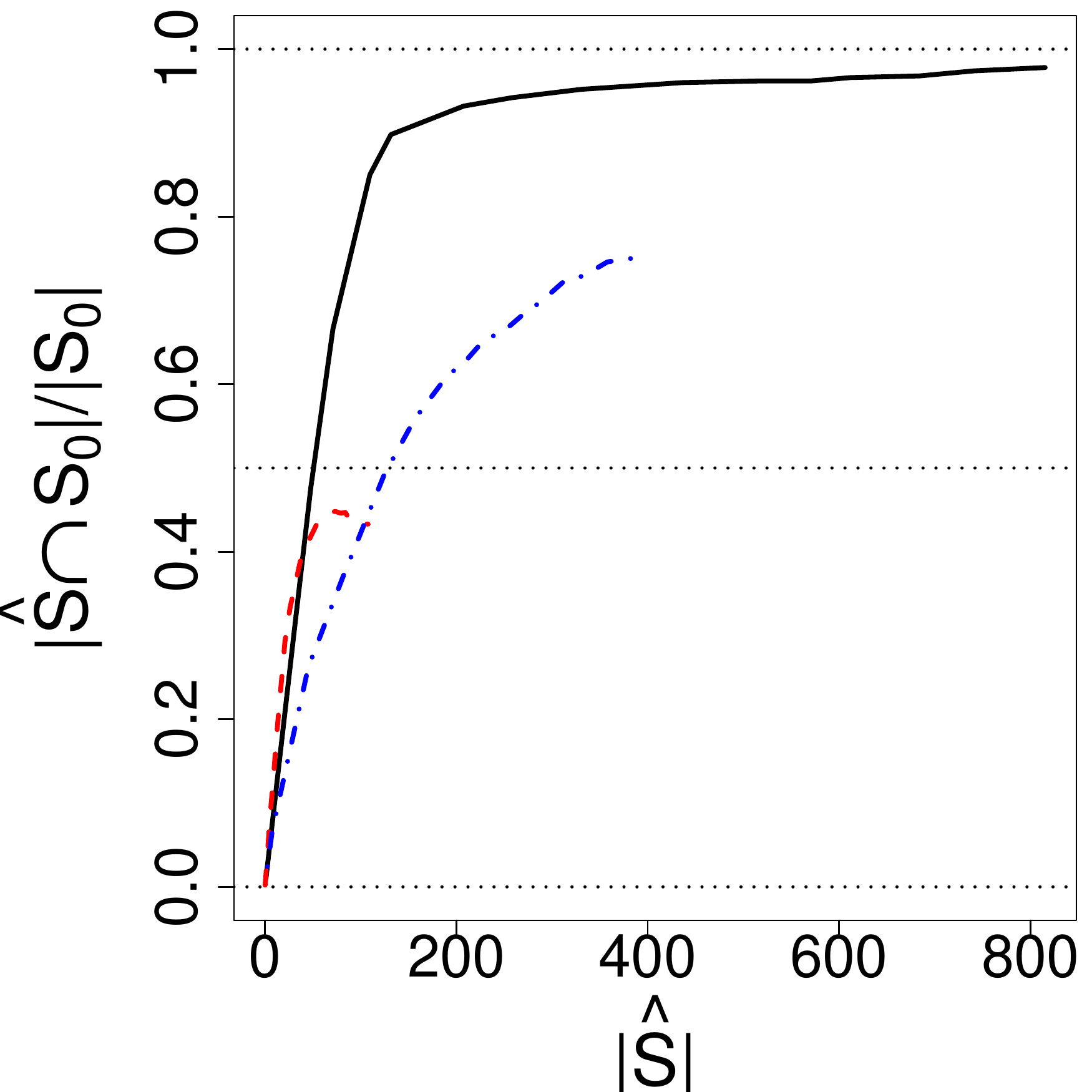}}
          \subfigure[(Ac),
          $\sigma=3$]{\includegraphics[scale=0.2]{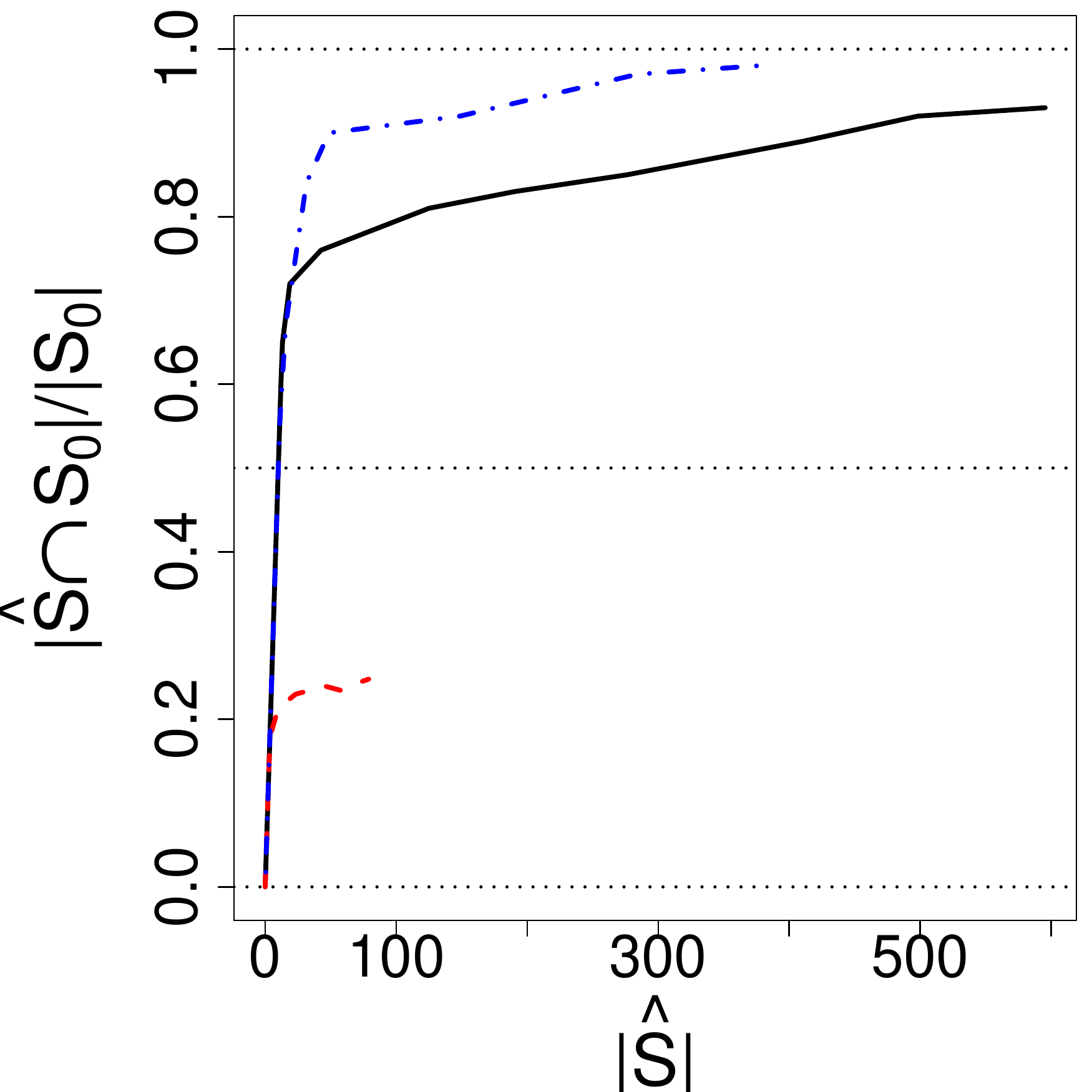}} 
          \subfigure[(Ad),
          $\sigma=3$]{\includegraphics[scale=0.2]{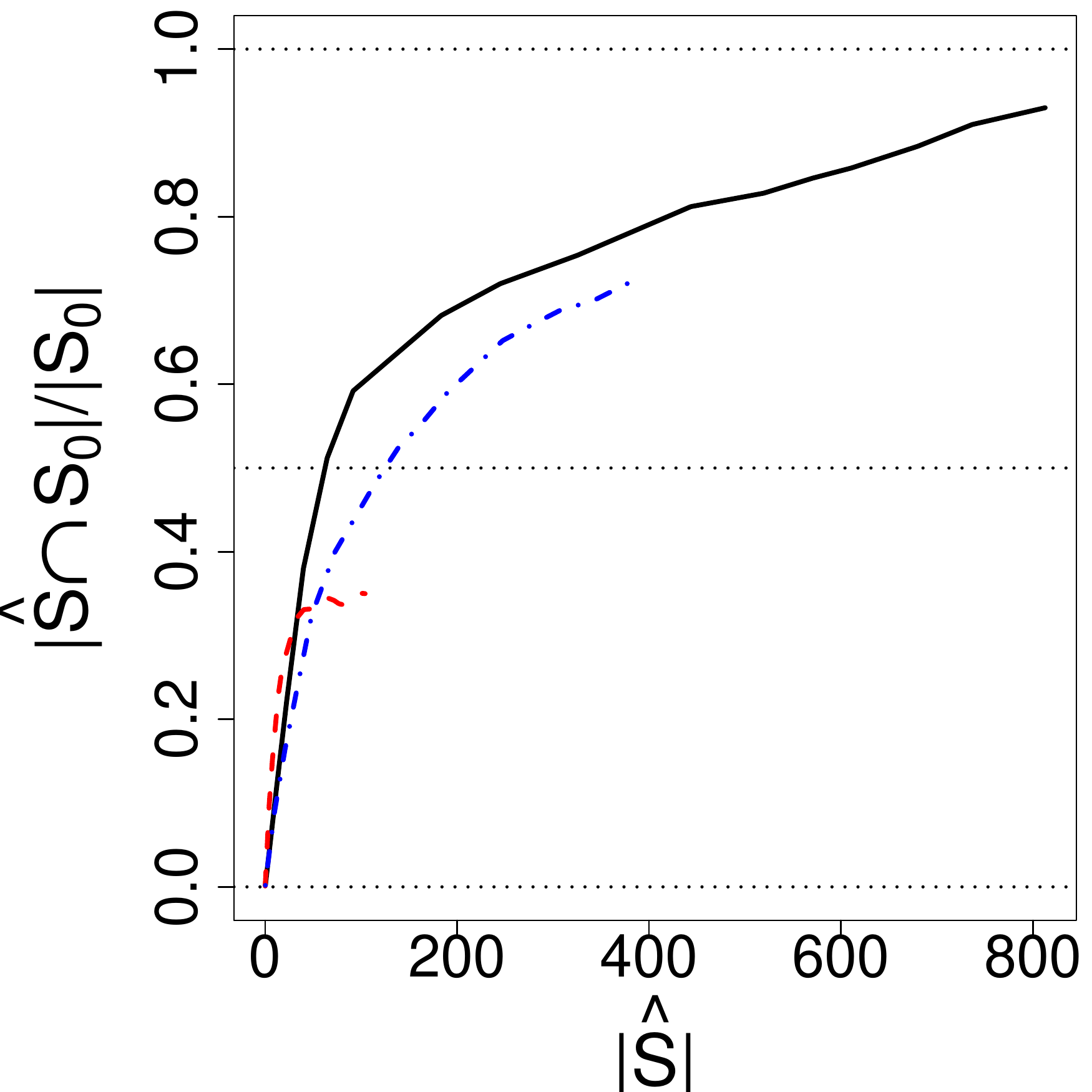}}              
}
\centerline{ 
          \subfigure[(Aa),
          $\sigma=12$]{\includegraphics[scale=0.2]{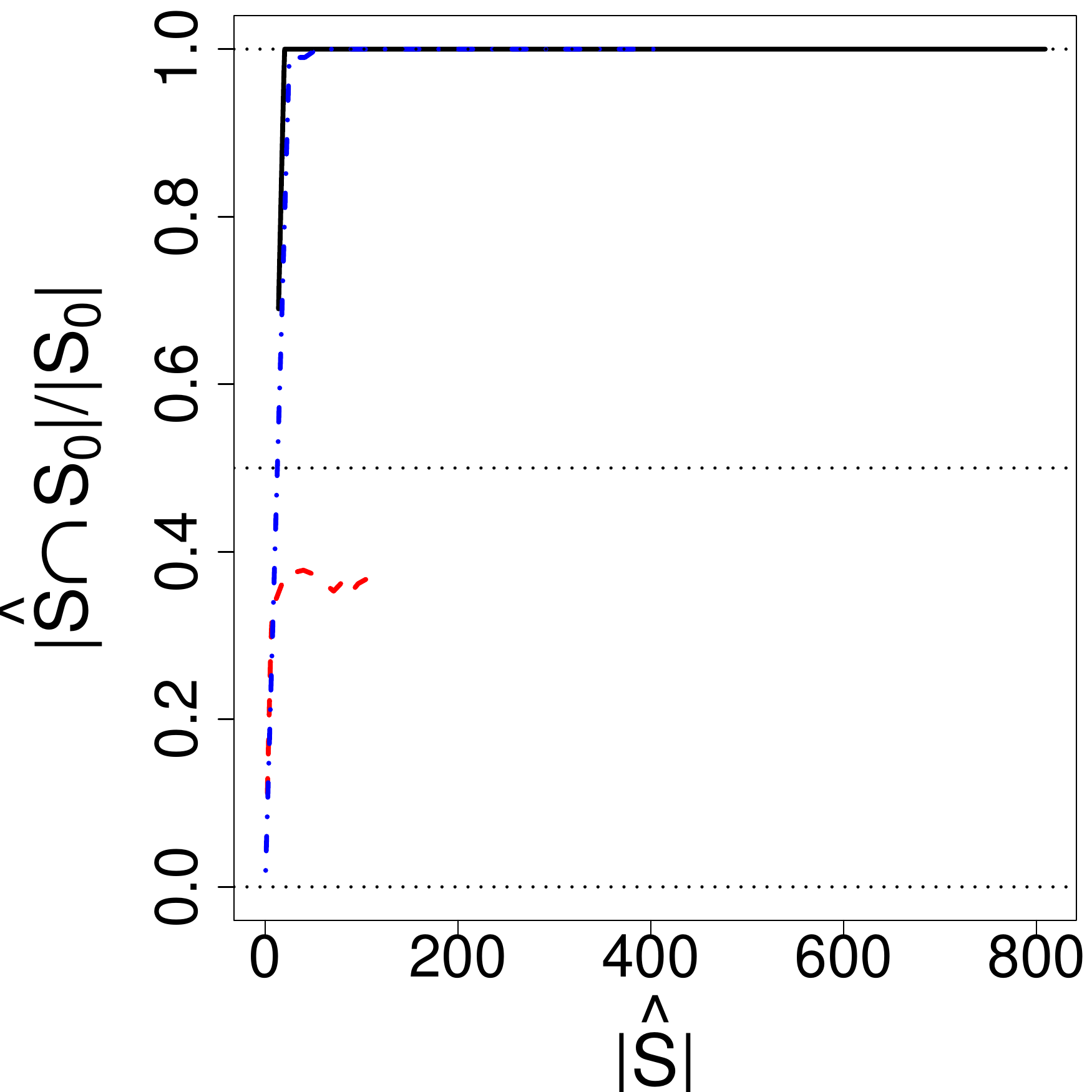}} 
          \subfigure[(Ab),
          $\sigma=12$]{\includegraphics[scale=0.2]{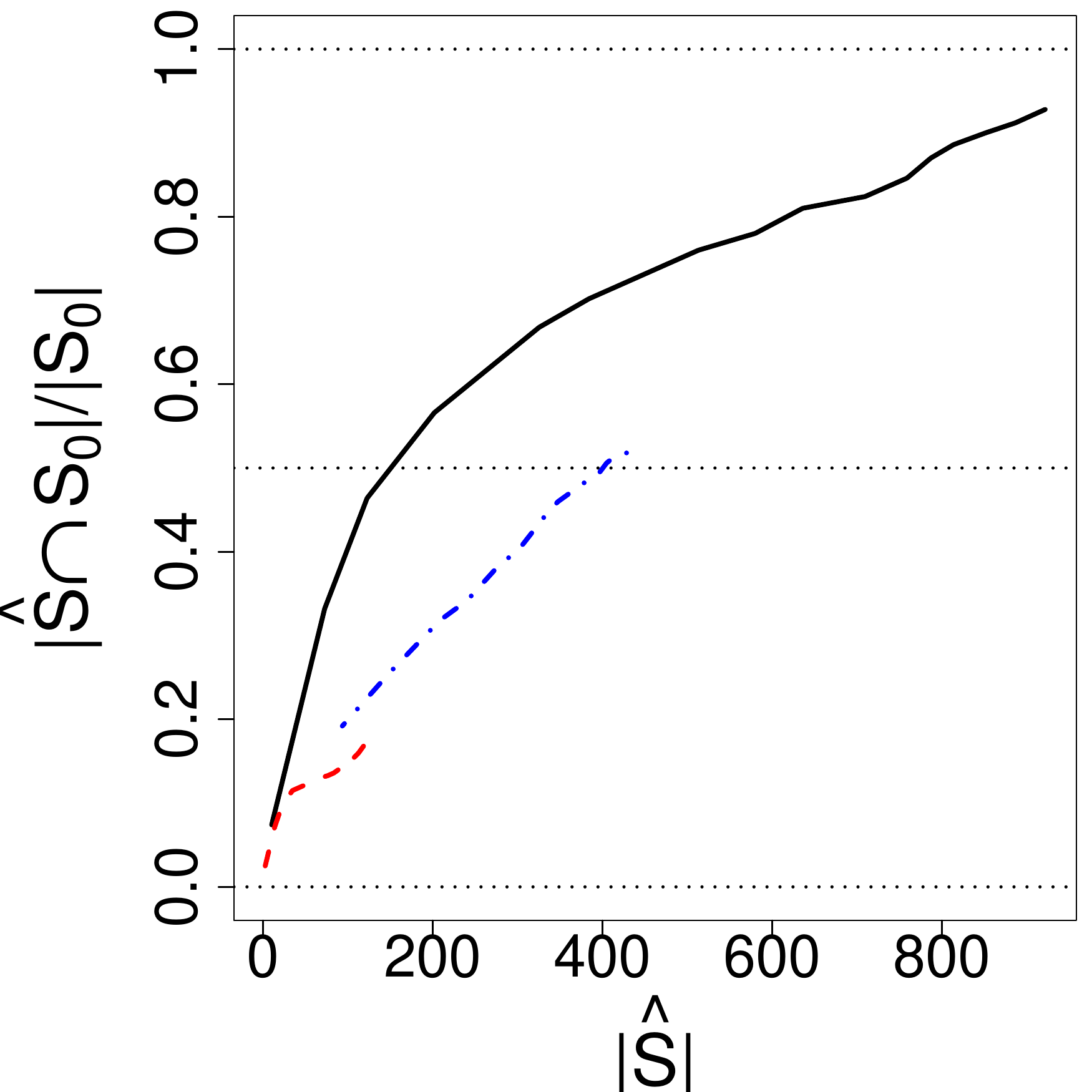}}
          \subfigure[(Ac),
          $\sigma=12$]{\includegraphics[scale=0.2]{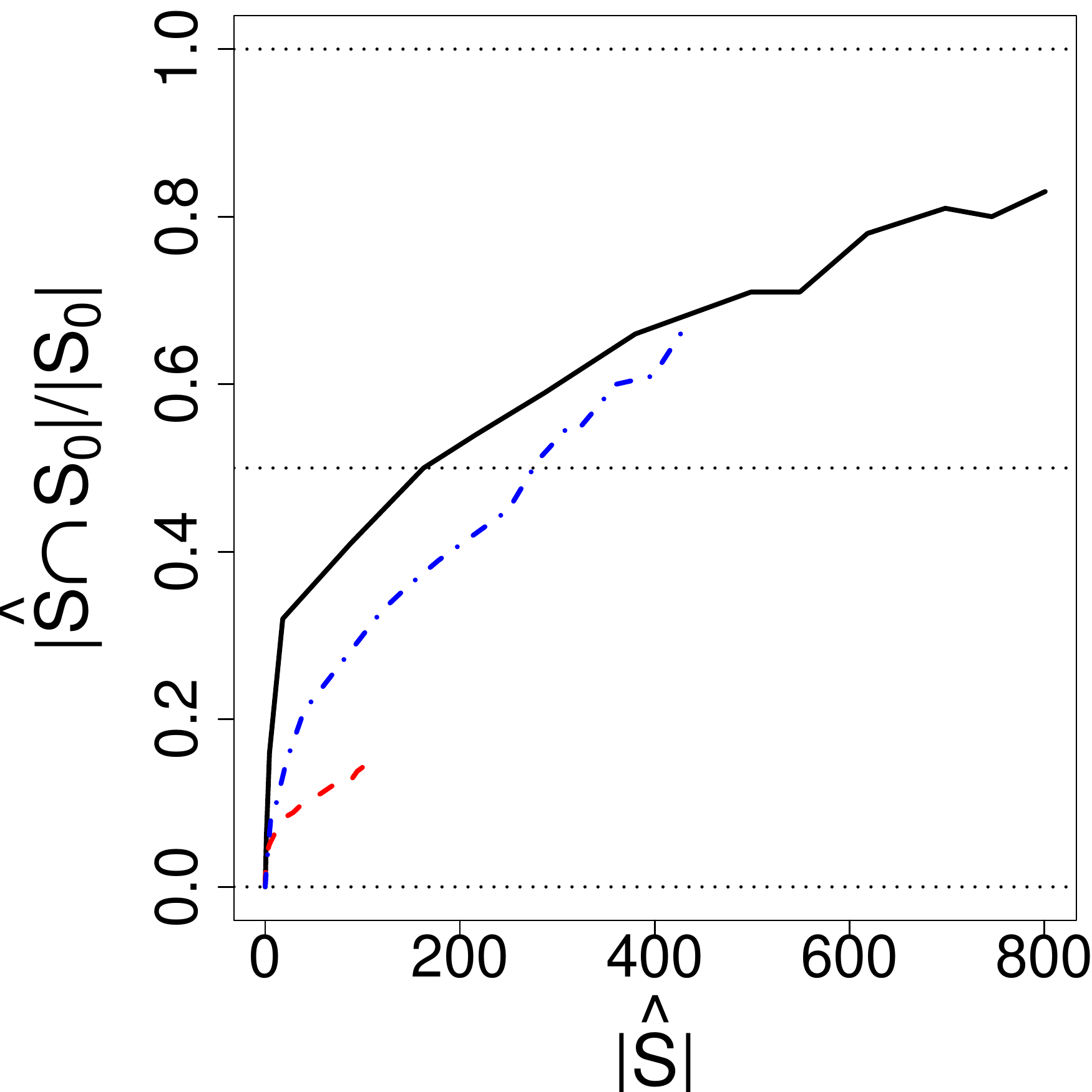}} 
          \subfigure[(Ad),
          $\sigma=12$]{\includegraphics[scale=0.2]{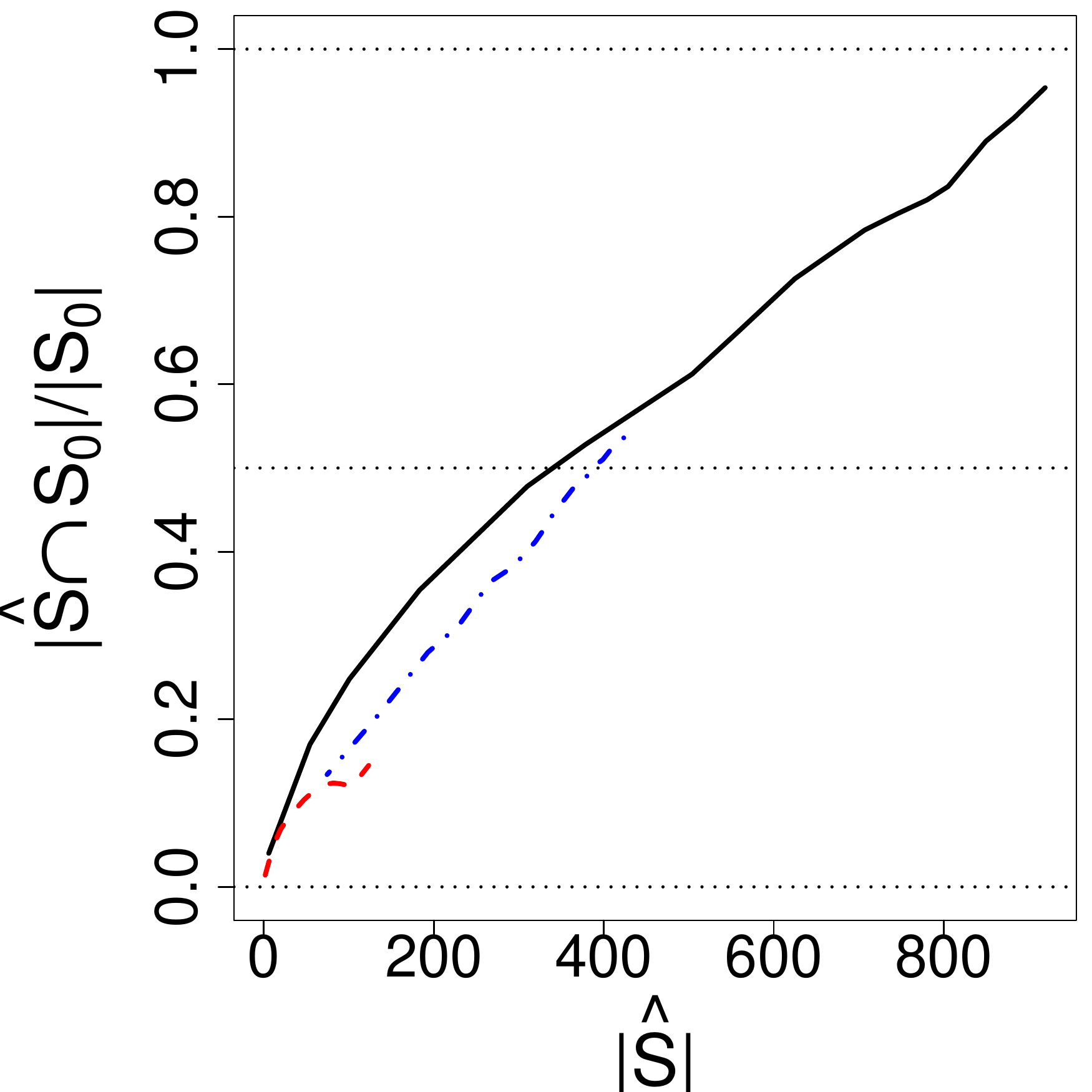}}              
}
        \caption{Plot of $\frac{|\hat{S} \cap
            S_0|}{|\hat{S}|}$ versus $|\hat{S}|$ for block diagonal
          model. Cluster representative Lasso (CRL, black solid line),
          cluster group Lasso (CGL, blue dashed-dotted line), and Lasso
          (red dashed line).}  
        \label{sigmaBlock}
\end{figure}

Table \ref{sB} and Figure \ref{sigmaBlock} describe the
simulation results.
\begin{table}[!htb]
\centering 
\begin{tabular}{c c cccc} 
\hline 
$\sigma$  & Method & (Aa) & (Ab) & (Ac) & (Ad) \\ [0.5ex] 
\hline 
 &CRL & 10.78 (1.61) & 15.57 (2.43) & 13.08 (1.65) & 15.39 (2.35) \\
3 & CGL & 14.97 (2.40) & 37.05 (5.21) & 13.34 (2.06) & 24.31 (6.50)\\
&Lasso & 11.94 (1.97) & 16.23 (2.47) & 12.72 (1.67) & 15.34 (2.53) \\[1ex]
\hline 
&CRL & 161.73 (25.74) & 177.90 (25.87) & 157.86 (20.63) & 165.30 (23.56) \\
12&CGL & 206.19 (29.97) & 186.61 (25.69) & 160.31 (23.04) & 168.26 (24.70) \\
&Lasso & 168.53 (25.88) & 179.47 (25.77) & 158.02 (20.31) & 166.50 (23.74) \\[1ex]
\hline 
\end{tabular}
\caption{MSE for the block diagonal model with standard deviations in brackets.} 
\label{sB} 
\end{table}
From Table \ref{sB} we see that over all the configurations, the
CGL method has lower predictive performance than the other two
methods. Comparing the two methods Lasso and CRL, we can not distinguish a
clear difference with respect to prediction. We also find that sign
switches of half of the active variables (Ac,Ad) do not have a negative
effect on the predictive performance of the CRL method (which in principle
could suffer severely from sign switches). The CGL
method even gains in predictive performance in (Ac) and (Ad) compared to
the no sign-switch configurations (Aa) and (Ab). 

From Figure \ref{sigmaBlock} we infer that for the block diagonal simulation
model the two methods CRL and CGL outperform the Lasso concerning variable
screening. Taking a closer look, the Cluster Lasso methods CRL and CGL
benefit more when having a lot of active variables in a cluster as in
settings (Aa) and (Ac). 

\subsubsection{Single block model}

We simulate the covariables $X \sim
\mathcal{N}_p(0,\Sigma_B)$ where 
\begin{equation*}
\Sigma_{B;i,j}=
\begin{cases}
  1,  & i=j,\\
0.9, & i,j \in \{1, \dots ,30\}\ \text{and}\ i \neq j,\\
  0, & \text{else}.
\end{cases}
\end{equation*}
Such a $\Sigma_B$ corresponds to a single group of strongly correlated
variables of size 30. The rest of the 970 
variables are uncorrelated. For the regression parameter $\beta^0$ we consider the
following configurations: 
\begin{itemize}

\item[(Ba)] $S_0 = \{1,2,\ldots ,15 \} \cup \{31,32,\ldots ,35 \}$ and 
for any $j \in S_0$ we sample
$\beta^0_j$ from the set $\{2/s_0, 4/s_0, \ldots ,2\}$ without
replacement (anew in each simulation run).

\item[(Bb)] $S_0 = \{1,2,\ldots ,5 \} \cup \{31,32,\ldots ,45 \}$ and 
for any $j \in S_0$ we sample
$\beta^0_j$ from the set  $\{2/s_0, 4/s_0, \ldots ,2\}$ without
replacement (anew in each simulation run).
 
\item[(Bc)] $\beta^0$ as in (Ba) but we switch the sign of half and randomly
  chosen active parameters (anew in each simulation run). 

\item[(Bd)] $\beta^0$ as in (Bb) but we switch the sign of half and randomly
  chosen active parameters (anew in each simulation run). 
\end{itemize}

\begin{figure}[!htb]
        \centerline{ 
          \subfigure[(Ba),
          $\sigma=3$]{\includegraphics[scale=0.2]{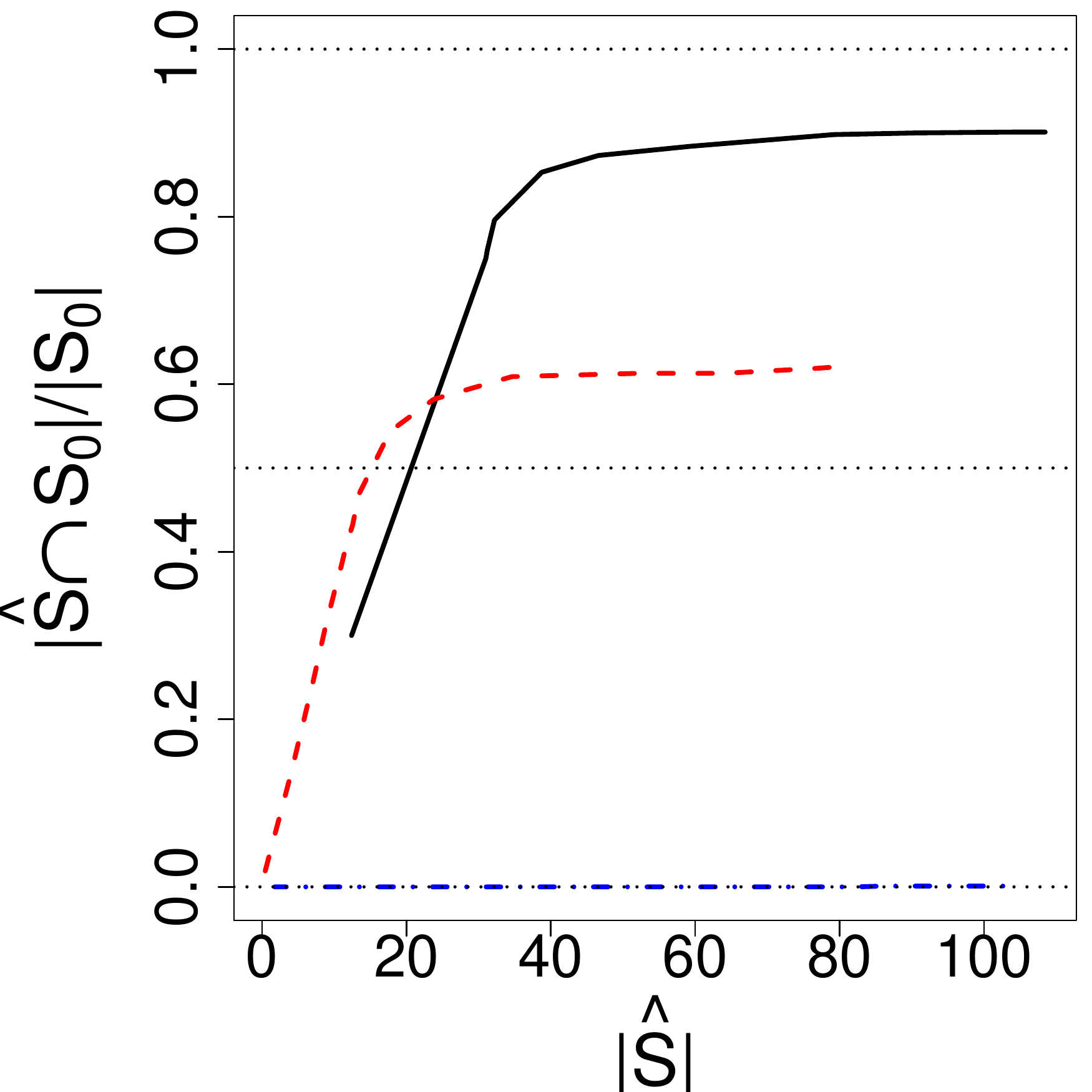}} 
          \subfigure[(Bb),
          $\sigma=3$]{\includegraphics[scale=0.2]{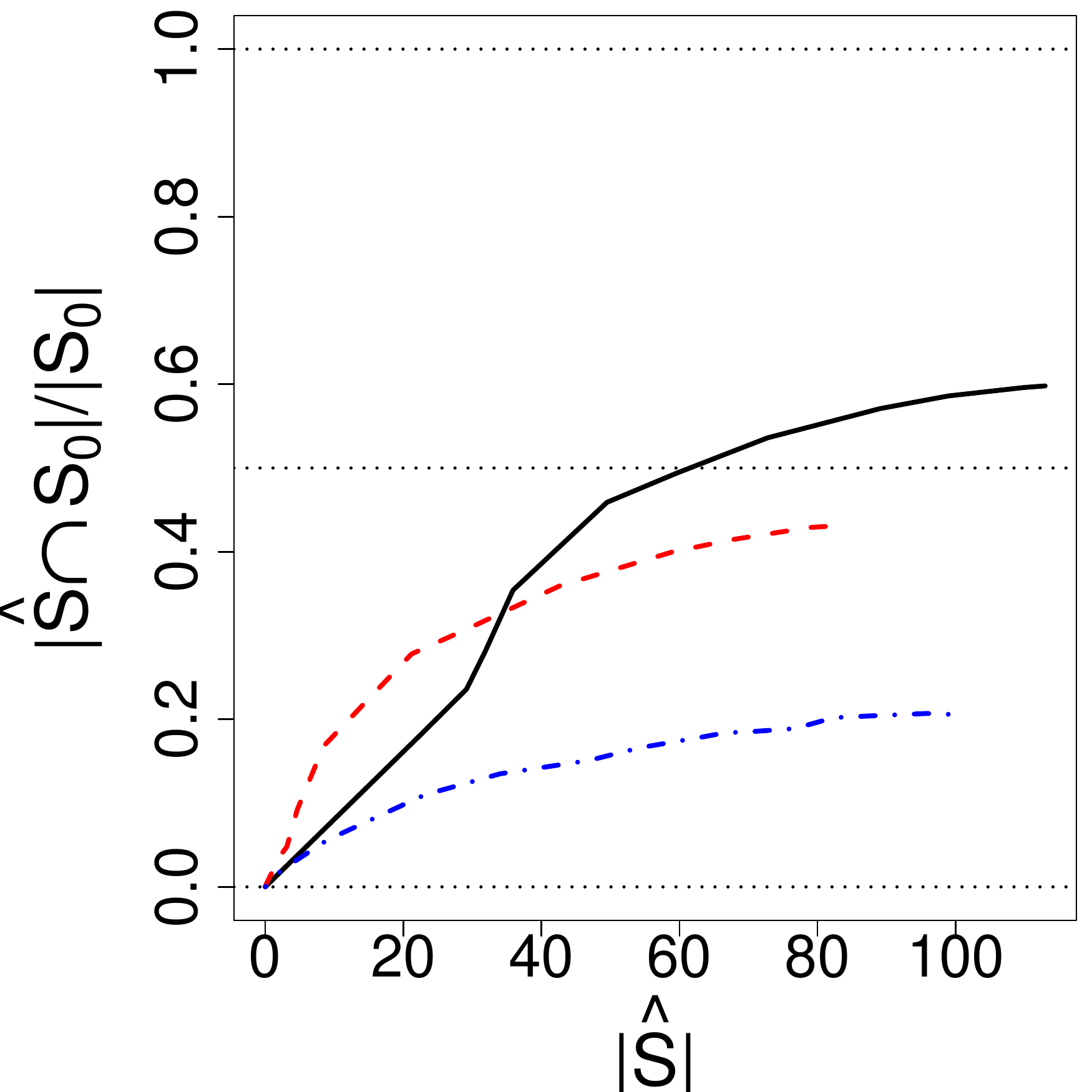}}
          \subfigure[(Bc),
          $\sigma=3$]{\includegraphics[scale=0.2]{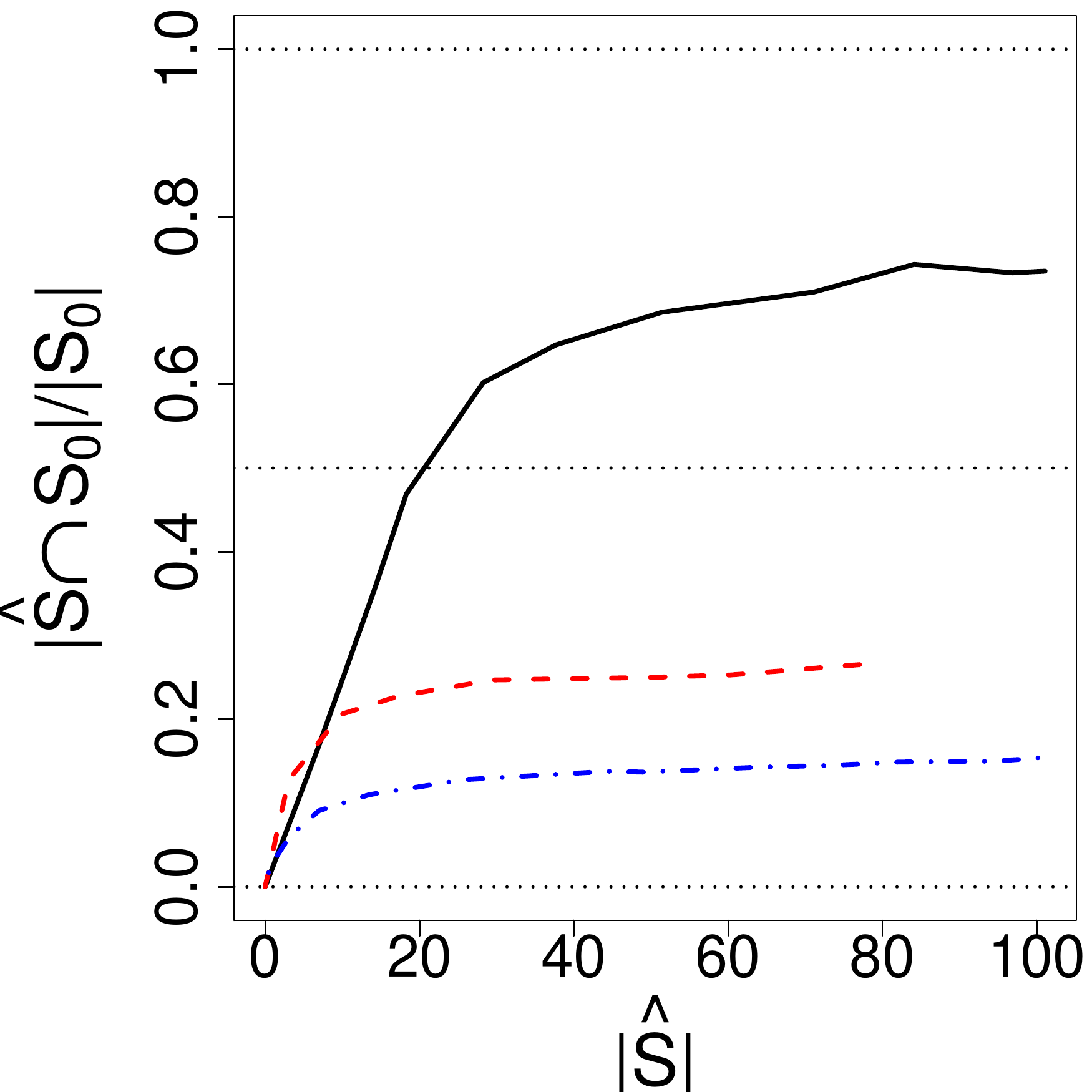}} 
          \subfigure[(Bd),
          $\sigma=3$]{\includegraphics[scale=0.2]{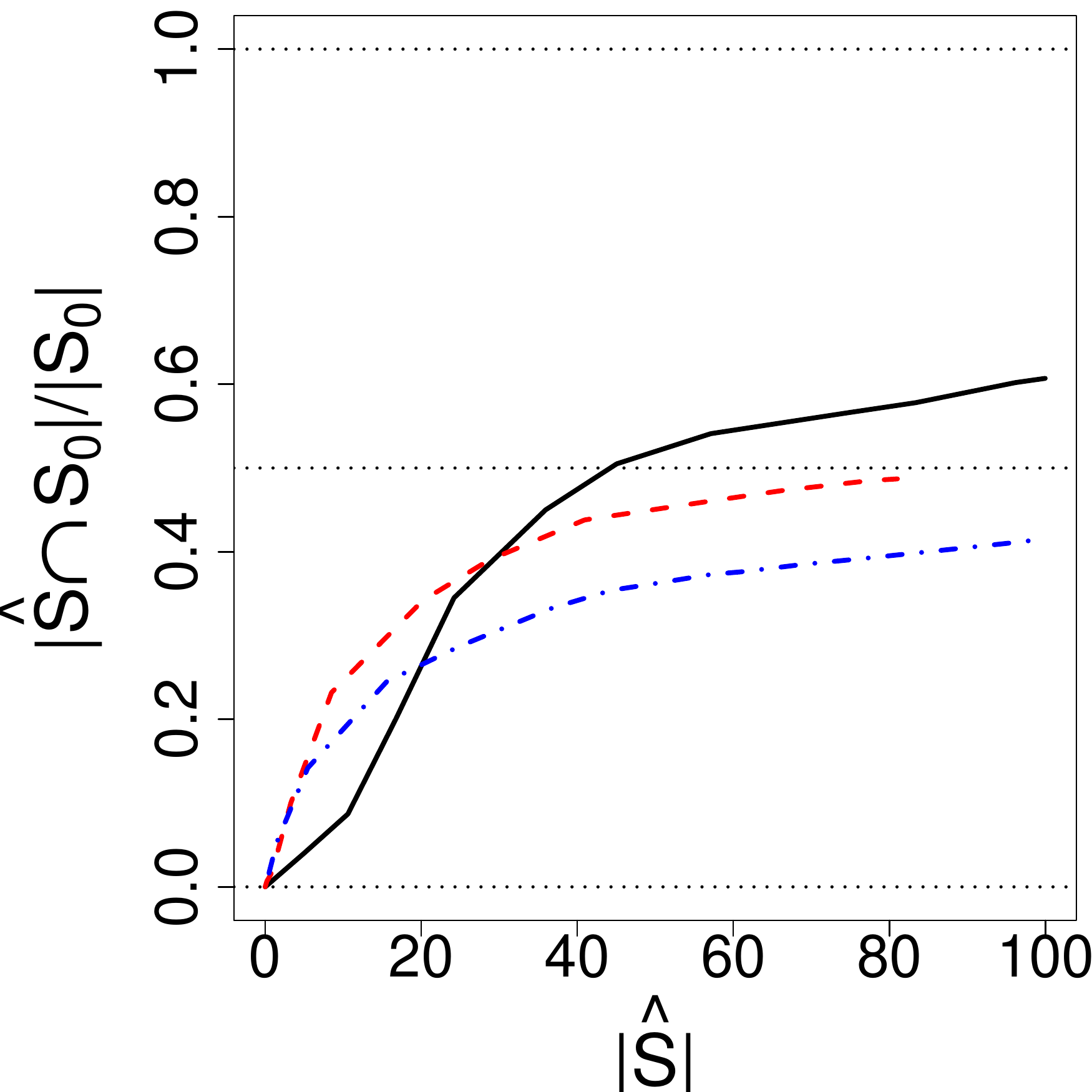}}              
}
\centerline{ 
          \subfigure[(Ba),
          $\sigma=12$]{\includegraphics[scale=0.2]{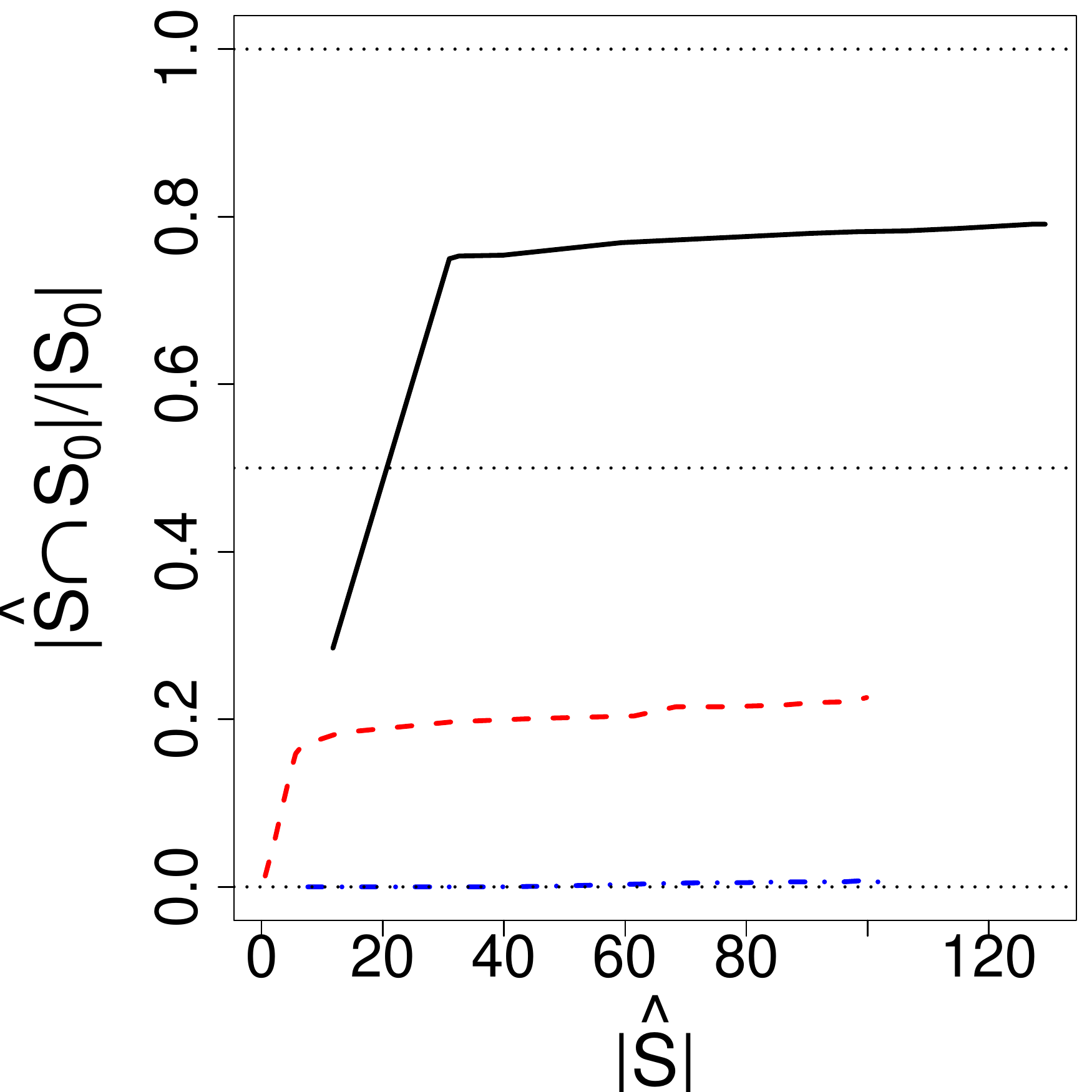}} 
          \subfigure[(Bb),
          $\sigma=12$]{\includegraphics[scale=0.2]{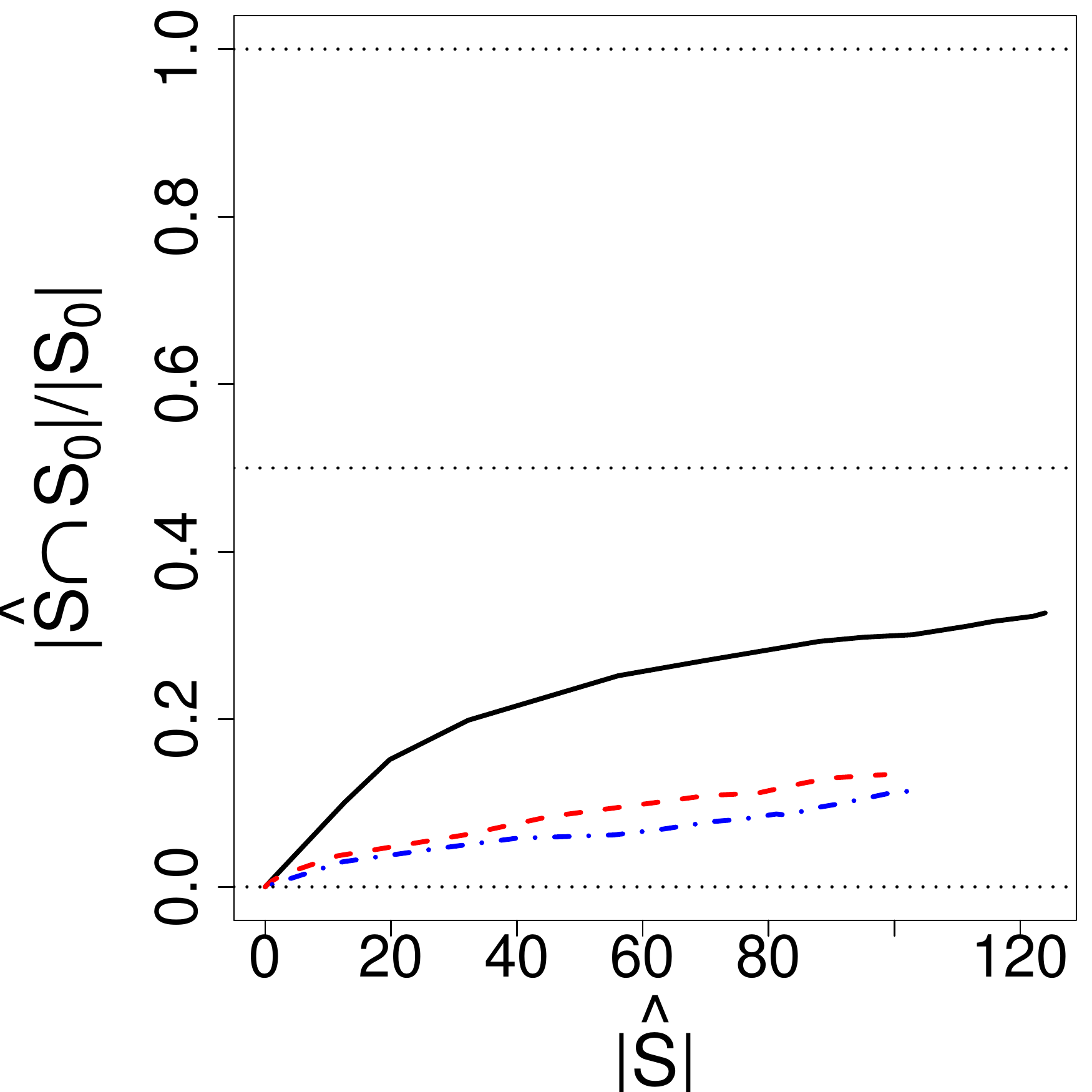}}
          \subfigure[(Bc),
          $\sigma=12$]{\includegraphics[scale=0.2]{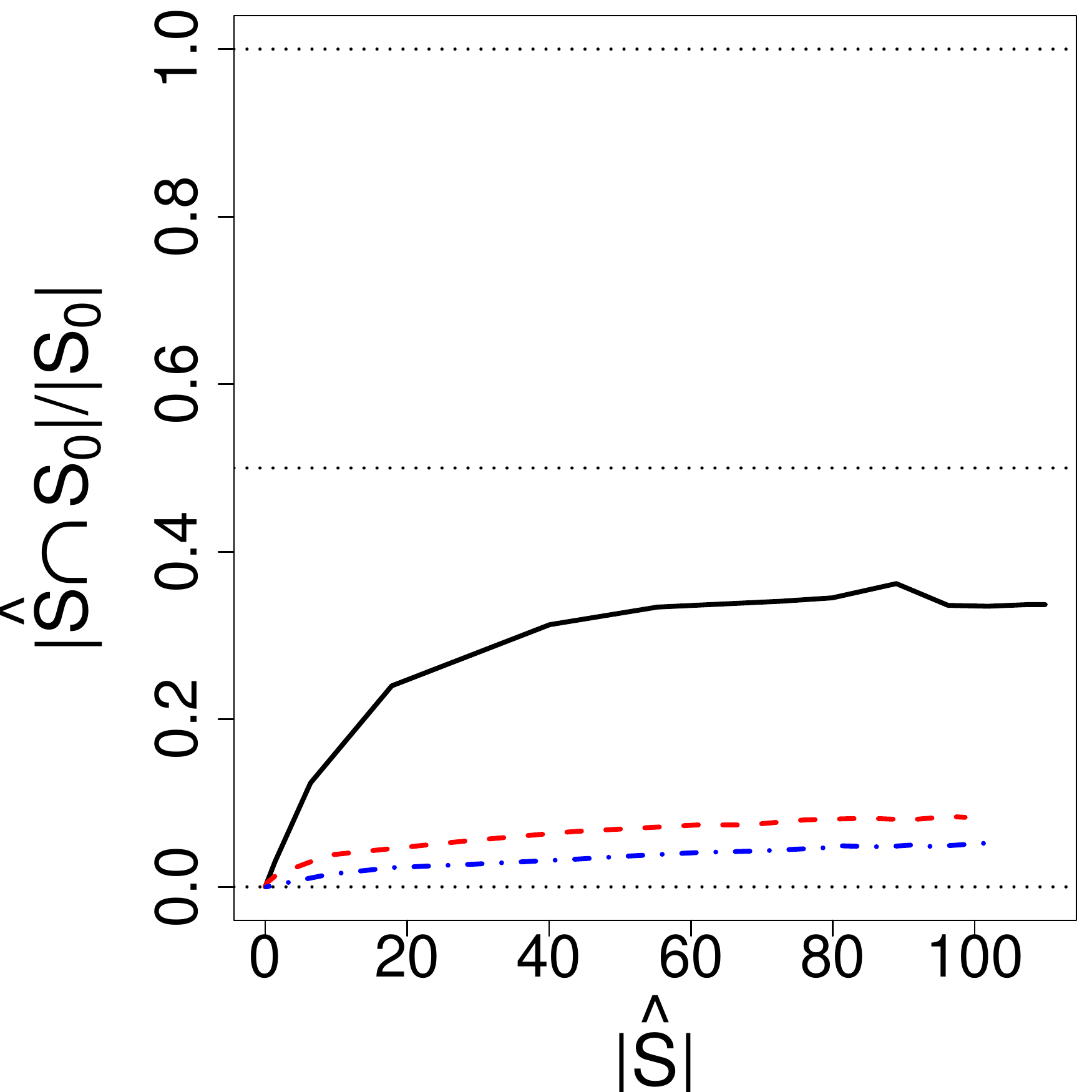}} 
          \subfigure[(Bd),
          $\sigma=12$]{\includegraphics[scale=0.2]{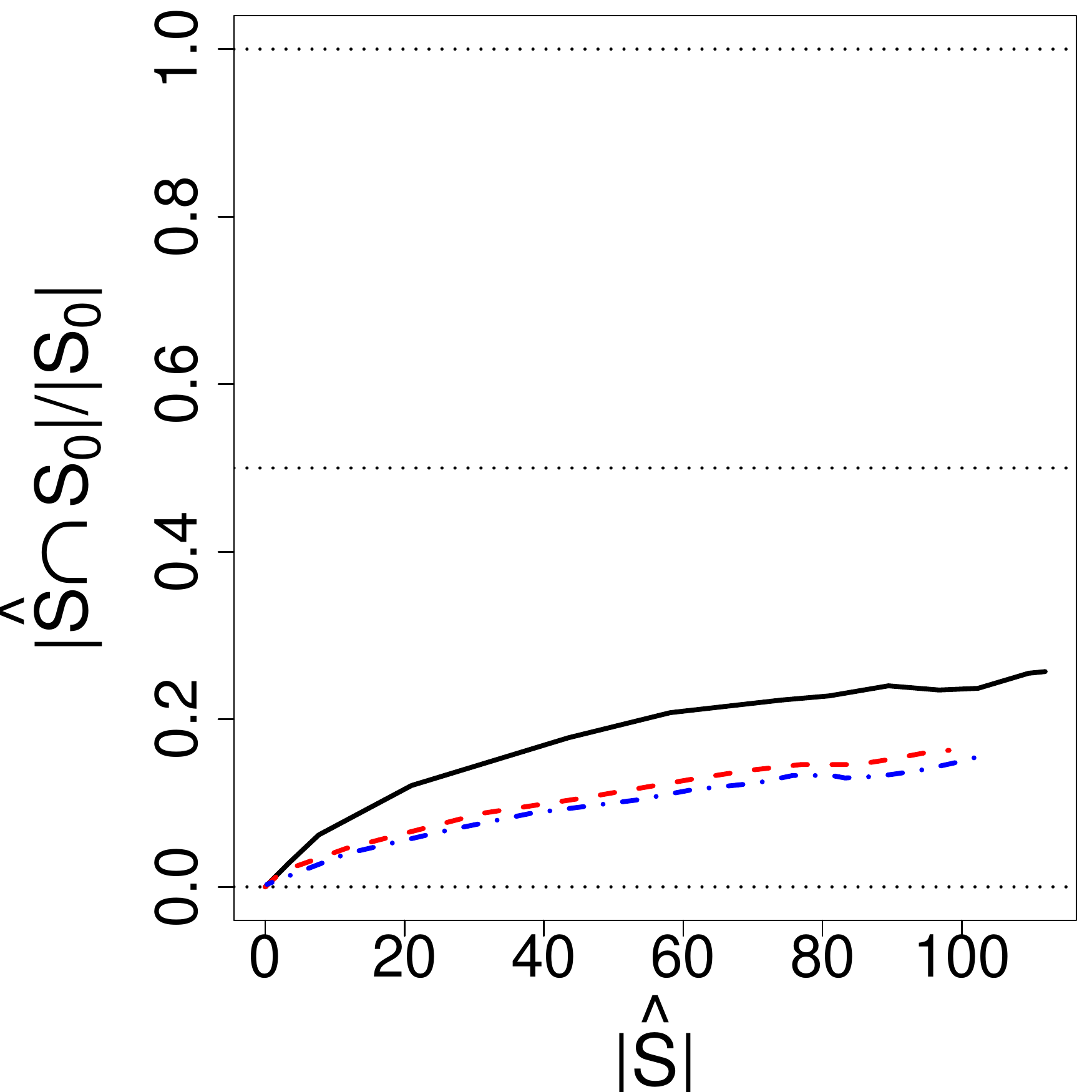}}              
}
        \caption{Plot of $\frac{|\hat{S} \cap
            S_0|}{|\hat{S}|}$ versus $|\hat{S}|$ for single block
          model. Cluster representative Lasso (CRL, black solid line), 
          cluster group Lasso (CGL, blue dashed-dotted line), and Lasso
          (red dashed line).} 
        \label{oneBlock}
\end{figure}

In the fist set-up (Ba), a major fraction of the active variables are in
the same block 
of highly correlated variables. In the second scenario (Bb), most of the
active variables are distributed among the independent variables. 
The remaining two
configurations (Bc) and (Bd) are modifications in terms of
random sign changes. 
The results are described in Table \ref{oB} and Figure \ref{oneBlock}.
\begin{table}[!htb] 
\centering  
\begin{tabular}{c c cccc} 
\hline 
$\sigma$  & Method & (Ba) & (Bb) & (Bc) & (Bd)\\ [0.5ex] 
\hline
 &CRL & 16.73 (2.55) & 27.91 (4.80) & 15.49 (2.93) & 22.17 (4.47) \\
3 &CGL & 247.52 (28.74) & 54.73 (10.59) & 21.37 (9.51) & 31.58 (14.17) \\
&Lasso & 17.13 (3.01) & 27.18 (4.51) & 15.02 (2.74) & 21.91 (4.48) \\[1ex]
\hline 
&CRL & 173.89 (23.69) & 181.62 (24.24) & 161.01 (23.19) & 175.49 (23.61) \\
12& CGL & 384.78 (48.26) & 191.26 (25.55) & 159.40 (23.88) & 174.49 (25.40)
\\
&Lasso & 173.37 (23.23) & 178.86 (23.80) & 160.55 (22.80) & 174.14 (23.14)
\\[1ex] 
\hline 
\end{tabular}
\caption{MSE for the single block model with standard deviations in brackets.}
\label{oB} 
\end{table}

In Table \ref{oB} we see that over all the configurations the
CRL method performs as well as the Lasso, and both of them outperform the
CGL. We again find that the CGL method gains in predictive performance when the
signs of the coefficient vector are not the same everywhere, and this
benefit is more pronounced when compared to the the block diagonal model.  

The plots in Figure \ref{oneBlock} for variable screening show that the CRL
method performs better than the Lasso for all of the configurations. The
CGL method is clearly inferior to the Lasso especially in the setting (Ba)
where the CGL seems to have severe problems in finding the true active
variables. 

\subsubsection{Duo block model}

\begin{figure}[!htb]
        \centerline{           
          \subfigure[(C),
          $\sigma=3$]{\includegraphics[scale=0.3]{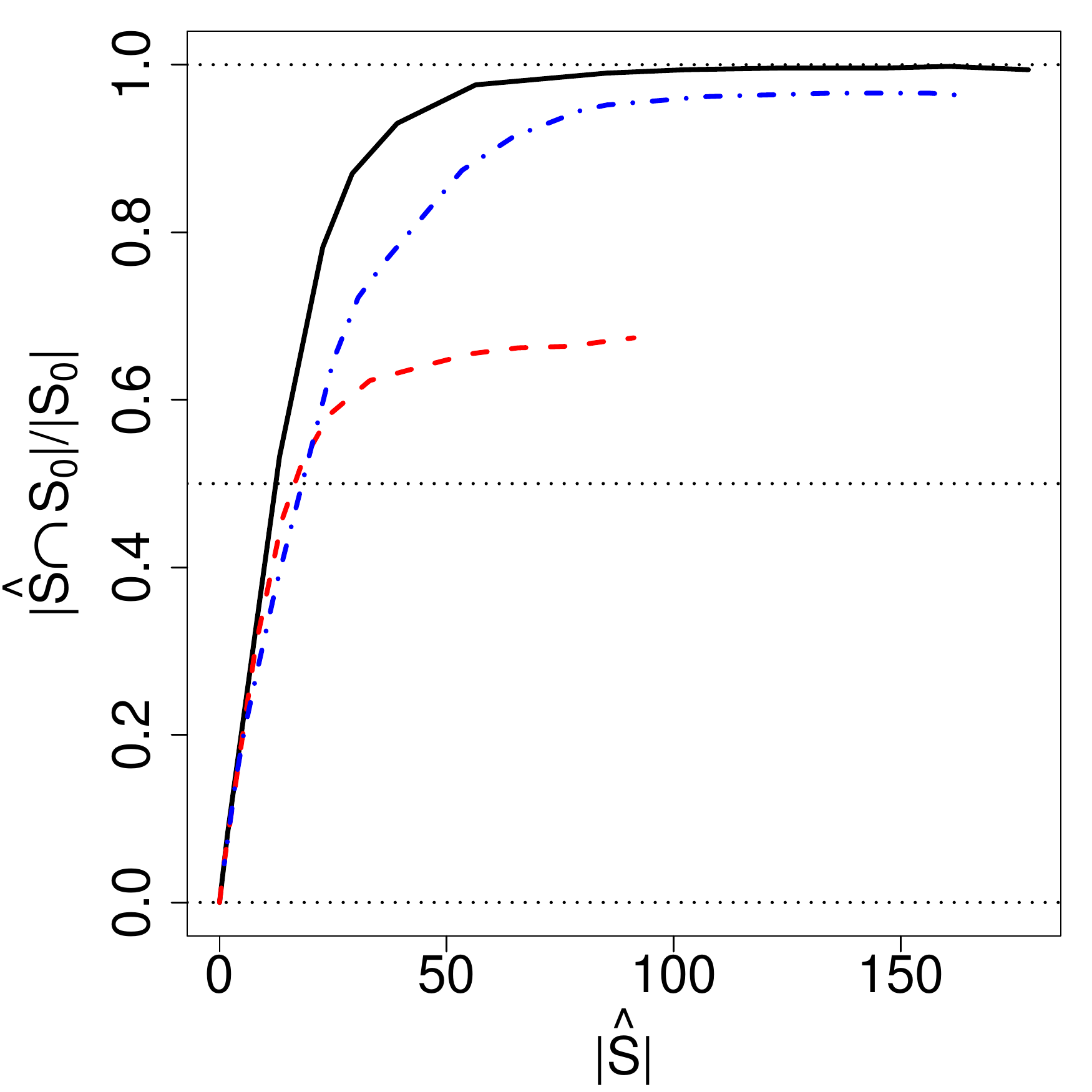}}
          \subfigure[(C),
          $\sigma=12$]{\includegraphics[scale=0.3]{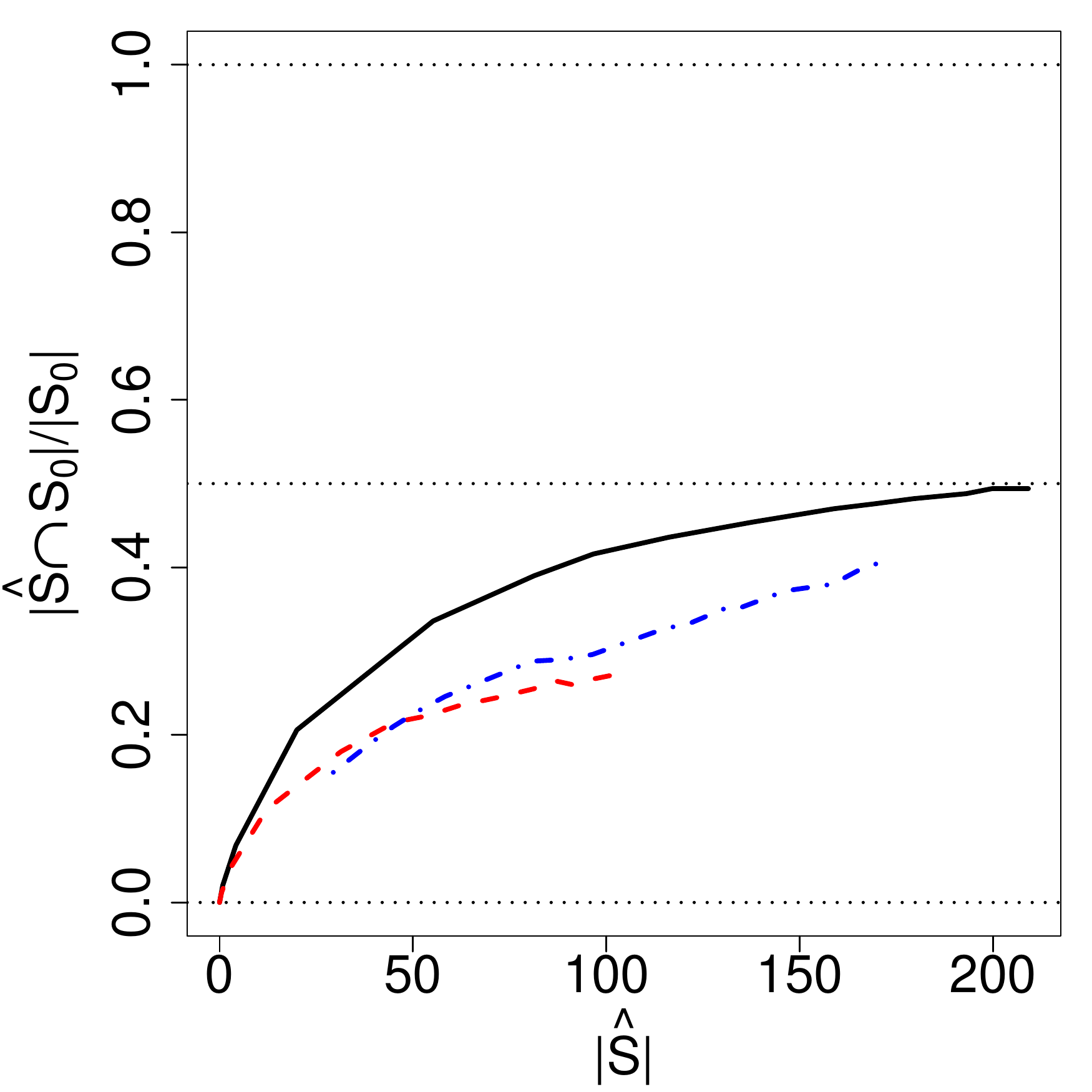}}              
}

        \caption{Plot of $\frac{|\hat{S} \cap
            S_0|}{|\hat{S}|}$ versus $|\hat{S}|$ for duo block
          model. Cluster representative Lasso (CRL, black solid line), 
          cluster group Lasso (CGL, blue dashed-dotted line), and Lasso
          (red dashed line).} 
        \label{duoBlock}
\end{figure}

We simulate the covariables $X$ according
to $\mathcal{N}_p(0,\Sigma_C)$ where $\Sigma_C$ is a block diagonal
matrix. We use the $2 \times 2$ block matrices 
\begin{equation*}
\Gamma = 
\begin{bmatrix} 
1 & 0.9 \\ 
0.9 & 1 
\end{bmatrix},
\end{equation*}
and the block diagonal of $\Sigma_C$ consists now of 500 such block matrices
$\Gamma$. In this setting we only look at one set-up for the parameter
$\beta^0$:
\begin{itemize}

\item[(C)] $S_0 = \{1,\ldots ,20\}$ with 
$\beta^0_{j}=
\begin{cases}
  2,  & j \in \{1,3,5,7,9,11,13,15,17,19\},\\
  \frac{\frac{1}{3}\sqrt{\frac{\log{p}}{n}}\sigma}{1.9}, & j \in \{2,4,6,8,10,12,14,16,18,20\}.
\end{cases}$
\end{itemize}
The idea of choosing the parameters in this way is given by the fact that
the Lasso would typically not select the variables from $\{2,4,6,\ldots,
20\}$ but selecting the other from $\{1,3,5,\ldots, 19\}$. 
The following Table \ref{dB} and
Figure \ref{duoBlock} show the simulation results for the duo block model.
\begin{table}[!htb]
\centering  
\begin{tabular}{c c c} 
\hline 
$\sigma$  & Method & (C) \\ [0.5ex] 
\hline  
 &CRL & 22.45 (4.26)  \\
3 &CGL & 32.00 (6.50)  \\
&Lasso & 22.45 (4.64)  \\[1ex]
\hline 
&CRL & 190.93 (25.45) \\
12&CGL & 193.97 (27.05)  \\
&Lasso & 190.91 (25.64) \\[1ex]
\hline  
\end{tabular}
\caption{MSE for duo block model with standard deviations in brackets.} 
\label{dB}
\end{table}

From Table \ref{dB} we infer that for the duo block model, all three
estimation methods have a similar prediction performance. Especially for
$\sigma=12$ we see no difference between the methods. But in terms of
variable screening, we see in Figure \ref{duoBlock} that the two techniques
CRL and CGL are clearly better than the Lasso. 

\subsection{Pseudo-real data}\label{realdata}

For the pseudo real data example described below, we also consider the CRL
  method with ordinary hierarchical clustering as detailed in Section
  \ref{sec.ohclust}. We denote the method by CRLcor.
 
We consider here an example with real data design matrix $\bx$ but synthetic
regression coefficients $\beta^0$ and simulated Gaussian errors
${\cal N}_n(0,\sigma^2I)$ in a linear model as in (\ref{mod1}).
For the real data design matrix $\bx$ we consider a data set about
riboflavin (vitamin B2) production by bacillus subtilis. That data has been
provided by DSM (Switzerland). The covariates are measurements of the logarithmic
expression level of $4088$ genes (and the response variable $\by$
is the logarithm of the riboflavin production rate, but we do not use it
here). The data consists of 
$n=71$ samples of genetically engineered mutants of bacillus
subtilis. There are different strains of bacillus subtilis which are 
cultured under different fermentation conditions, which makes the
population rather heterogeneous.

We reduce the dimension to $p = 1000$ covariates which have largest
empirical variances and choose the size of the active set as $s_0 = 10$. 
\begin{itemize}
\item[(D1)] $S_0$ is chosen as a randomly selected
variable $k$ and the nine covariates which have highest absolute
correlation to variable $k$ (anew in each simulation run). For each $j \in
S_0$ we use $\beta^0_j  = 1$.

\item[(D2)] $S_0$ is chosen as one randomly selected entry
    in each of the five biggest clusters of both clustering methods (using
    either Algorithm \ref{alg1} or hierarchical clustering as described in
    Section \ref{sec.ohclust}) 
    resulting in $s_0 = 10$ (anew in each simulation run). For each $j \in
    S_0$ we use $\beta^0_j  = 1$. 
\end{itemize}

\begin{figure}[!htb]
        \centerline{          
          \subfigure[(D1),
          $\sigma=3$]{\includegraphics[scale=0.3]{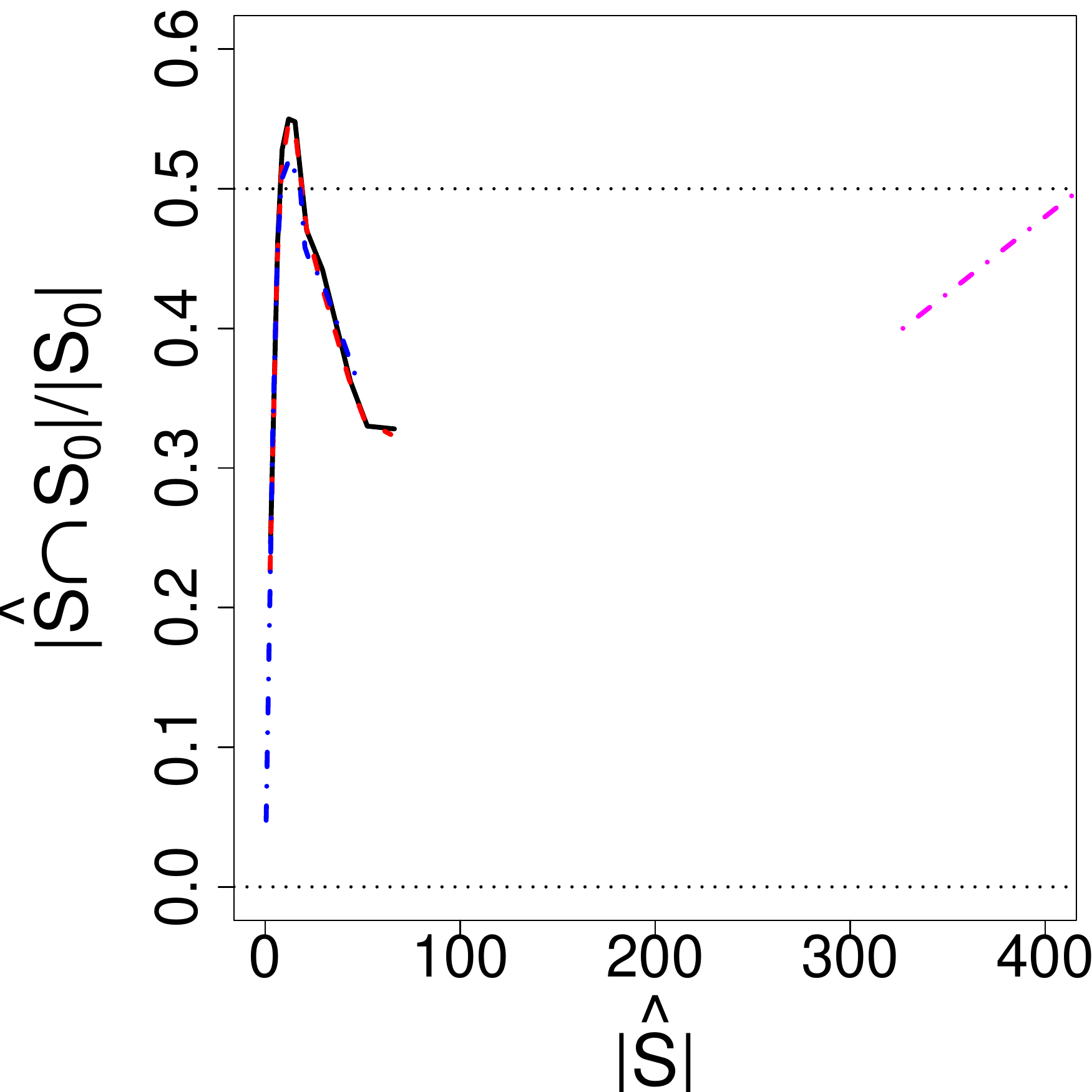}}
          \subfigure[(D1),
          $\sigma=12$]{\includegraphics[scale=0.3]{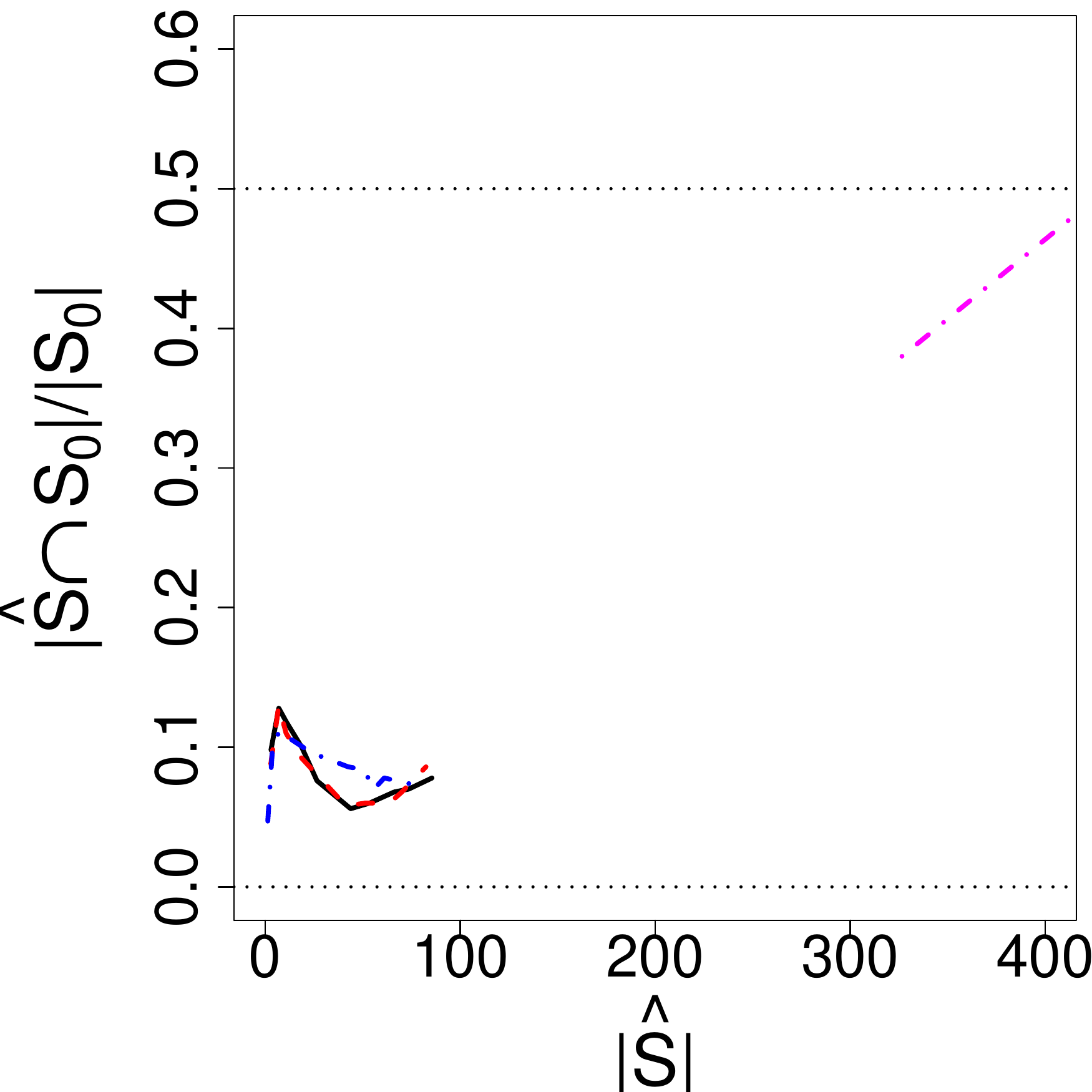}}              
}
        \centerline{          
          \subfigure[(D2),
          $\sigma=3$]{\includegraphics[scale=0.3]{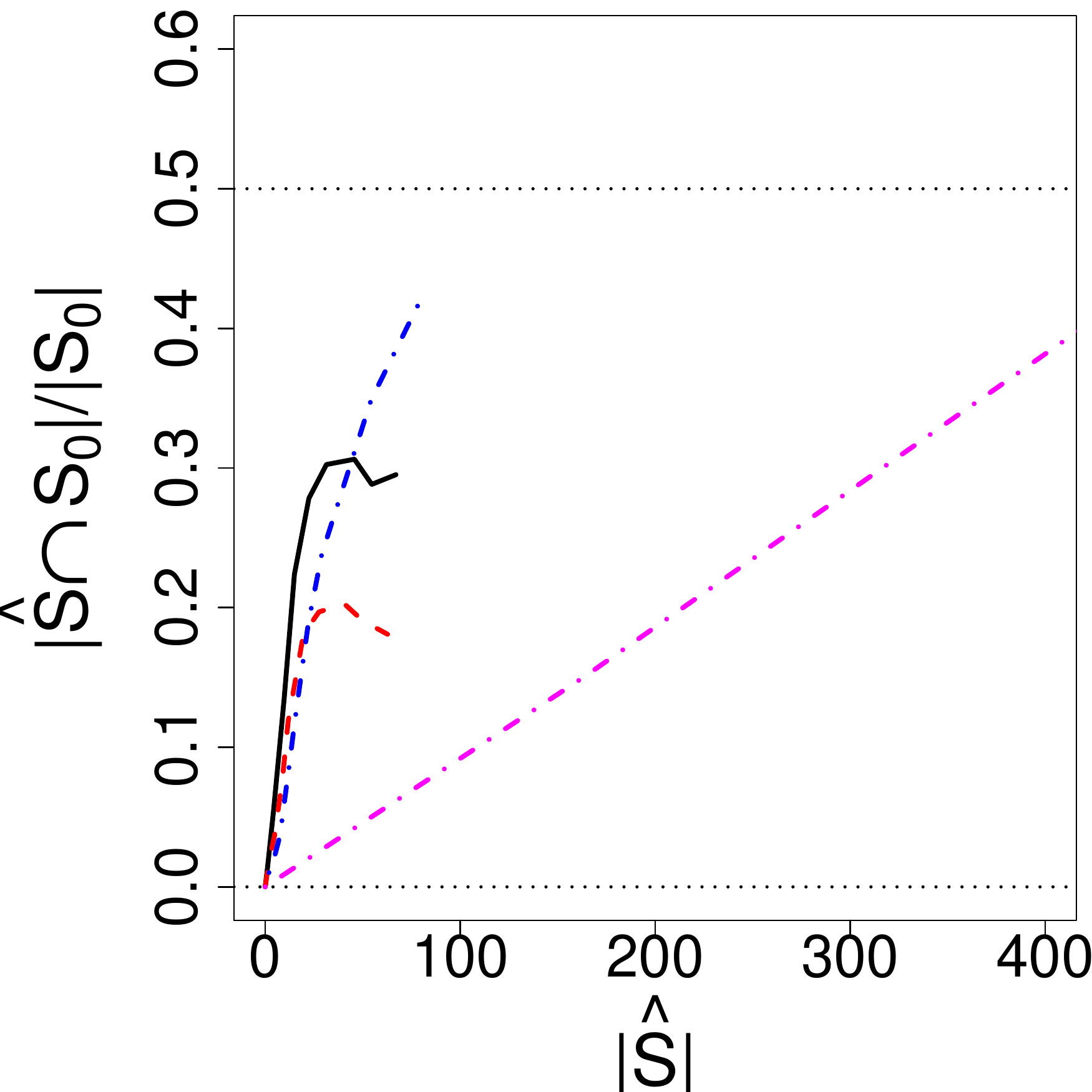}}
          \subfigure[(D2),
          $\sigma=12$]{\includegraphics[scale=0.3]{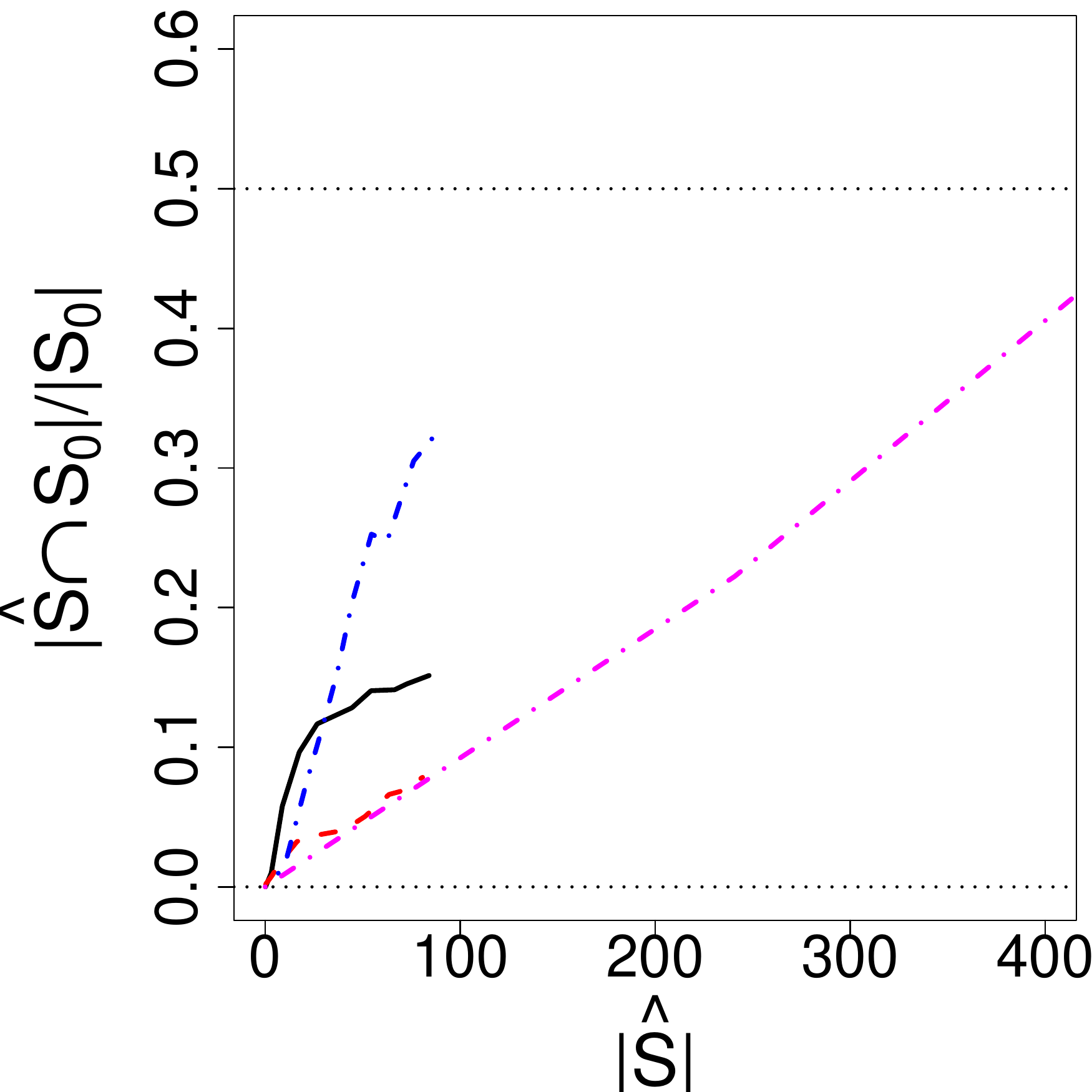}}              
}
        \caption{Plots of $\frac{|\hat{S} \cap
            S_0|}{|S_0|}$ versus $|\hat{S}|$ for the pseudo real riboflavin
          data. Cluster representative Lasso (CRL, black solid line),  
          cluster group Lasso (CGL, blue dashed-dotted line), Lasso
          (red dashed line) and CRLcor (magenta dashed-dotted line).}  
        \label{riboflavin}
\end{figure}

The results are given in Table \ref{riboTable}
and Figure \ref{riboflavin}, based on 50 independent simulation runs. 

\begin{table}[!htb]
\centering  
\begin{tabular}{c c c c}  
\hline 
$\sigma$  & Method & (D1) & (D2) \\ [0.5ex] 
\hline  
 &CRL & 2.47 (0.94) & 2.99 (0.72) \\
3 &CGL & 2.36 (0.93) & 3.13 (0.74) \\ 
&Lasso & 2.47 (0.94) & 2.96 (0.60)\\
&CRLcor & 39.02 (25.15) & 7.08 (2.76)\\[1ex]
\hline 
&CRL & 19.62 (10.11) & 14.80 (4.91)\\
12& CGL & 17.49 (9.28) & 14.90 (5.44)\\
&Lasso & 19.63 (10.00) & 15.66 (4.84)\\
&CRLcor & 50.40 (27.68) & 15.46 (5.74)\\[1ex]
\hline 
\end{tabular}
\caption{Prediction error for the pseudo real riboflavin
  data with standard deviations in brackets.} 
\label{riboTable}
\end{table}

Table \ref{riboTable} shows that we do not really gain any predictive power
when using the proposed cluster lasso methods CRL or CGL: this finding is
consistent with the reported results for simulated data in Section
\ref{simulationstudy}. The method CRLcor, using standard hierarchical
clustering based on correlations (see Section \ref{sec.ohclust}) performs very
poorly: the reason is that the automatically chosen number of clusters
results in a partition with one very large cluster, and the representative mean
value of such a very large cluster seems to be
inappropriate. Using the group Lasso for such a partition (i.e.,
  clustering) is ill-posed as well since the group size of such a large
  cluster is larger than sample size $n$. 

Figure \ref{riboflavin} shows a somewhat different picture for variable
screening. For the setting (D1), all methods except CRLcor perform
similarly, but for (D2), the
two cluster Lasso methods CRL and CGL perform better than plain
Lasso. Especially for the low noise level
$\sigma=3$ case, we see a substantial performance gain of the CRL and CGL
compared 
to the Lasso. Nevertheless, the improvement over plain Lasso is less
pronounced than for quite a few of the simulated models in Section
\ref{simulationstudy}. The CRLcor method is again performing very poorly:
the reason is the same as mentioned above for prediction while in addition,
if the large cluster is selected, it results in a large contribution of the
cardinality $|\hat{S}|$.  

\subsection{Summarizing the empirical results}

We clearly see that in the pseudo real data example and most of the
simulation settings, the cluster Lasso techniques (CRL and CGL) outperform
the Lasso in 
terms of variable screening; the gain is less pronounced for the pseudo
real data example. Considering prediction, the
CRL  and the Lasso display similar performance while the CGL is not keeping
up with them. Such a deficit of the CGL 
method seems to be caused for cases where we have many non-active variables
in an active group, leading to an efficiency loss: it might be repaired by
using a sparse group Lasso 
\citep{friedetal10}. The difference between the clustering methods, 
Algorithm \ref{alg1} and standard hierarchical clustering based on
correlations (see Section \ref{sec.ohclust}), is essentially
nonexistent for the simulation models in Section \ref{simulationstudy} while
for the pseudo real data example in Section \ref{realdata}, the
disagreement is huge and our novel Algorithm \ref{alg1} leads to much
better results.  

\section{Conclusions}\label{sec.conclusions}

We consider estimation in a high-dimensional linear model with strongly
correlated variables. In such a setting, single variables
cannot (or are at least very difficult to) be identified. We propose to
group or cluster the 
variables first and do subsequent estimation with the Lasso for
cluster-representatives (CRL: cluster representative Lasso) or with the
group Lasso using the structure of the inferred clusters (CGL: Cluster
group Lasso). Regarding the first step, we 
present a new bottom-up agglomerative clustering algorithm which aims for
small canonical correlations between groups: we prove that it finds an
optimal solution, that it is statistically consistent, and we give a simple
rule for selecting the  
number of clusters. This new algorithm is motivated by the natural idea to
address the problem of almost linear dependence between variables, but if
preferred, it
can be replaced by another suitable clustering procedure. 

We present some theory which: (i) shows that canonical correlation based
clustering leads to a (much) improved compatibility constant for the
cluster group Lasso; and (ii) addresses bias and detection issues
when doing subsequent estimation on cluster representatives, e.g. as with
the CRL method. Regarding the second issue (ii), one favorable scenario is
for (nearly) uncorrelated clusters with potentially 
many active variables in a cluster: the bias due to working with
cluster representatives is small if the within group correlation is high,
and detection is good if the regression coefficients within a group do not
cancel. The other beneficial setting is for clusters with at most one active
variable per cluster but the between cluster correlation does not need to
be very small: if the cluster size is large or the correlation within the
clusters is large, the bias due to cluster representatives is small and
detection works well. We note that large cluster sizes cannot be properly
handled by the cluster group Lasso while they can be advantageous for the
cluster representative Lasso; instead of the group Lasso, one should take
for such cases a sparse group Lasso \citep{friedetal10} or a smoothed group Lasso
\citep{meieretal09}. Our theoretical analysis sheds light when and
why estimation with cluster representatives works well and leads to
improvements, in comparison to the plain Lasso. 

We complement the theoretical analysis with various empirical results which
confirm that the cluster Lasso methods (CRL and CGL) are particularly
attractive for improved variable screening in comparison to the plain
Lasso. In view of the fact that variable screening and dimension
reduction (in terms of the original variables) is one of the main
applications of Lasso in high-dimensional data analysis, the cluster 
Lasso methods are an attractive and often better alternative for this task.  

\section{Proofs}\label{sec.proofs}
 
\subsection{Proof of Theorem \ref{th1}}

We first show an obvious result. 
\begin{lemma}\label{lemm1}
Consider a partition ${\cal G} = \{G_1,\ldots ,G_q\}$ which satisfies
(\ref{crit.cc}). Then, for every $r,\ell \in \{1,\ldots ,q\}$ with $r \neq
\ell$: 
\begin{eqnarray*}
\hat{\rho}_{\mathrm{can}}(J_1,J_2) \le \tau\ \mbox{for all subsets}\ J_1
\subseteq G_r,\ J_2 \subseteq G_{\ell}.
\end{eqnarray*}
\end{lemma}
The proof follows immediately from the inequality
\begin{eqnarray*}
\hat{\rho}_{\mathrm{can}}(J_1,J_2) \le \hat{\rho}_{\mathrm{can}}(G_r,G_{\ell}).
\end{eqnarray*}
\hfill$\Box$ 

For proving Theorem \ref{th1}, the fact that we obtain a solution
satisfying (\ref{crit.cc}) is a 
straightforward consequence of the definition of the algorithm which
continues to merge groups until all canonical correlations between groups
are less or equal to $\tau$. 

We now prove that the obtained partition is the finest clustering
with $\tau$-separation. Let $\hat{\cal G}(\tau)=\{G_1,\ldots,G_q\}$ be an
arbitrary clustering with $\tau$-separation and $\hat{\cal G}_b,
b=1,\ldots,  b^*$, 
be the sequence of partitions generated by the algorithm (where $b^*$ is
the stopping (first) iteration where $\tau$-separation is reached). We need
to show that $\hat{\cal G}_{b^*}$ is a finer partition of $\{1,\ldots,p\}$ than
$\hat{\cal G}(\tau)$. Here, the meaning of ``finer than'' is not strict, i.e.,
including ``equal to''.   
To this end, it suffices to prove by induction that $\hat{\cal G}_{b}$ is finer
than $\hat{\cal G}(\tau)$ for $b=1,\ldots ,b^*$.  
This is certainly true for $b=1$ since the algorithm begins with the finest
partition of $\{1,\ldots,p\}$. Now assume the induction condition that
$\hat{\cal G}_{b}$ is finer than $\hat{\cal G}(\tau)$ for $b<b^*$. The algorithm 
computes $\hat{\cal G}_{b+1}$ by merging two members, say $G'$ and $G''$, of
${\cal G}_{b}$ such that  
$\hat{\rho}_{\mathrm{can}}(G',G'') > \tau$. Since $\hat{\cal G}_{b}$ is
finer than $\hat{\cal G}(\tau)$, there must exist members $G_{j_1}$ and  
$G_{j_2}$ of $\hat{\cal G}(\tau)$ such that $G'\subseteq G_{j_1}$ and
$G''\subseteq G_{j_2}$. This implies  
$\hat{\rho}_{\mathrm{can}}(G_{j_1},G_{j_2})\ge
\hat{\rho}_{\mathrm{can}}(G',G'') > \tau$, see Lemma 
\ref{lemm1}. Since
$\hat{\rho}_{\mathrm{can}}(G_j,G_k)\le \tau$ for all $j\neq k$, we must 
have $j_1=j_2$. Thus, the algorithm merges two subsets of a common member
(namely $G_{j_1} = G_{j_2}$) of $\hat{\cal G}(\tau)$.  
It follows that $\hat{\cal G}_{b+1}$ is still finer than $\hat{\cal
  G}(\tau)$. \hfill$\Box$  

\subsection{Proof of Theorem \ref{th-a}}

For ease of notation, we abbreviate a group index $G_r$ by $r$. 
The proof of Theorem \ref{th-a} is based on the following bound for the maximum 
difference between the sample and population correlations of linear
combination of variables. Define  
\begin{eqnarray*}
\Delta_{r,{\ell}} = \max_{u\neq 0,v\neq 0}\big|\hrho(\bx^{(r)}u,\bx^{(\ell)}v) - \rho(u^TX^{(r)},v^TX^{(\ell)})\big|. 
\end{eqnarray*}
\begin{lemma}\label{lm-consistency} 
Consider $\bx$ as in (\ref{Gaussian}) 
and ${\cal G}^0 = \{G_1,\ldots, G_q\}$ a partition of $\{1,\ldots,p\}$. 
Let $\Sigma_{r,{\ell}}=\Cov(X^{(r)},X^{(\ell)})$, $t>0$ and
$d_r=\rank(\Sigma_{r,r})$.   
Define $\Delta^*_{r,{\ell}}$ by (\ref{eq:a1}). Then, 
\begin{eqnarray*}
\PP[\max_{1\le j<k\le q} (\Delta_{r,{\ell}} - \Delta^*_{r,{\ell}}) \ge 0 ] \le \exp(-t). 
\end{eqnarray*} 
\end{lemma}
Proof of Lemma \ref{lm-consistency}. 
Taking a new coordinate system if necessary, 
we may assume without loss of generality that $\Sigma_{r,r} = I_{d_r}$ and $\Sigma_{{\ell},{\ell}}=I_{d_{\ell}}$. 
Let $\hSigma_{r,{\ell}}$ be the sample versions of $\Sigma_{r,{\ell}}$. 
For $r \neq \ell$, we write a linear regression model $\bx^{(\ell)} = \bx^{(r)}\Sigma_{r,{\ell}} 
+ \bep^{(r,{\ell})}V_{r,{\ell}}^{1/2}$ such 
that $\bep^{(r,{\ell})}$ is an $n \times d_{\ell}$ matrix of i.i.d.\,$N(0,1)$ entries independent of $\bx^{(r)}$ and 
$\|V_{r,{\ell}}^{1/2}\|_{(S)}\le 1$, where $\|\cdot\|_{(S)}$ is the spectrum norm. 
This gives 
$\hSigma_{r,{\ell}} = \hSigma_{r,r}\Sigma_{r,{\ell}} + (\bx^{(r)})^T\bep^{(r,{\ell})}V_{r,{\ell}}^{1/2}/n$. 
Let $U_rU_r^T$ be the SVD of the projection $\bx^{(r)}\hSigma_{r,r}^{-1}(\bx^{(r)})^T/n$, 
with $U_r\in \R^{n\times d_r}$. 
Since $\bx^{(r)}$ is independent of $\bep^{(r,{\ell})}$, $U_r^T\bep^{(r,{\ell})}$ is a $d_r\times d_{\ell}$ 
matrix of i.i.d. $N(0,1)$ entries. Let 
\begin{eqnarray*}
\Omega_r =\big\{\|\hSigma_{r,r}^{1/2} - I_{d_r}\|_{(S)}\le t_r\}, \ 
\Omega_{r,{\ell}} =\big\{\|U_r^T\bep^{(r,{\ell})}/\sqrt{n}\|_{(S)} \le t_r\wedge t_{\ell}\big\}. 
\end{eqnarray*}
It follows from (\ref{eq:a1}) and Theorem II.13 of \citet{davidson01} that
\begin{eqnarray*} 
\max\big\{\PP[\Omega_r^c], \PP[\Omega_{r,{\ell}}^c]\big\} 
\le 2 \PP [N(0,1) > \sqrt{2(t+\log(q(q+1)))}] \le 2e^{-t}/\{q(q+1)\}. 
\end{eqnarray*}
In the event $\Omega_r$, we have 
\begin{eqnarray*}
\Delta_{r,r} 
&=& \max_{\|u\|_2=\|v\|_2=1}\Big|\frac{u\hSigma_{r,r}v}
{\|\hSigma_{r,r}^{1/2}u\|_2\|\hSigma_{r,r}^{1/2}v\|_2}- u^Tv\Big|
\cr &\le& \|\hSigma_{r,r}-I_{d_r}\|_{(S)}/(1-t_r)^2
+ \max_{\|u\|_2=\|v\|_2=1}\Big|\frac{u^Tv}
{\|\hSigma_{r,r}^{1/2}u\|_2\|\hSigma_{r,r}^{1/2}v\|_2}- u^Tv\Big|
\cr &\le& \{(1+t_r)^2-1\}/(1-t_r)^2+1/(1-t_r)^2-1
\cr & = & \Delta^*_{r,r}. 
\end{eqnarray*}
For $r\neq \ell$, the variable change $u\to \hSigma_{r,r}^{-1/2}u$ gives 
\begin{eqnarray*}
\Delta_{r,{\ell}} = \max_{\|u\|_2=\|v\|_2=1}\Big|u^T\hSigma_{r,r}^{-1/2}\Big(\frac{\hSigma_{r,{\ell}}}{\|\hSigma_{\ell,\ell}^{1/2}v\|_2}
- \frac{\Sigma_{r,{\ell}}}{\|\hSigma_{r,r}^{-1/2}u\|_2}\Big)v\Big|. 
\end{eqnarray*}
In the event $\Omega_r\cap \Omega_{\ell}\cap\Omega_{r,{\ell}}$, we have 
\begin{eqnarray*}
\Delta_{r,{\ell}} &\le& \max_{\|u\|_2=\|v\|_2=1}
\Big\{\big|u\hSigma_{r,r}^{-1/2} (\hSigma_{r,{\ell}}-\Sigma_{r,{\ell}})v\big|/\|\hSigma_{\ell,\ell}^{1/2}v\|_2
\cr && \qquad + \Big|u^T\hSigma_{r,r}^{-1/2}\Sigma_{r,{\ell}}v(1/\|\hSigma_{\ell,\ell}^{1/2}v\|_2- 1/\|\hSigma_{r,r}^{-1/2}u\|_2)\Big|\Big\}
\cr &\le& \max_{\|u\|_2=\|v\|_2=1}
\Big\{\|\hSigma_{r,r}^{-1/2} (\hSigma_{r,{\ell}}-\Sigma_{r,{\ell}})\big\|_{(S)}/\|\hSigma_{\ell,\ell}^{1/2}v\|_2
 + \Big|\|\hSigma_{r,r}^{-1/2}u\|_2/\|\hSigma_{\ell,\ell}^{1/2}v\|_2- 1\Big|\Big\}
\cr &\le & \|(\hSigma_{r,r}^{1/2} - \hSigma_{r,r}^{-1/2})\Sigma_{r,{\ell}}
+ \hSigma_{r,r}^{-1/2}(\bx^{(r)})^T\bep^{(r,{\ell})}V_{r,{\ell}}^{1/2}/n\|_{(S)}/(1-t_{\ell})
\cr && + 1/\{(1-t_r)(1-t_{\ell})\}-1
\cr &\le & \big(\|\hSigma_{r,r}^{1/2} - \hSigma_{r,r}^{-1/2}\|_{(S)}
+ \|U_r^T\bep^{(r,{\ell})}/\sqrt{n}\|_{(S)}\big)/(1-t_{\ell}) + (t_r+t_{\ell})/\{(1-t_r)(1-t_{\ell})\}
\cr &\le& \{1/(1-t_r)-(1-t_r)+t_r\wedge t_{\ell}\}/(1-t_{\ell})+ (t_r+t_{\ell})/\{(1-t_r)(1-t_{\ell})\}
\cr &= & (1-(1-t_r)^2+(1-t_r)(t_r\wedge t_{\ell})+t_r+t_{\ell})/\{(1-t_r)(1-t_{\ell})\}. 
\end{eqnarray*}
Thus, for $t_r\le t_{\ell}$, we find $\Delta_{r,{\ell}}\le \Delta^*_{r,{\ell}}$. 
Since $\Delta_{r,{\ell}}=\Delta_{\ell,r}$, the above bounds hold simultaneously in the intersection of 
all $\Omega_r$ and those $\Omega_{r,{\ell}}$ with either $t_r<t_{\ell}$ or $r<\ell$ for $t_r=t_{\ell}$. 
Since there are totally $q+q(q-1)/2 = q(q+1)/2$ such events, 
$\PP[\Delta_{r,{\ell}}\le\Delta^*_{r,{\ell}}\ \forall r,{\ell}] \ge 1 -
\exp(-t)$. $\hfill\Box$  

\bigskip\noindent
{\bf Proof of Theorem \ref{th-a}.} 
It follows from (\ref{th-a-1}) that ${\cal G}^0$ is the finest population clustering with $\tau$-separation 
for all $\tau_-\le\tau\le\tau_+$. 
Since $\rho_{can}(G_r,G_{\ell}) = \max_{u,v}\rho(u^TX^{(r)},v^TX^{(\ell)})$ 
and $\hrho_{can}(G_r,G_{\ell}) = \max_{u,v}\hrho(\bx^{(r)}u,\bx^{(\ell)}v)$, 
(\ref{th-a-1}) and Lemma \ref{lm-consistency} implies that with at least probability $1-e^{-t}$, 
the inequalities 
\begin{eqnarray*}
&& \hrho_{can}(G_r,G_{\ell})\le \rho_{can}(G_r,G_{\ell})+\Delta^*_{r,{\ell}} \le\tau_-, 
\cr && \max_{k_1<k_2}\hrho_{can}(G_{r;k_1},G_{r;k_2}) 
\ge \max_{k_1<k_2}\rho_{can}(G_{r;k_1},G_{r;k_2}) - \Delta^*_{r,r} \ge \tau_+,
\end{eqnarray*}
hold simultaneously for all $r<\ell$ and nontrivial partitions $\{G_{r;k}, k\le q_r\}$ of $G_r, r=1,\ldots,q$. 
In this case, ${\cal G}^0$ is also the finest sample clustering with $\tau$-separation 
for all $\tau_-\le\tau\le\tau_+$. The conclusion follows from Theorem
\ref{th1}.\hfill$\Box$ 

\subsection{Proof of Theorem \ref{groupcompatibility.lemma}}

We write for notational simplicity $S_0 := S_{0, {\rm Group}}$ and $s_0 :=
| S_{0, {\rm Group}} |$. Note first that for all $r$,   
$$\| \bx^{(G_r)} \beta_{G_r} \|_2^2 = n \beta_{G_r}^T \hat \Sigma_{r,r} \beta_{G_r}  = n \|\gamma_{G_r} \|_2^2 , $$
where $\gamma_{G_r} := \Sigma_{r,r}^{1/2} \beta_{G_r}  $.
Moreover, 
$$\| \bx^{(S_0)} \beta_{S_0} \|_2^2 = n \gamma_{S_0}^T \hat R_{S_0} \gamma_{S_0} . $$
It follows that
\begin{equation}\label{eigenvalue.equation}
\sum_{r \in S_0} \|\bx^{(G_r)} \beta_{G_r} \|_2^2 =n  \| \gamma_{S_0} \|_2^2 $$ $$ \le n \gamma_{S_0}^T \hat R_{S_0} \gamma_{S_0} / \hat \Lambda_{\rm min}^2
= \|X^{(S_0)} \beta_{S_0} \|_2^2 / \hat \Lambda_{\rm min}^2. 
\end{equation}

Furthermore,  
 $$  \biggl  | \beta_{S_0}^T (\bx^{(S_0)})^T \bx^{(S_0^c)} \beta_{S_0^c} \biggr |=  n\biggl  | \sum_{r \in S_0} \sum_{\ell \in S_0^c}
 \beta_{G_r} ^T \hat \Sigma_{r,\ell} \beta_{G_\ell}^T \biggr | $$
 $$ =n \biggl | \sum_{r \in S_0} \sum_{\ell \in S_0^c} \gamma_{G_r}^T \hat \Sigma_{r,r}^{-1/2} \hat \Sigma_{r,\ell} \hat \Sigma_{\ell,\ell}^{-1/2}
 \gamma_{G_\ell}  \biggr | \le n \sum_{r \in S_0} \sum_{\ell \in S_0^c}  \hat \rho_{\rm can} (G_r, G_\ell) 
 \| \gamma_{G_r} \|_2 \|\gamma_{G_\ell} \|_2   $$
 $$ =  \sum_{r \in S_0} \sum_{\ell \in S_0^c}  \hat \rho_{\rm can} (G_r, G_\ell) 
 \|\bx^{(G_r)} \beta_{G_r} \|_2 \|\bx^{(G_\ell)} \beta_{G_\ell}  \|_2   $$ $$=  \sum_{r \in S_0} \sum_{\ell \in S_0^c}  { \hat \rho_{\rm can} (G_r, G_\ell) } {\bar m \over
 \sqrt {m_r m_\ell} } 
 \|\bx^{(G_r)} \beta_{G_r} \|_2 \sqrt {m_r \over  \bar m }  \|\bx^{(G_\ell)} \beta_{G_\ell} \|_2 \sqrt {m_\ell \over \bar m } 
 $$
$$ \le \rho \| \bx^{(S_0)} \beta_{S_0} \|_{2,1} \| \bx^{(S_0^c)} \beta_{S_0^c} \|_{2,1} . $$
Hence, for all $\beta$ satisfying $\| \beta_{S_0^c} \|_{2,1} \le 3 \| \beta_{S_0} \|_{2,1} $, we have
\begin{equation}\label{inproduct.equation}
  \biggl  | \beta_{S_0}^T (\bx^{(S_0)})^T \bx^{(S_0^c)} \beta_{S_0^c} \biggr | \le 3 \rho \| \beta_{S_0} \|_{2,1}^2 .
  \end{equation}

Applying the Cauchy-Schwarz inequality and (\ref{eigenvalue.equation}) gives
$$
 \| \beta_{S_0}  \|_{2,1}^2  := \biggl ( \sum_{r \in S_0} \|\bx^{(G_r)} \beta_{G_r} \|_2  \sqrt {m_r/ \bar m}  \biggr )^2 
$$
$$ \le { \sum_{r \in S_0} m_r \over \bar m } \sum_{r \in S_0} \|\bx^{(G_r)} \beta_{G_r} \|_2^2 \le { \sum_{r \in S_0} m_r \over \bar m } 
{ \| \bx^{(S_0)} \beta_{S_0} \|_2^2 \over  \hat \Lambda_{\rm min}^2 } = {s_0 \bar m_{S_0} \over \bar m }{ \| \bx^{(S_0)} \beta_{S_0} \|_2^2 \over \hat  \Lambda_{\rm min}^2 }. 
 $$
 Insert this in (\ref{inproduct.equation}) to get
  $$ \biggl  | \beta_{S_0}^T (\bx^{(S_0)})^T \bx^{(S_0^c)} \beta_{S_0^c}
  \biggr | \le  \rho {3 s_0 \bar m_{S_0} \over \bar m \hat \Lambda_{\rm
      min}^2 } \| \bx^{(S_0)} \beta_{S_0} \|_2^2. $$ 
  Use the assumption that
$$ \rho {3 s_0 \bar m_{S_0} \over \bar m \hat \Lambda_{\rm min}^2 }  < 1 ,$$
and apply Lemma 6.26 in \cite{pbvdg11} to conclude that
$$\| \bx^{(S_0)} \beta_{S_0}\|_2^2 \le  \biggl (1- \rho {3 s_0 \bar m_{S_0} \over \bar m \hat \Lambda_{\rm min}^2 } \biggr )^{-2} \|\bx \beta \|_2^2 . $$

Hence,  
$$
 \| \beta_{S_0}  \|_{2,1}^2 \le 
 {s_0 \bar m_{S_0} \over \bar m }{ \| \bx^{(S_0)} \beta_{S_0} \|_2^2 \over
   \hat  \Lambda_{\rm min}^2 }  
 $$ $$ \le\biggl (1- \rho {3 s_0 \bar m_{S_0} \over \bar m \hat
   \Lambda_{\rm min}^2 } \biggr )^{-2} 
 \biggl ( {s_0 \bar m_{S_0} \over \bar m } \biggr ) { \| \bx \beta \|_2^2
   \over \hat  \Lambda_{\rm min}^2 }  . $$ 
This leads to the first lower bound in the statement of the theorem. The
second lower bound follows immediately by the incoherence assumption for
$\rho$. Furthermore, it is not difficult to see that $\hat \Lambda_{\rm
  min}^2 \ge (1- | S_{0, {\rm Group}}| \rho_{S_{0, {\rm Group}}} )$, and using
the incoherence assumption for $\rho_{S_{0, {\rm Group}}}$ leads to strict
positivity.\hfill$\Box$

\subsection{Proof of Proposition \ref{prop0}}
We can invoke the analysis given in \citet[Th.6.1]{pbvdg11}. The
slight deviations involve: (i) use that $\EE[\eta_i|\bz] = 0$; (ii): due to the
Gaussian assumption 
$\Var(\eta_i|\bz)$ is constant equaling $\Var(\eta_i|\bz) =
\EE[\Var(\eta_i|\bz)] = \Var(\eta_i) = \xi^2 = \sigma^2 + \EE[(\mu_X -
\mu_Z)^2]$; and (iii): the probability bound in \citet[Lem.6.2]{pbvdg11}
can be easily obtained for non-standardized variables when multiplying
$\lambda_0$ with $\|\hat{\sigma}_{\bz}\|_{\infty}$. The issues (ii)
and (iii) explain the factors appearing in the definition of
$\lambda_0$.\hfill$\Box$ 

\subsection{Proof of Proposition \ref{prop2}}

Because of uncorrelatedness of $\bar{X}^{(r)}$ among $r = 1,\ldots ,q$, we
have: 
\begin{eqnarray*}
\gamma^0_r &=& \frac{\Cov(Y,\bar{X}^{(r)})}{\Var(\bar{X}^{(r)})} =
|G_r|^{-1} \frac{\sum_{j \in G_r} \Cov(Y,X^{(j)})}{\Var(\bar{X}^{(r)})}\\
&=& |G_r|\frac{\sum_{j \in G_r} \beta_j^0 \Var(X^{(j)}) + \sum_{i \neq j} \beta^0_i
  \Cov(X^{(i)},X^{(j)})}{\sum_{\ell \in G_r}(\Var(X^{(\ell)}) + \sum_{i \neq
    \ell} \Cov(X^{(i)},X^{(\ell)}))}.
\end{eqnarray*}
Define 
\begin{eqnarray}\label{weights}
w_j = \frac{\Var(X^{(j)}) + \sum_{i \neq j} \Cov(X^{(i)},X^{(j)})}{\sum_{\ell \in G_r}(\Var(X^{(\ell)}) + \sum_{i \neq
    \ell} \Cov(X^{(i)},X^{(\ell)}))}.
\end{eqnarray}
Then, $\sum_{j \in G_r} w_j =1$ and $\gamma^0_r = \sum_{j \in G_r} w_j
\beta^0_j$.

The statement 1 follows immediately from (\ref{weights}). Regarding
statement 2, we read off from (\ref{weights}) that 
$w_j \equiv |G_r|^{-1}$ for all $j \in G_r$ and hence $\gamma^0_r = \sum_{j \in
  G_r} \beta_j^0$.\hfill$\Box$

\subsection{Proof of Proposition \ref{prop3}}
We have
\begin{eqnarray*}
\gamma^0_r = \frac{\Cov(Y,\bar{X}^{(r)}|\{\bar{X}^{(\ell)};\ \ell \neq
  r\})}{\Var(\bar{X}^{(r)}|\{\bar{X}^{(\ell)};\ \ell \neq r\})},
\end{eqnarray*}
since for Gaussian distributions, partial covariances equal conditional
covariances \citep[cf.]{baba04}. For the numerator, we have:
\begin{eqnarray*}
& &\Cov(Y,\bar{X}^{(r)}|\{\bar{X}^{(\ell)};\ \ell \neq r\}) = |G_r|^{-1} \sum_{j \in
  G_r} \Cov(Y,X^{(j)}|\{\bar{X}^{(\ell)};\ \ell \neq r\})\\
&=&|G_r|^{-1} \sum_{i,j \in G_r} \beta^0_i
\Cov(X^{(i)},X^{(j)}|\{\bar{X}^{(\ell)};\ \ell \neq r\})\\
&+& |G_r|^{-1} \sum_{j \in
  G_r} \sum_{i \notin G_r} \beta^0_i \Cov(X^{(i)},X^{(j)}|\{\bar{X}^{(\ell)};\
\ell \neq r\})\\
&=&|G_r|^{-1} \sum_{i,j \in G_r} \beta^0_i
\Cov(X^{(i)},X^{(j)}|\{\bar{X}^{(\ell)};\ \ell \neq r\}) + \Gamma_r,\ |\Gamma_r|
\le \|\beta^0\|_1 \nu.
\end{eqnarray*}
for the denominator, we have:
\begin{eqnarray*}
\Var(\bar{X}^{(r)}|\{\bar{X}^{(\ell)};\ \ell \neq r\}) = |G_r|^{-2} \sum_{i,j \in
  G_r} \Cov(X^{(i)},X^{(j)}|\{\bar{X}^{(\ell)};\ \ell \neq r\}).
\end{eqnarray*}
Defining 
\begin{eqnarray}\label{weights2}
w_j = \frac{\sum_{i \in G_r} \Cov(X^{(i)},X^{(j)}|\{\bar{X}^{(\ell)};\ \ell \neq r\})}{\sum_{i,\ell \in
  G_r} \Cov(X^{(i)},X^{(\ell)}|\{\bar{X}^{(\ell)};\ \ell \neq r\})},
\end{eqnarray}
we obtain $\gamma^0_r = |G_r| \sum_{j \in G_r} w_j \beta^0_j + \Delta_r$ with
$|\Delta_r \le \|\beta^0\|_1\nu/C$. The other statement follows using
(\ref{weights2}) and as statement 1. in Proposition \ref{prop2}.\hfill$\Box$

\subsection{Proof of Proposition \ref{prop0a}}
Write 
\begin{eqnarray*}
Y = \bar{X}^T \gamma^0 + \eta = U^T\tilde{\beta}^0 + \eps = \bar{X}^T
\tilde{\beta}^0 - W \tilde{\beta}^0 + \eps.
\end{eqnarray*}
Therefore,
\begin{eqnarray*}
\bar{X}(\tilde{\beta}^0 - \gamma^0) = \eta - \eps + W^T \tilde{\beta}^0.
\end{eqnarray*}
Taking the squares and expectation on both sides, 
\begin{eqnarray*}
(\tilde{\beta}^0 - \gamma^0)^T \Cov(\bar{X}) (\tilde{\beta}^0 - \gamma^0) = \EE[(
\eta - \eps)^2] +  \EE|W^T \tilde{\beta}^0|^2 = B^2 + \EE|W^T
\tilde{\beta}^0|^2 \le 2 \EE|W^T \tilde{\beta}^0|^2,
\end{eqnarray*}
where the last inequality follows from (\ref{bias2}). Since $\Cov(\bar{X})
= \Cov(U) + \Cov(W)$, we have that $\lambda_{\mathrm{min}}^2(\Cov(\bar{X}))
\ge \lambda_{\mathrm{min}}^2(\Cov(U))$. This completes the
proof.\hfill$\Box$ 

\bibliographystyle{apalike} 
\bibliography{references}

\end{document}